\documentclass[twocolumn]{aa}  

\usepackage{graphicx}
\usepackage{float}
\usepackage{txfonts}
\usepackage{longtable}
\usepackage{supertabular}
\usepackage{booktabs}
\usepackage{lscape}
\usepackage{hhline}
\usepackage{natbib}
\bibpunct{(}{)}{;}{a}{}{,}
\usepackage{color}
\usepackage{adjustbox}

\newcommand{\ngc}  {NGC\,6334}
\newcommand{\kms}  {km~s$^{-1}$}
\newcommand{\et}   {et al.}
\newcommand{\jpb}  {Jy~beam$^{-1}$}
\newcommand{\hii}  {\ion{H}{ii}}
\newcommand{\mo}   {$M_\odot$}

\newcommand{\phn}  {\phantom{0}}
\newcommand{\phnn} {\phantom{00}}

\begin{document} 

   \title{Physical properties of the star-forming clusters in \ngc}

   \subtitle{A study of the continuum dust emission with ALMA}

   \author{M.~Sadaghiani
          \inst{1}
          \and \'A.~S\'anchez-Monge \inst{1}
          \and P.~Schilke \inst{1}
          \and H.~B.~Liu \inst{2}
          \and S.~D.~Clarke \inst{1}
          \and Q.~Zhang \inst{3}
          \and J.~M.~Girart \inst{4,5}
          \and D.~Seifried \inst{1}
          \and A.~Aghababaei \inst{1}
          \and H.~Li \inst{6}
          \and C.~Ju\'arez \inst{4}
          \and K.~S.~Tang \inst{7}}

   \institute{I. Physikalisches Institut, Universit{\"a}t zu K\"oln, Z\"ulpicher Str. 77, D-50937 K\"oln, Germany\\      
              \email{sadaghiani@ph1.uni-koeln.de}
              \and Academia Sinica Institute of Astronomy and Astrophysics, 11F of Astro-Math Bldg, 1, Sec. 4, Roosevelt Rd, Taipei 10617, Taiwan
              \and Harvard-Smithsonian center for Astrophysics, 60 Garden St., Cambridge, MA 02138 USA
              \and Institut de Ci\`encies de l'Espai, (CSIC), Campus UAB, Carrer de Can Magrans S/N, E-08193 Cerdanyola del Vall\`es, Catalonia
              \and Institut d'Estudis Espacials de Catalunya (IEEC), E-08034 Barcelona, Catalonia
              \and The Chinese University of Hong Kong, Shatin, N.T., Hong Kong, China 
              \and University of Illinois, 1002 W Green St, Urbana, IL 61801, USA}

   \date{Received ; accepted }

 
\abstract
{}
{We aim to characterise certain physical properties of high-mass star-forming sites in the \ngc\ molecular cloud, such as the core mass function (CMF), spatial distribution of cores, and mass segregation. }
{We used the Atacama Large Millimetre/sub-millimetre Array (ALMA) to image the embedded clusters \ngc-I and \ngc-I(N) in the continuum emission at 87.6~GHz. We achieved a spatial resolution of 1300~au, enough to resolve different compact cores and fragments, and to study the properties of the clusters.}
{We detected 142 compact sources distributed over the whole surveyed area. The ALMA compact sources are clustered in different regions. We used different machine-learning algorithms to identify four main clusters: \ngc-I, \ngc-I(N), \ngc-I(NW), and \ngc-E. The typical separations between cluster members range from 4\,000 au to 12\,000~au. These separations, together with the core masses (0.1--100 \mo), are in agreement with the fragmentation being controlled by turbulence at scales of 0.1 pc. We find that the CMFs show an apparent excess of high-mass cores compared to the stellar Initial Mass Function. We evaluated the effects of temperature and unresolved multiplicity on the derived slope of the CMF. Based on this, we conclude that the excess of high-mass cores might be spurious and due to inaccurate temperature determinations and/or resolution limitations. We searched for evidence of mass segregation in the clusters and we find that clusters \ngc-I and \ngc-I(N) show hints of segregation with the most massive cores located in the centre of the clusters. }
{We searched for correlations between the physical properties of the four embedded clusters and their evolutionary stage (based on the presence of \hii\ regions and infrared sources). \ngc-E appears as the most evolved cluster, already harboring a well-developed \hii\ region. \ngc-I is the second-most evolved cluster with an ultra-compact \hii\ region. \ngc-I(N) contains the largest population of dust cores distributed in two filamentary structures and no dominant \hii\ region. Finally, \ngc-I(NW) is a cluster of mainly low-mass dust cores with no clear signs of massive cores or \hii\ regions. We find a larger separation between cluster members in the more evolved clusters favoring the role of gas expulsion and stellar ejection with evolution. The mass segregation, seen in the \ngc-I and \ngc-I(N) clusters, suggests a primordial origin for \ngc-I(N). In contrast, the segregation in \ngc-I might be due to dynamical effects. Finally, the lack of massive cores in the most evolved cluster suggests that the gas reservoir is already exhausted, while the less evolved clusters still have a large gas reservoir along with the presence of massive cores. In general, the fragmentation process of \ngc\ at large scales (from filament to clump, i.e. at about 1~pc) is likely governed by turbulent pressure, while at smaller scales (scale of cores and sub-fragments, i.e. a few hundred~au) thermal pressure starts to be more significant.}

\keywords{ISM: star formation -- stars: massive -- radio continuum: ISM -- ISM: clouds -- ISM: individual objects: \ngc-I, \ngc-I(N)}

\maketitle
%

%
\section{Introduction}\label{sec:introduction}

Most stars in our Galaxy do not form in isolation. Instead, they tend to form within clusters containing high-mass stars and to be embedded in molecular clouds \citep[e.g.][]{McKee2007, Bressert2012, Tan2014, Rathborne2015}. In this picture, massive embedded clusters (i.e. protoclusters), with gas and dust masses larger than 10$^3$ \mo~and $\mathrm{H}_2$ densities exceeding 10$^4$~cm$^{-3}$, are believed to be fundamental units of star formation \citep[e.g.][]{Lada2003}.
Massive young protoclusters are the primary laboratories for constraining the physical processes of the formation process of (high-mass) stars.
These structures, also called massive clumps, have typical sizes of 1~pc, and they are formed via the fragmentation of the parent cloud during its gravitational collapse. As evolution proceeds, these clumps fragment to form clusters with several hundreds or thousands of members. The cluster members present a rich diversity, from cold dense cores, to chemically-rich, hot molecular cores and ultracompact \hii~regions \citep[see][for a review]{Tan2014}. Investigating the physical nature of clusters and their environment based on observational constraints leads us to construct an empirical foundation upon which theoretical models of star formation are established \citep[e.g.][]{Klessen2004, Lee2016}. In order to further explore the properties of massive protoclusters, we have conducted detailed ALMA observations of the nearby star-forming region \ngc.\\
\indent \ngc\ is a high-mass star-forming complex that lies in the Carina-Sagittarius arm within the Galactic plane at a relatively nearby distance of 1.3$\pm$0.3~kpc \citep{Chibueze2014}. \ngc\ is located in the southern sky as a 100~pc giant filamentary cloud that has been the target of multiple studies at different wavelengths \citep[e.g.][]{Persi2008, Russeil2013}. Radio and infrared observations reveal several active star-forming sites at various stages of evolution distributed along the densest inner part of the filament. The first five main regions that have been identified in the infrared received the names \ngc-I to \ngc-V \citep[see][]{Persi2008}. All the regions but one are associated with \hii regions \citep[e.g.][]{Rodriguez1982}. In addition to the five main sources, a number of dense condensations have been further identified along the filament. One of the more prominent of these sources is located to the north of \ngc-I and is named \ngc-I(N). In the past few years, a number of works have focused on the study of the properties of the filaments, as well as on individual star-forming regions. At large scales, the physical properties of the whole molecular cloud have been studied via \textit{Herschel} continuum observations by \citet{Russeil2013}. The study of the velocity field and dynamic status of the cloud have shown collapse along the main axis of the filament as well as mass accretion rates on the level of a few 10$^{-5}$ to 10$^{-3}$ \mo~yr$^{-1}$ \citep[see e.g.][]{Zernickel2013, Sanchez2015}. The magnetic field in the region has been studied via polarimetric observations, revealing a magnetic field orientation perpendicular to the long axis of the main filament \citep[e.g.][]{Zhang2014,Li2015,Juarez2017}.\\
\indent In this paper, we focus on the study of two of the most prominent star-forming regions in the filamentary cloud: \ngc-I and \ngc-I(N). The region \ngc-I, located in the northeastern part of the inner and denser filament, contains a cluster of stars visible in near-infrared \citep{Tapia1996, Seifahrt2008} and a cometary ultracompact \hii\ region \citep[also known as \ngc-F;][]{depree1995, Carral2002}. Submillimeter Array (SMA) observations at 1.3~mm with an angular resolution of 2\arcsec\ reveal a cluster of four sources in a Trapezium-like arrangement \citep{Hunter2006}, of which two are classified as hot cores due to their rich chemistry and high temperatures of above 100~$\mathrm{K}$ (MM1 and MM2). In addition to the previously known sources, with 1.3~mm ALMA observations at 0.17\arcsec, 5 new millimetre sources have been detected by \citet{Brogan2016}. Furthermore, the observations reveal that MM1 is resolved into seven components concentrated within a radius of 1\,000~au. The second hot core, MM2, shows a lower level of multiplicity, being resolved into two components. \\
\indent The embedded source \ngc-I(N), located to the north of \ngc-I, contains a group of Class~I and Class~II methanol masers \citep{Walsh1998}. SMA observations at 1.3~mm with a resolution of 2\arcsec\ reveal a bright millimetre source which is composed of four components \citep{Brogan2009}. \citet{Hunter2014} report new observations with the SMA and the VLA at angular resolutions of 0.5\arcsec which reveal a richer cluster with a total of 25 members. All components have spectral energy distributions (SEDs) consistent with dust, except for one which is associated with an \hii\ region. The detection of line emission from molecules such as as CH$_3$CN allows for the identification of a number of hot cores in the central part of the cluster.\\
\indent In this paper, we present new high-angular resolution ALMA observations at 3~mm which reveal a large population of new millimetre sources in \ngc-I and \ngc-I(N). Our results suggest that the extent of the clusters is larger than previously thought. In \S~\ref{sec:observations}, we describe our ALMA observations and the process of data reduction. In \S~\ref{sec:results}, we illustrate the morphology of the continuum emission toward the observed region and identify the compact sources as well as their counterparts at infrared, radio, and X-ray wavelengths. 
In \S~\ref{sec:clusters}, we study the distribution of the dense cores throughout the observed region and identify four well-defined clusters. In \S~\ref{sec:physical-properties}, the physical properties of the clusters, such as the core mass function and the level of mass segregation, are derived.
In \S~\ref{sec:discussion}, we discuss the implication of our findings relevant to cluster formation. Finally, in \S~\ref{sec:summary} we summarise the main results and present the conclusions.

\begin{figure*}[ht!]
\begin{center}
\includegraphics[width=1.009\textwidth]{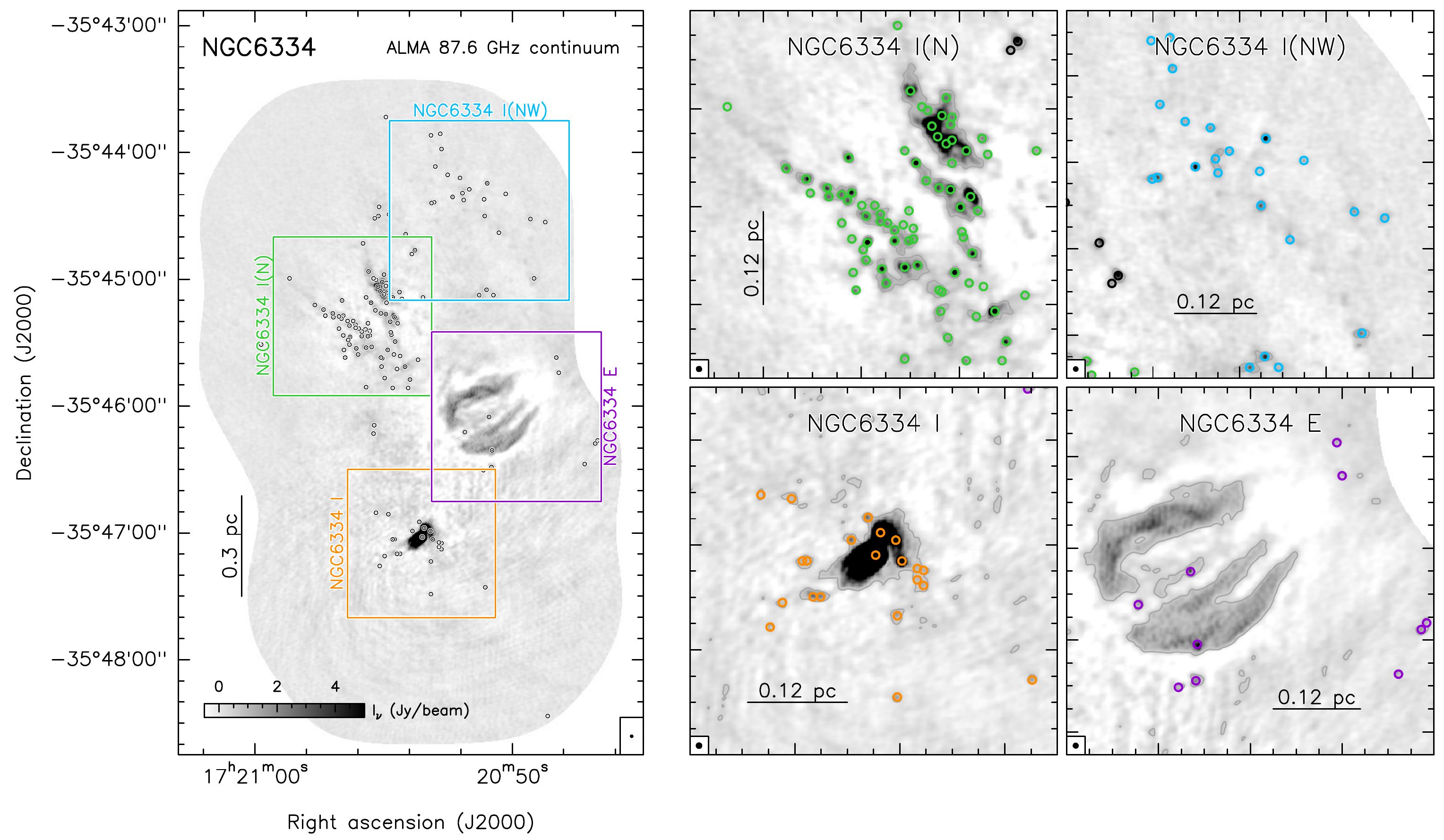}
\caption{\textit{Left}: ALMA 87.6~GHz continuum emission map of \ngc. The colored rectangles mark relevant regions studied in more detail throughout the paper. Their names, from north to south, are \ngc-I(NW), \ngc-I(N), \ngc-E and \ngc-I.
\textit{Right}: Close-up views of four selected regions. In all panels, the circles denote the position of the compact sources identified using SExtractor (see \S~\ref{sec:source-identification}, and Table~\ref{tab:catalog}). The color of the circles in the right panels indicates the association of each source with a cluster (see \S~\ref{sec:cluster-identification}). The synthesised beam of 1\arcsec\ is shown in the bottom left or right corners of each panel.}
\end{center}
\label{fig:alma3mm}
\end{figure*}

%
\section{ALMA observations and data reduction}\label{sec:observations}

\ngc-I and \ngc-I(N) were observed with ALMA (Atacama Large Millimetre/Sub-millimetre Array; \citealt{alma2015}) during its cycle 3 from April 2016 to September 2016, under project number 2015.1.00230.S. We used the main array (i.e. 12m-size antennas) consisting of 40-43 antennas, and the Atacama Compact Array (ACA) with up to 8--9 7m-size antennas. We covered an area of $120\arcsec\times120\arcsec$ around each source, using 17- and 7-pointing mosaics for the main array and ACA, respectively. The centres of the mosaics were set to $\alpha$(J2000)=$17^\mathrm{h}20^\mathrm{m}53.\!\!^\mathrm{s}0$, $\delta$(J2000)=$-35^{\circ}46^{\prime}59.\!\!^{\prime\prime}0$ for \ngc-I, and $\alpha$(J2000)=$17^\mathrm{h}20^\mathrm{m}54.\!\!^\mathrm{s}0$, $\delta$(J2000)=$-35^{\circ}45^{\prime}06.\!\!^{\prime\prime}0$ for \ngc-I(N). This coverage results in a small overlap between the two mosaics.\\
\indent The ALMA correlator was configured to cover specific frequency ranges within the band~3 of ALMA. A broad spectral window (total bandwidth of 1875~MHz) centred at the frequency 87.6~GHz was used to be sensitive to the continuum emission. Twelve additional units with a bandwidth of 58.59~MHz and a spectral resolution of 141~kHz (or 0.4~\kms) were tuned at the frequencies of specific molecular transitions of dense gas tracers (e.g. HCN, HCO$^+$, CH$_3$CCH), ionised gas tracers (e.g. H42$\alpha$) and shock tracers (e.g. SiO). In the current paper, we present the results obtained from the analysis of the continuum emission. The analysis of the molecular lines will be presented in a forthcoming paper.\\
\indent The ALMA data were calibrated using the ALMA calibration pipeline available in CASA version~4.7.2. Flux calibration was obtained through observations of the bright quasar J1733$-$1304 (with a flux of 3.63~Jy at 91.5~GHz, and a spectral index of $-0.53$) and the planet Neptune. The gains were calibrated by interleaved observations of the quasar J1717$-$3342 (flux 0.639~Jy at 91.5~GHz, with spectral index $-0.32$). The gain calibrators were observed every eight minutes. The bandpass response was obtained by observing the bright quasars J1617$-$5848 and J1924$-$2914. After the calibration was applied, the line-free channels of the broad spectral window were identified and used to create the continuum images. The continuum emission in source \ngc-I is bright enough to self-calibrate the data. Three iterations of phase-only self-calibration on the ALMA-12m data led to a considerably improved image. The \texttt{CLEAN}ing process was done with the task \texttt{TCLEAN} using multi-scale synthesis, and setting the robust parameter of Briggs equal to 0.5 as compromise between resolution and sensitivity. The resulting images were restored with a synthesised beam of $1\farcs6\times1\farcs2$ (original beam size), and a circular beam of $1\farcs0$ (hereafter super-resolution image) equivalent to a linear resolution of 1300~au. The super-resolution image is better suited to identify and distinguish continuum compact sources in crowded environments.\\
\indent In order to recover spatial scales that are filtered out in the 12m-ALMA dataset, we combined the 12m-ALMA data with the ACA data. In the first step, the ACA-only continuum image was used as a model in the \texttt{CLEAN}ing process of the 12m-ALMA data, following the so-called model-assisted imaging method. This allows for a better convergence and results in an image with fewer artifacts. In a second step, the ACA-only and the 12m-ALMA continuum images were combined using the \texttt{FEATHER} task within CASA. The final continuum images have a RMS noise level that varies from 0.06~m\jpb\ to 0.3~m\jpb\ in signal-free regions and regions close to the brightest sources, respectively. The dynamic range of the image, measured as the peak brightness in \ngc-I and I(N) to the RMS noise in the noisy region of the combined image is 1100. The combined image is used to study the extended structure of the continuum emission.

%
\section{Results}\label{sec:results}

In the left panel of Fig.~\ref{fig:alma3mm}, we show the ALMA 87.6~GHz (or 3~mm) continuum emission towards \ngc-I and \ngc-I(N). The emission is dominated by bright sources distributed over the whole surveyed region. The brightest source is located towards the south and coincides with the cometary, ultracompact \hii\ region of \ngc-I \citep[e.g.][]{depree1995, Carral2002}. About $60\arcsec$ to the north-northwest, we identify an extended shell-like feature that corresponds to the more evolved compact \hii\ region \ngc-E \citep[e.g.][]{Carral2002, Rodriguez2003}. In the north-east direction, and about $100\arcsec$ north of \ngc-I, we identify a number of compact sources at the location of \ngc-I(N) distributed along two elongated structures, reminiscent of filaments. Finally, a number of fainter, compact sources appear distributed over the whole region, with a larger concentration to the north-west of \ngc-I(N).\\
\indent In the following, we present a catalogue of ALMA continuum compact sources, we search for counterparts of these sources at other wavelengths, and report on physical properties such as mass and density.

\begin{figure*}[h!]
\begin{center}
\includegraphics[scale=0.3]{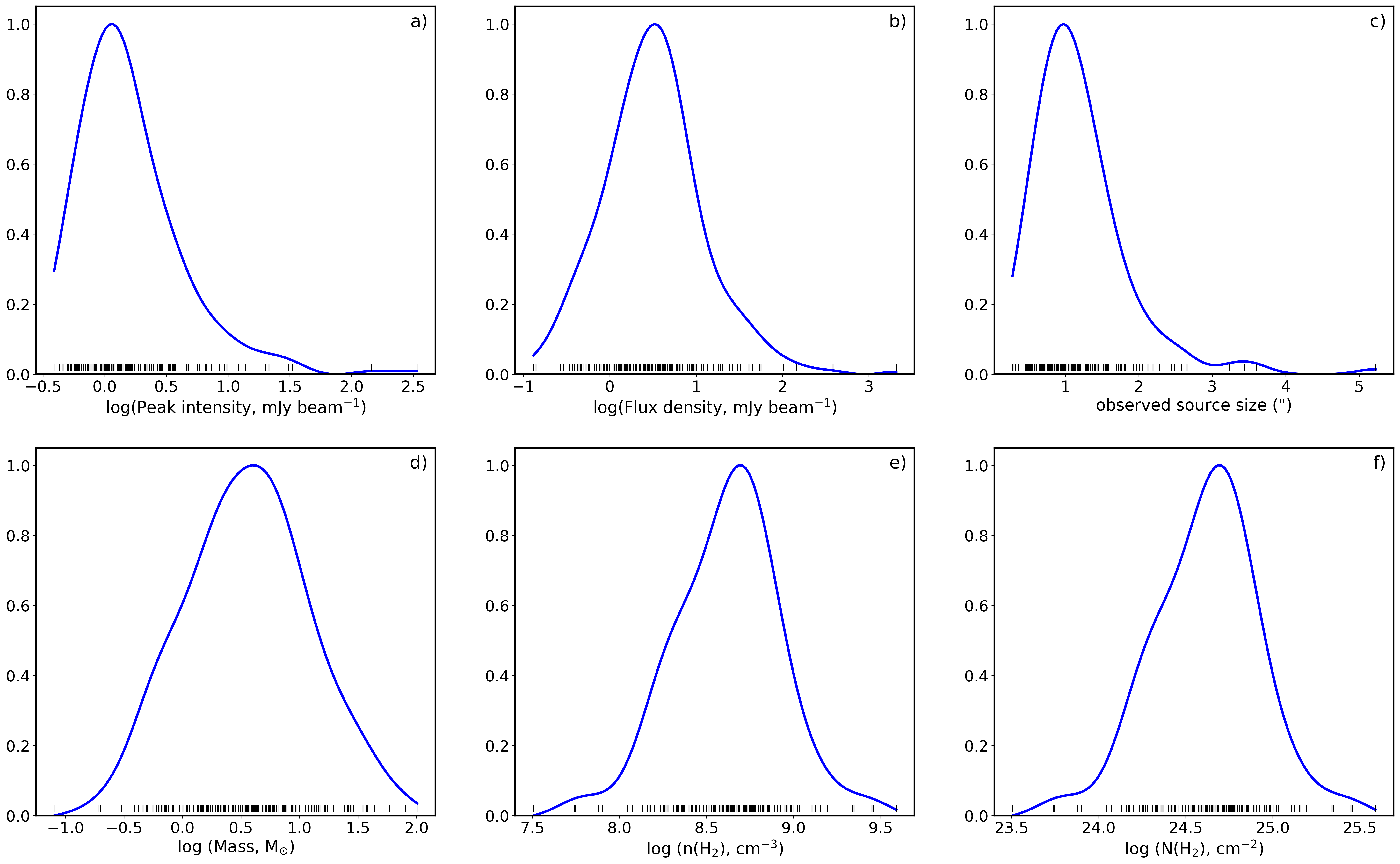}
\caption{Distribution of (\textit{a}) peak intensity; (\textit{b}) flux density; (\textit{c}) observed size; (\textit{d}) dust and gas mass; (\textit{e}) H$_2$ volume density; and (\textit{f}) H$_2$ column density for the 142 ALMA continuum sources detected toward \ngc. The dark-blue solid lines correspond to the KDE (kernel density estimate) built from the observed properties which are marked in the bottom of each panel, just above the x-axis.}
\label{fig:distribution}
\end{center}
\end{figure*}

\begin{figure*}[h!]
\begin{center}
\includegraphics[scale=0.57]{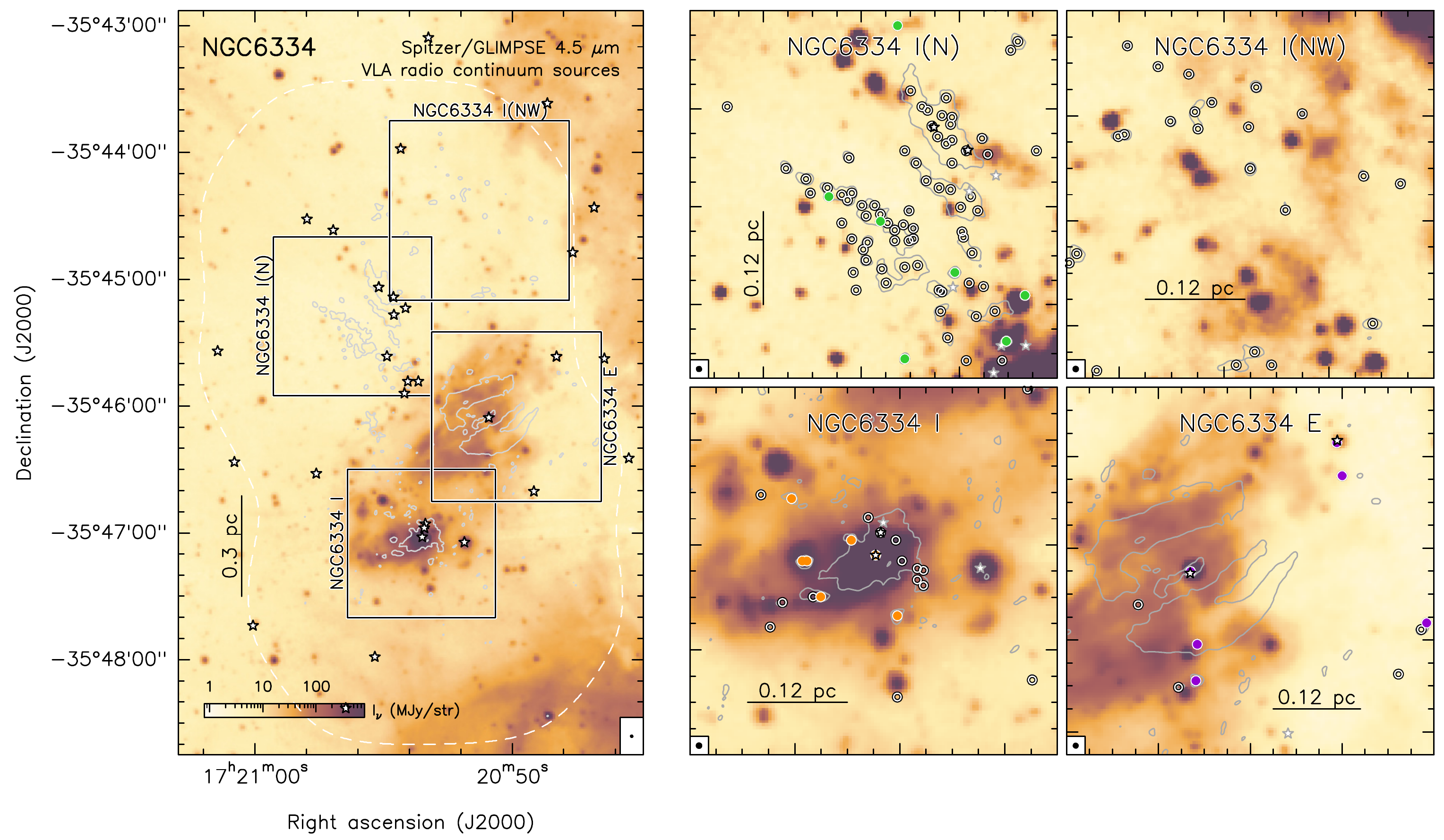}
\caption{\textit{Left}: \textit{Spitzer} 4.5~$\mathrm{\mu m}$ image in color and ALMA continuum emission at 87.6~GHz in grey contours. The contour level corresponds to 0.5 Jy beam$^{-1}$. The radio continuum sources observed with VLA at 4--8~GHz \citep{Medina2018} are shown as black stars. The rectangles mark the position of the relevant regions also marked in Fig~\ref{fig:alma3mm}.
\textit{Right}: Close-up view of the four regions marked with rectangles in the left panel. The circles indicate the position of the ALMA continuum sources (see also Fig.~\ref{fig:alma3mm}). The colored filled circles correspond to ALMA sources with counterparts at infrared wavelengths \citep[from the catalogue of][]{Willis2013}. The stars indicate the position of VLA sources as in the left panel. Black stars mark those VLA sources with an ALMA counterpart at 87.6 GHz. Grey stars indicate VLA sources with no counterparts in our catalogue.}
\label{fig:counterpart}
\end{center}
\end{figure*}


%
\subsection{Continuum source identification}\label{sec:source-identification}

We have used the \texttt{SExtractor} package\footnote{See \url{http://www.astromatic.net/software/sextractor}} \citep[see][]{Bertin1996} via its \texttt{python} wrapper interface \texttt{sewpy}\footnote{Available at \url{https://sewpy.readthedocs.io/en/latest/}}, to identify and build a catalogue of compact sources from our ALMA continuum image. Via a series of steps, the algorithm determines the background level (given by the user as a detection threshold) and whether pixels belong to the background or to objects. For sources that overlap in the image, the algorithm deblends them by searching for intensity variations larger than a given threshold. Finally, sources must have an extension of more than a given minimum number of pixels to be included in the final catalogue. For each source, \texttt{SExtractor} determines, among other parameters, the position, spatial extent and integrated flux. For the ALMA image of \ngc, we used a detection threshold of $6\sigma$, where $\sigma$ is the rms noise level, and a deblending threshold of $24\sigma$. The rms noise level is calculated based on the standard deviation of the pixel values in a background mesh throughout different regions of the map. Therefore, the inhomogeneity in the distribution of the noise is taken into account automatically and a higher value of $\sigma$ is estimated for the noisy regions.
The minimum number of pixels necessary to consider a detection as a source is 5 pixels with a pixel size of 0.2\arcsec. The flux determined by \texttt{SExtractor} corresponds to the counts of all pixels above the threshold minus the background level. This value is, later on, divided by the number of pixels of the synthesised beam to take into account that the units of the image are \jpb.
The observed geometric parameters (i.e. shape and size) of the sources are computed from the intensity-weighted moments of the pixels that form the source. A semi-major and semi-minor axis, together with a position angle, are derived assuming that the sources can be described by ellipses.\\
\indent A total of 185 sources were automatically identified by \texttt{SExtractor}. After visual inspection, we manually removed those located in regions of the map with a higher rms noise level (e.g. close to \ngc-I) and associated with \texttt{CLEAN}ing artifacts. The final catalogue of ALMA continuum sources contains 142 objects. We listed them, ordered by increasing right ascension in Table~\ref{tab:catalog}, and marked them in Fig.~\ref{fig:alma3mm} (see also a finding chart for the sources in Fig.~\ref{fig:finding-charts}). The peak intensities and flux densities of the identified compact sources range from 0.4 to 340~m\jpb\ and from 0.1 to 2073~mJy, respectively. The observed sizes (diameters) range from $0.3\arcsec$ to $5.2\arcsec$, corresponding to 390~au and 6760~au. In the top panels of Fig.~\ref{fig:distribution}, we show the distribution of peak intensities, flux densities, and observed sizes.

\begin{figure*}[ht!]
\begin{center}
\includegraphics[height=1.2\columnwidth, clip]{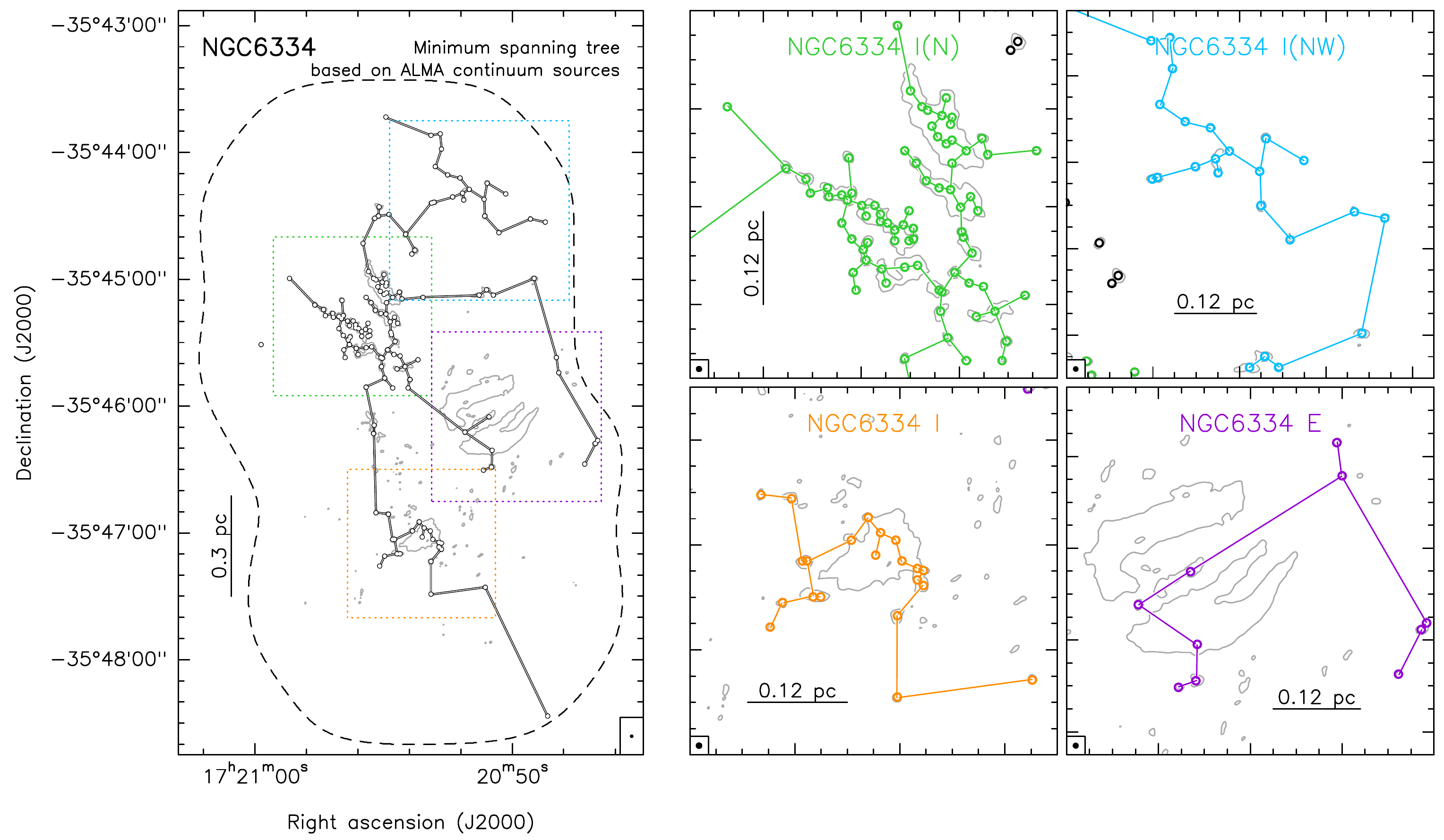} 
\caption{\textit{Left}: Minimum spanning tree (MST) for the ALMA continuum sources. The positions of the dense cores are indicated with circles, while the lines denote the tree. The rectangles mark the positions of the relevant regions also marked in Fig.~\ref{fig:alma3mm}. \textit{Right}: Minimum spanning tree for each of the four clusters identified in \ngc. The MSTs for each region/cluster are built considering only those members that belong to each cluster (see Table~\ref{tab:catalog}). The ALMA continuum emission at 87.6~GHz is shown in grey contours. The contour level corresponds to 0.5 Jy beam$^{-1}$.}
\label{fig:MST}
\end{center}
\end{figure*}

%
\subsection{Counterparts at different wavelengths}\label{sec:counterparts}

We searched for counterparts of the ALMA continuum sources (see \S~\ref{sec:source-identification} and Fig.~\ref{fig:alma3mm}) at different wavelengths from radio to X-rays, and also at millimetre wavelengths when observed with higher angular resolution.\\
\indent The Very Large Array (VLA) has been used by different authors to evaluate the radio-continuum content of \ngc. Among others, \citet{Hunter2014} and \citet{Brogan2016} have studied at high angular resolution (0.17\arcsec--0.5\arcsec) the centres of \ngc-I and \ngc-I(N) at different frequencies from 6 to 230~GHz. More recently, \citet{Medina2018} published a catalogue of radio continuum sources extracted from VLA 4--8~GHz images that covers the whole area that we observed with ALMA. In Fig.~\ref{fig:counterpart}, we mark the location of the radio-continuum sources with star symbols. We consider that an ALMA source is associated with a radio-continuum source if the distance between the two sources is less than the spatial extent of the sources. In total, six out of the 142 ALMA sources are also visible at centimeter wavelengths. Two of them are located in \ngc-I, two in the region of \ngc-I(N), and the remaining two sources are associated with the shell-like \hii\ region \ngc-E. Column~(7) of Table~\ref{tab:catalog} indicates these correspondences. The radio-continuum emission found towards the ALMA sources indicate the presence of thermal free-free radiation (see \S~\ref{sec:mass}), which can be due to \hii\ regions or shocked gas from thermal radiojets \citep[see e.g.][]{Kurtz2005,Sanchez2013a,Anglada2018}.\\
\indent We used the IRAC and MIPSGAL Spitzer catalogue of infrared sources published by \citet{Willis2013} to search for counterparts at infrared wavelengths. The catalogue contains more than 700\,000 sources detected at wavelengths from 3.8~$\mu$m up to 24~$\mu$m toward the entire \ngc~complex. Among others, $\sim$2000 sources are located towards the region observed with ALMA.
They are classified according to their evolutionary status as class~I or class~II objects. The color image in the panels of Fig.~\ref{fig:counterpart} shows the Spitzer 4.5~$\mu$m emission. We determine that an ALMA source is coincident with an infrared source if their separation is less than the distance given by the source size of the ALMA source and the mean full-width half-maximum of the IRAC images, which corresponds to $2\arcsec$. In the right panels of Fig.~\ref{fig:counterpart}, we mark the location of ALMA sources which have infrared counterparts with filled circles and we list the associations in column~(8) of Table~\ref{tab:catalog}. In total, we find 24 infrared counterparts for the ALMA sources. The comparison of our ALMA catalogue with the \citet{Willis2013} catalogue allows us to determine the evolutionary state of the continuum sources. We consider that those ALMA sources without an infrared counterpart correspond to deeply embedded cores (i.e. class~0 objects).\\
\indent We also investigated whether the ALMA sources are associated with X-ray emission. For this, we used the Chandra catalogue published by \citet{Feigelson2009} which contains $\sim$1600 sources. Among others, $\sim$140 X-ray sources are located towards the region observed with ALMA. We do not find any counterpart for the ALMA sources within separations of $0.75\arcsec$ (corresponding to half the sum of the resolutions of the Chandra and ALMA images).\\
\indent Finally, we compared the ALMA compact sources identified in our super-resolution continuum images, with millimetre continuum objects identified in both \ngc-I and \ngc-I(N) at higher angular resolutions from the literature (i.e. $<1.0\arcsec$). For \ngc-I, \citet{Brogan2016} resolve two of our ALMA sources (numbers 43 and 44 in Table~\ref{tab:catalog}) into multiple fragments (the first into 7 and the second into 2) when observed at an angular resolution of $0.17\arcsec$ at 20, 42 and 100~GHz. Similarly, a fraction of the sources detected in the region close to \ngc-I(N) were previously reported by \citet{Hunter2014} using SMA and VLA observations from 5 to 345~GHz. The continuum emission at 230~GHz from one of the SMA sources is clearly resolved into three components. Two of these components are also detected in our ALMA super-resolution continuum map (numbers 92 and 95 in Table~\ref{tab:catalog}).\\
\indent In summary, out of the 142 ALMA compact sources, six have a radio continuum counterpart, 24 are associated with infrared emission, and two show both radio and infrared emission. This results in 112 of the total catalogue only detected at millimetre wavelengths, likely tracing a cold dust population.

%
\subsection{Physical properties of the continuum sources}\label{sec:mass}

In the following, we determine physical properties (mass, density, and column density) for the identified compact sources. The low number of millimetre continuum sources associated with a radio-continuum counterpart suggests that most of the emission detected with ALMA at 87.6~GHz originates from cold dust. The dust and gas mass can be determined by
\begin{equation}\label{eq:mass}
M_{\textsubscript{d+g}}=\frac{S_{\nu}D^{2}}{B_{\nu}(T_\mathrm{d})\kappa_{\nu}}, 
\end{equation}
where $S_{\nu}$ is the integrated flux density of dust at 87.6~GHz, $D$ is the distance to the source \citep[1.3~kpc;][]{Chibueze2014}, $B_{\nu}(T_\mathrm{d})$ is the Planck function at a dust temperature $T_\mathrm{d}$, and $\kappa_{\nu}$ is the absorption coefficient per unit of total mass density at frequency $\nu$. We assume optically thin emission, a gas to dust mass ratio of 100, and we adopt a dust mass opacity coefficient of 0.899~cm$^2$~g$^{-1}$ at 230~GHz \citep{Ossenkopf1994} corresponding to grains with thin ice mantles, which we extrapolate to 0.131~cm$^2$~g$^{-1}$ at 87.6~GHz assuming a power-law dependence with a $\beta$ power-law index of 2 \citep{Shirley2011}. For sources with a radio-continuum counterpart, we determine the contribution of thermal free-free ionised gas emission to the ALMA flux. For this, we assume two different spectral indices\footnote{The spectral index $\alpha$ is defined as $S_\nu\propto\nu^\alpha$, where $S_\nu$ is the flux density and $\nu$ the frequency.} for the thermal free-free emission: $-0.1$ representing the case of optically thin emission, and $+2$ for the optically thick case \citep[see e.g.][]{Panagia1975, Kurtz2005}. After extrapolating the flux at radio wavelengths to the frequency of ALMA with the two spectral indices, we obtain a lower and upper limit to the contribution of ionised gas at 87.6~GHz. Assuming that the emission is optically thick, we find that for four ALMA sources (numbers 5, 15, 46 and 67 in Table~\ref{tab:catalog}) the total observed flux could originate from the thermal free-free emission, while the flux of the other two sources is partially contaminated (35\% for source 44 and 52\% for source 95). Finally, the temperature $T_\mathrm{d}$ of the cores is assumed to be 20~K when the source is only detected in the ALMA image and has no counterparts, 50~K if it has an infrared or radio counterpart, and 100~K if the source has been associated with a chemically-rich hot molecular core \citep[see e.g.][]{Kurtz2005, Cesaroni2005}. These temperatures are selected on the basis of previous observations of dense cores \citep[see e.g.][]{Sanchez2013c}. 
For some sources with temperature measurements \citep[see][]{Hunter2014}, we used those temperatures. The temperatures used in the mass determination are listed in Table~\ref{tab:catalog}. The effect of the inaccurate temperature estimation on the determination of the mass is discussed in \S~\ref{sec:T bias}.\\
\indent Assuming the cores to be spheres of radius $R$, the volume density $n_\mathrm{H_2}$ is determined as
\begin{equation}\label{eq:density}
n_{\mathrm{H}_{2}}=\frac{1}{\mu m_\mathrm{H}}\frac{M_\mathrm{d+g}}{(4/3)\pi R^{3}},
\end{equation}
where $\mu$ is the mean molecular mass per hydrogen atom (equal to 2.3), $m_\mathrm{H}$ is the hydrogen mass, and $R$ is the radius of the core assumed to be the average of the semi-minor and semi-major axis of the source size listed in Table~\ref{tab:catalog}. Finally, the column density, $N_{\mathrm{H}_{2}}$, is calculated using 
\begin{equation}\label{eg:column-density}
N_{\mathrm{H}_{2}}=\int_{\text{line of sight}}n_{\mathrm{H}_{2}} dl, 
\end{equation}
where $l$ corresponds to the size of the core. The volume and column densities of the identified sources are listed in Table~\ref{tab:catalog}.

\begin{figure*}[htb!] 
\begin{center}
\includegraphics[scale=0.37]{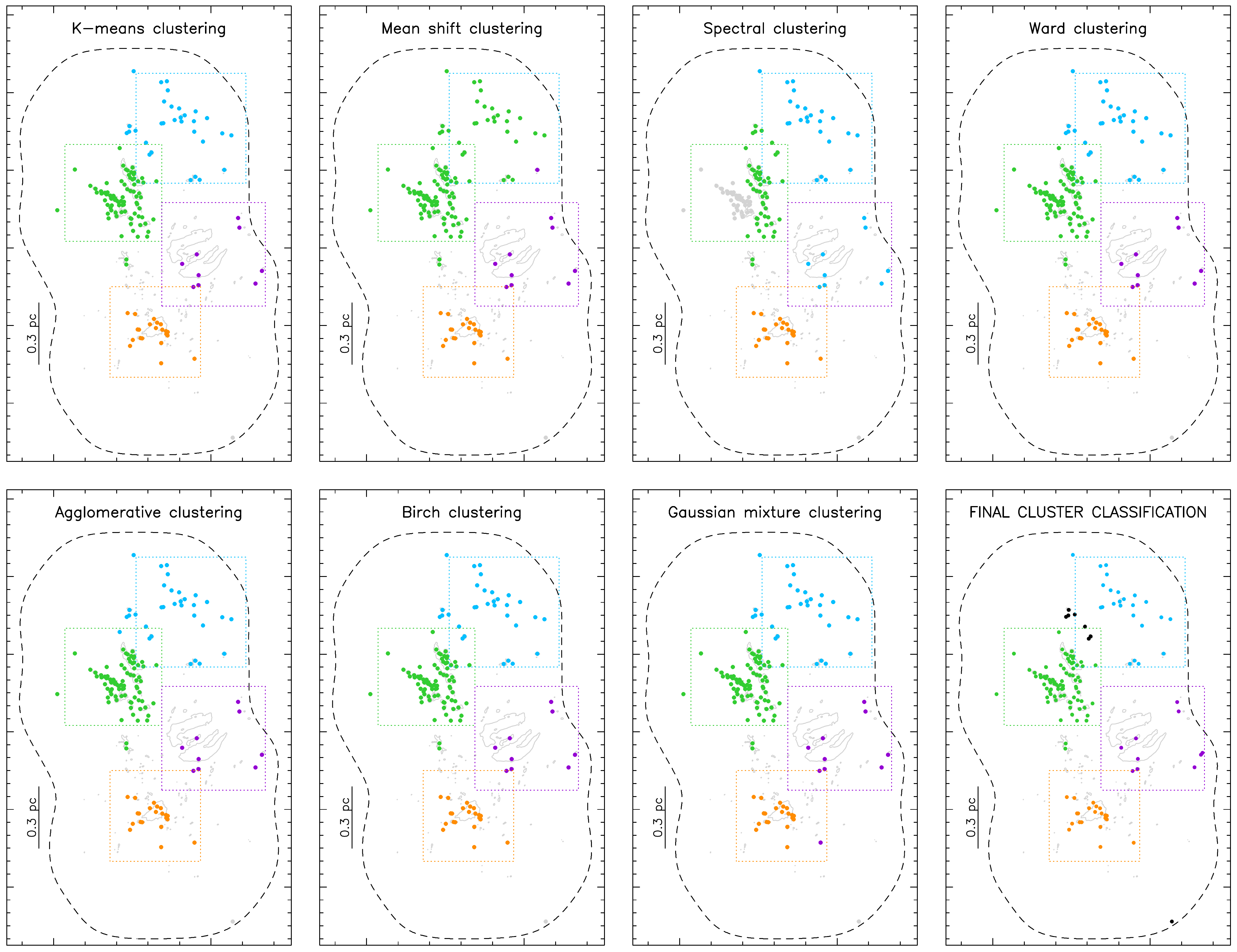}
\caption{Each panel shows area of NGC6334 observed with ALMA (see Fig.~\ref{fig:alma3mm}) and colored circles correspond to the ALMA continuum sources detected in region (see Table~\ref{tab:catalog}). Each panel correspond to one of the seven clustering algorithms used to identify clusters in the region (see \S~\ref{sec:cluster-identification}). The colors of the circles depict the different clusters to which each continuum source belongs according to each algorithm. The last panel shows the final association of each source with one of the four clusters: light blue for \ngc-I(NW), lime for \ngc-I(N), violet for \ngc-E and orange for \ngc-I. The black circles in the last panel correspond to the sources not assigned to any cluster. In all panels, the rectangles mark the position of the relevant regions also marked in Fig.~\ref{fig:alma3mm}.}
\label{fig:clustering}
\end{center}
\end{figure*}

\begin{table*}[!ht]
\caption{Clustering algorithms and basic input parameters}
\label{tab:clustering}
\begin{tabular}{l l l}
\hline\hline
\noalign{\smallskip}
Method & Input parameters & Metric \\
\hline
\noalign{\smallskip}
K-means                        & data points, number of clusters                              & distance between points \\
Mean-shift                     & data points                                                  & distance between points \\
Spectral clustering            & data points, number of clusters, number of neighbours        & graph distance \\
Ward hierarchical clustering   & data points, number of clusters, linkage                     & distance between points \\
Agglomerative clustering       & data points, number of clusters, linkage, affinity           & any pairwise distance \\
Gaussian mixtures              & data points, number of mixture components                    & Mahalanobis distances to centres \\
Birch                          & data points, number of clusters, threshold, branching factor & Euclidean distance between points \\
\hline
\end{tabular}
\end{table*}

In the bottom panels of Fig.~\ref{fig:distribution}, we present the distribution of masses, volume densities and column densities. The dust and gas mass ranges from $<$ 1~\mo\ up to a hundred \mo, with mean and median values of 8~\mo\ and 4~\mo. The most massive object (source 44 with 100~\mo) is located in the vicinity of \ngc-I. The H$_2$ volume densities are in the range $10^{7}$--$10^{9}$~cm$^{-3}$, with the three most massive cores (source 107, 87 and 84) having densities above $10^{9}$~cm$^{-3}$. The H$_2$ column densities of the brightest sources are above 10$^{25}$~cm$^{-2}$ at the observed scales of $\sim$1000~au. It is worth noting that for 90 cores, we have assumed a temperature of 20~K or 50~K. An error in the temperature of 10--20~K results in an error in the mass of about 40--50\%. While other parameters such as the dust opacity may introduce uncertainties in the mass estimation, the temperature remains as a dominant factor when comparing masses of cores that belong to the same cluster (see more details in \S~\ref{sec:T bias} and \S~\ref{sec:resolution bias}). 

%
\section{Clusters in \ngc}\label{sec:clusters}

The objects identified in the ALMA image of \ngc\ (see Fig.~\ref{fig:alma3mm}) do not appear uniformly distributed over the whole region. As discussed above, they seem to cluster in different groups around, e.g. \ngc-I, \ngc-I(N) and \ngc-E among others. In order to evaluate the distribution and internal structure of the cores in \ngc, we used the $Q$ parameter presented by \citet{Cartwright2004}. The $Q$ parameter is a dimensionless measure of the presence of smooth over-densities and fractal structures in a population of data points. The parameter is calculated based on the construction of a minimum spanning tree (MST), which is defined as a network of lines without closed loops that connects all the data points with the minimum path length. We constructed the MST for the cores in \ngc, using the \texttt{python} package \texttt{NetworkX}\footnote{Available at \url{https://networkx.github.io/}} which implements the MST according to Kruskal's algorithm \citep{Kruskal1956}. The MST of the whole region is shown in the left panel of Fig.~\ref{fig:MST}. We estimate the $Q$ parameter as the ratio of the normalised mean length of the branches or segments of the MST, $\bar{m}$, to the correlation length, $\bar{s}$, which corresponds to the mean projected separation between data points normalised by the cluster radius \citep[see][]{Hunter2014}. The cluster radius is defined as the distance from the mean position of all cluster members to the furthest core from this point. A value of $Q\lesssim0.8$ indicates the presence of fractal sub-structures in the data set, while values $Q\gtrsim0.8$ correspond to a centrally concentrated structure with a radial large-scale density gradient \citep[see][]{Cartwright2004}.\\
\indent The MST for the catalogue of 142 ALMA continuum sources is shown in Fig.~\ref{fig:MST} (left panel) as black solid lines. We calculate $\bar{m}$ and $\bar{s}$ to be $7\arcsec$ ($\sim0.04$~pc) and $76\arcsec$ ($\sim0.5$~pc), respectively. Following Eq.~(1) of \citet{Hunter2014}, and considering a radius of the cluster (or region) of $194\arcsec$ ($\sim1.2$~pc), we obtain $Q=0.60\pm0.1$. This low value confirms the presence of fractal structures or clusters within the cores identified in \ngc. To estimate the uncertainty of the $Q$ parameter, we removed 20\% of the cores available in the sample randomly and measured the $Q$ parameter for the remaining cores. Repeating this process 100 times, we obtained a distribution of the measured values of $Q$. A Gaussian fit to this distribution resulted in a standard deviation $\sigma$=0.1, that corresponds to the uncertainty of the Q parameter. We note that the linear extent of a cluster is not enough to fully characterise the statistics of fractal structures, and more sophisticated measures are required \citep[e.g. $Q^+$;][]{Jaffa2017}. However, due to the low statistics of our sample, these methods will not lead to reliable results. \\
\indent In the following, we identify the different clusters in the \ngc-I and \ngc-I(N) regions and study their properties.

%
\subsection{Cluster identification}\label{sec:cluster-identification}

We used different statistical algorithms to identify clusters of sources within the catalogue of 142 ALMA compact objects. A total of seven algorithms, implemented in the \texttt{python} package \texttt{scikit-learn}\footnote{See more information in the website \url{http://scikit-learn.org/stable/modules/clustering.html}} \citep{scikit-learn}, were considered. Basic details on the different methods are listed in Table~\ref{tab:clustering}, and a more detailed description is presented in Appendix~\ref{app:clustering}.\\
\indent In addition to the coordinates of the data points, all the considered algorithms except the mean-shift approach, require the user to indicate the expected number of clusters $K$ as an input parameter. In order to evaluate the optimal number of clusters, we used two different approaches and cross-checked the consistency of their results (see \S~\ref{app:number-of-clusters} for details). The first one is known as the elbow-point analysis which measures the average distance between the group members and the centre of the group for different number of clusters. The second approach known as Silhouette analysis measures the consistency between the data points in a cluster. As shown in Figs.~\ref{fig:K-values} and \ref{fig:silhouette}, both methods suggest that the number of clusters in our region is between 2 and 6, with $K=4$ the most probable value.\\
\indent In Fig.~\ref{fig:clustering}, we color the 142 objects according to the cluster to which they belong for each one of the seven algorithms considered \footnote{We exclude source 6 which is located at the very bottom of the surveyed
area and appears spatially disconnected from the other sources.}. In general, there is a good agreement between the different methods. The southern compact sources appear always grouped in one cluster clearly associated with \ngc-I (see orange circles). In all methods but one, the compact objects associated with the filamentary structures seen towards \ngc-I(N) are classified as one single group (see lime circles). Only for the case of spectral clustering the sources belonging to the most eastern filament appear classified as an independent group (see grey circles). The remaining sources, mainly located to the north-west of \ngc-I(N) appear in all cases grouped as an additional cluster (see blue circles). Finally, the sources in the vicinity of \ngc-E seem to form a loose association with just a few members. In total we consider four well-defined clusters which we name: \ngc-I, \ngc-I(N), \ngc-I(NW) and \ngc-E (see last panel Fig.~\ref{fig:clustering}). We calculated a degree of membership for each core, by counting how many of the seven algorithms assign the considered core to one of the four clusters. This gives us the percentage of association to a cluster, or confidence level, that is listed in Table~\ref{tab:catalog}. If this value is lower than 60\%, we consider that we can not assign the corresponding core to one of the clusters. Further confirmation for the association of each core to a cluster can be obtained in forthcoming works by determining the kinematics of the members. Based on the actual criterion, seven sources are excluded from the defined clusters.
The right panels of Fig.~\ref{fig:alma3mm} show a close-up view of the four identified clusters, with the sources colored according to their cluster membership. In Table~\ref{tab:clusters}, we list the numbers of members of each cluster. \ngc-I(N) is the most populated with almost 80 members.

\begin{figure}[t!]
\centering
\includegraphics[width=1\columnwidth,clip]{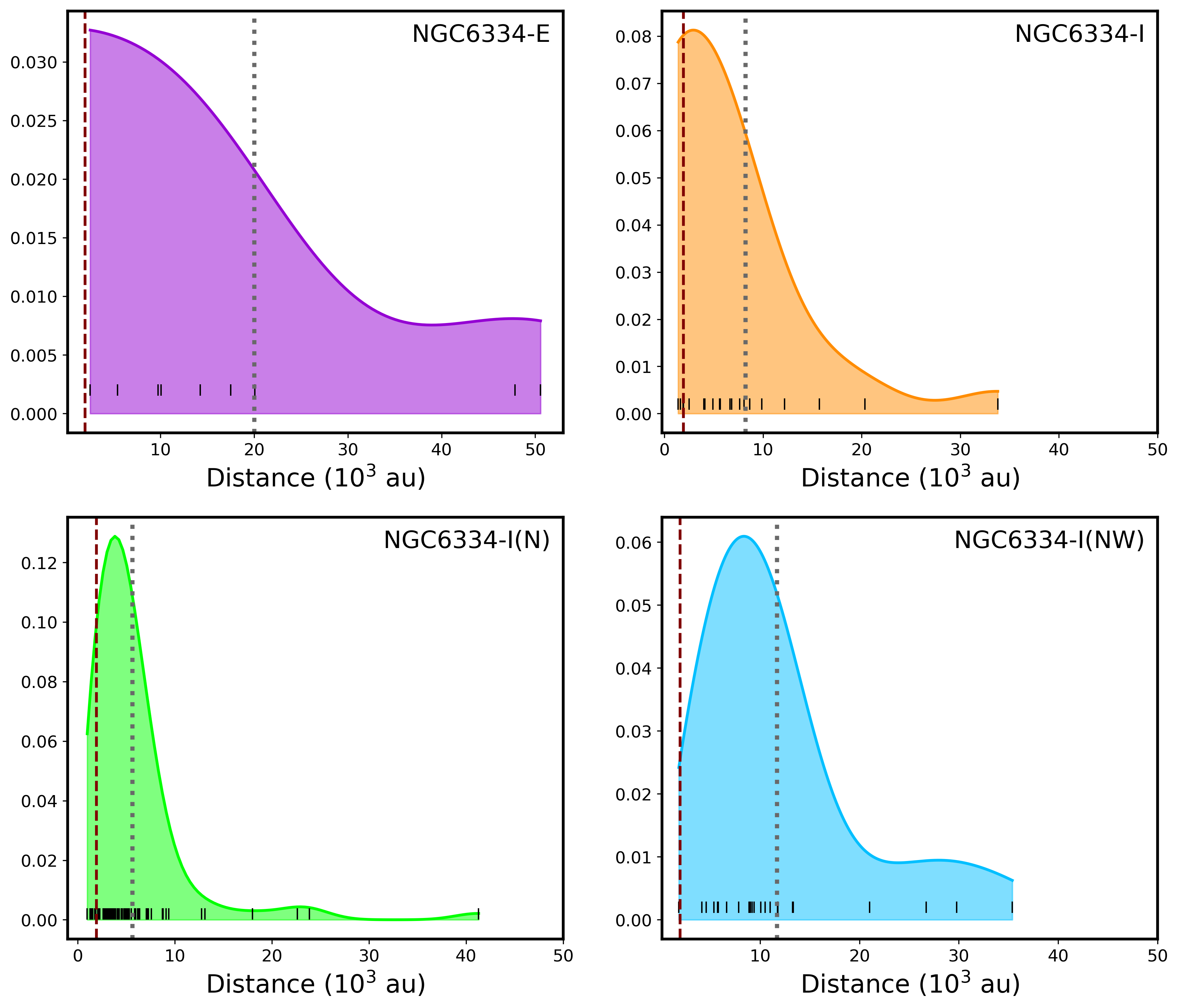}
\caption{KDE of the projected separation between nearest neighbours for \ngc-E, \ngc-I, \ngc-I(N) and \ngc-I(NW).
In each panel, the red vertical dashed line shows the 1\,300~au resolution limit of the ALMA observations, while the vertical grey dotted line denotes the median distance between the cores in the cluster. 
The short vertical lines at the bottom of each panel indicate the observed separations between members of each cluster.}
\label{fig:KDEs-MST}
\end{figure}

%
\subsection{Spatial structure of the clusters in \ngc}\label{sec:spatial-clusters}

In the right panels of Fig.~\ref{fig:MST}, we present the connecting segments of the MST for each one of the four main clusters identified in  \S~\ref{sec:cluster-identification}. We used the results of the MST to evaluate the distances between nearest neighbours, $d_\mathrm{nn}$. The distributions (or Kernel Density Estimates, KDEs) of the separations for unique pairs of ALMA sources belonging to each cluster are shown in Fig.~\ref{fig:KDEs-MST}, while the mean $d_\mathrm{nn}$, the median $d_\mathrm{nn}$ and the peak of the KDE are listed in Table~\ref{tab:clusters}. The largest separations between sources are found in clusters \ngc-E and \ngc-I(NW), with typical distances of about 10\,000~au. On the other side, clusters \ngc-I and \ngc-I(N) have similar peak separations (about 4\,000~au), although with 1.5-2 times larger mean and median separations for cluster \ngc-I compared to \ngc-I(N). This is also seen in the KDEs shown in Fig.~\ref{fig:KDEs-MST}, with a broader distribution for cluster~I. Taken all together, this suggests that the members of I(N) are more uniformly distributed over the whole extend of the cluster, with typical separations of about 4\,500~au. The radius of the clusters is determined as the distance from the geometric centre of the cluster to the furthest member. In Table~\ref{tab:clusters}, we list the radius of the four main clusters, which are in the range from 48\,000~au to 83\,000~au. Making use of the separations and the sizes of the clusters, we followed Eq.~(1) of \citet{Hunter2014} to determine the $Q$ parameter for each cluster. We obtain values of 0.58$\pm$0.2, 0.72$\pm$0.3 , 0.79$\pm$0.2 and 0.64$\pm$0.3 for \ngc-E, I, I(N) and I(NW), respectively. Except for \ngc-E, the $Q$ parameter for the other three clusters is not significantly less than 0.8, due to large uncertainties. The $Q$ value for cluster \ngc-E suggests the possible presence of sub-structure (due to a $Q$ parameter significantly less than 0.8). However, we note that the total number of sources in that cluster (10) limits the conclusions that can be derived from this analysis. Larger statistics in each cluster is required to more reliably characterise their internal structure.\\
\indent In the second approach, we explored the internal structure of the clusters using the two-point correlation function $\zeta(r)$. The function measures the excess probability of finding a pair of points, as a function of separation, in comparison to the expectation from a random distribution.The two-point correlation function (TPCF) is defined as \citep{Landy1993}:
\begin{equation}
\zeta(r)=\frac{DD(r)-2DR(r)+RR(r)}{RR(r)},
\end{equation}
where $DD(r)$ is the observed distribution of pair separations, $RR(r)$ is the same but for a set of points which are randomly produced, and $DR(r)$ is the distribution of separations when one point of the pair is selected from the observed sample and the other from a random sample. The $\zeta(r)=0$ indicates a random distribution, while positive and negative amplitudes in  $\zeta(r)$ correspond to separations that are respectively exceeding or not those of a random distribution. To measure $\zeta(r)$, we used the \texttt{FragMent} Python/C library introduced in \citet{Clarke2019}\footnote{Available at \url{https://github.com/SeamusClarke/FragMent}}. The algorithm considers the separation between every pair of fragments as a single data point and convolves each point with a Gaussian kernel. We compute the random separation distributions ($DR(r)$ and $RR(r)$) by producing 10\,000 instances of randomly placed points for each cluster to obtain adequate statistics. Finally, we computed an uncertainty estimate of the TPCF through the relation \citep{Sawangwit2011}
\begin{equation}
\sigma_{\zeta}=\sqrt{\frac{1+\zeta(r)}{DD(r)}}.
\end{equation}

\begin{figure}[t!]
\centering
\includegraphics[width=1\columnwidth,clip]{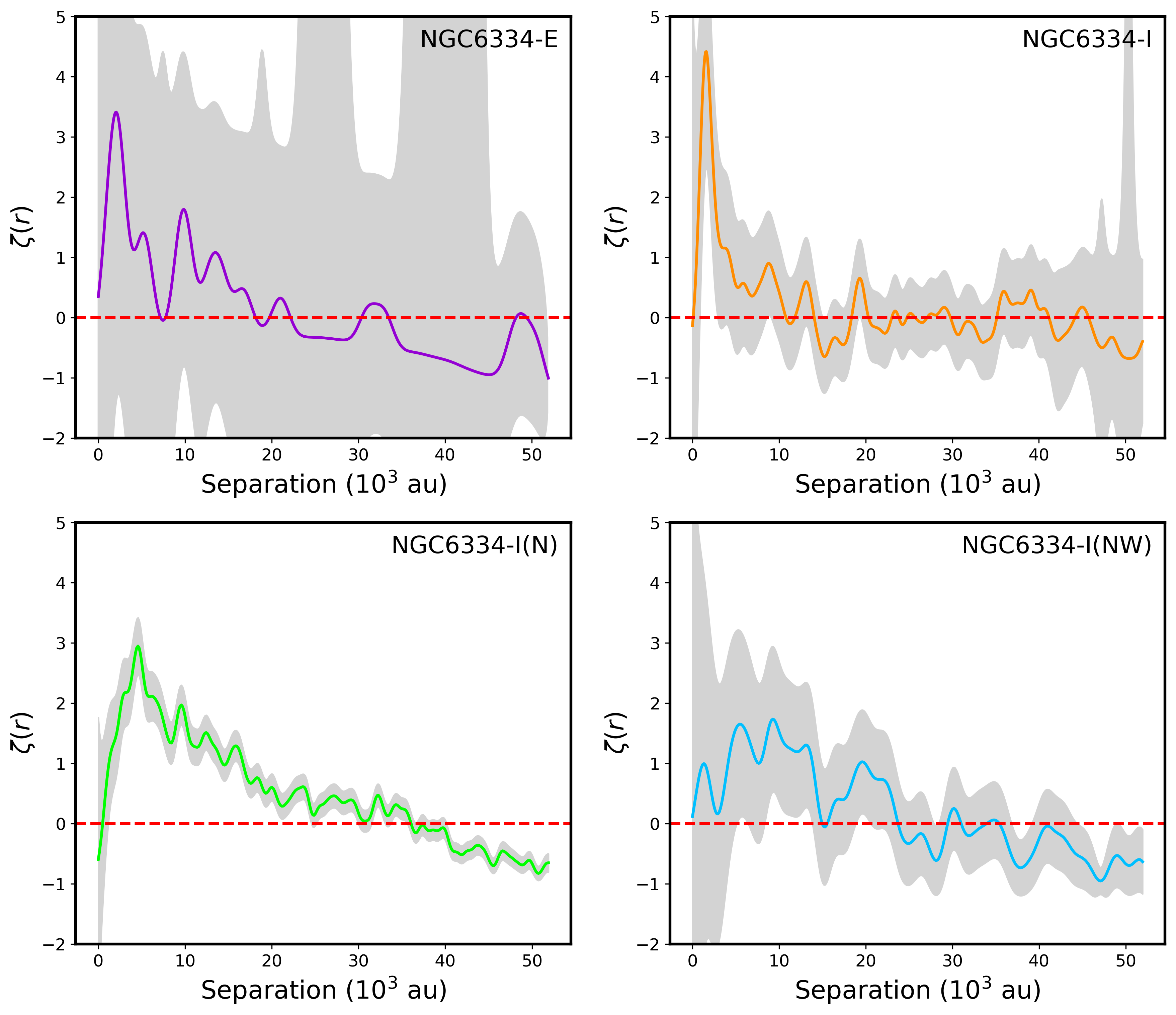}
\caption{Two-point correlation function of ALMA continuum sources in \ngc-E, \ngc-I, \ngc-I(N) and \ngc-I(NW). The gray areas show the $\mathrm{1}$~$\sigma$ confidence intervals. The dashed horizontal line at the zero level corresponds to a uniform distribution.}
\label{fig:TPCF}
\end{figure}

The TPCFs of each cluster are shown in Fig.~\ref{fig:TPCF}. The TPCF of \ngc-I(NW) seems to have a few random peaks with no preferred length scale. \ngc-E and \ngc-I show strong peaks at 1500~au. However, the large uncertainty in \ngc-E prevents us from deriving firm conclusions in this cluster. Finally, a significant excess of separations at 4\,000~au is observed for \ngc-I(N) which indicates a shift to larger separations compared to \ngc-I. \\
\indent In summary, the $Q$ parameter analysis suggests that the clusters are centrally concentrated structures, with some sub-structure in \ngc-E and maybe in \ngc-I(N).
The median separation between the cores is the largest value in \ngc-E, indicating that the cores are widely distributed. In contrast, \ngc-I(N) with the smallest median separation exhibits a tighter distribution. The comparison of the TPCFs indicates that \ngc-I is more centrally concentrated than \ngc-I(N) that has most of the cores embedded in two filamentary-like structures.


\begin{table*}[ht!]
\caption{Cluster members, central coordinates, and distances between nearest neighbours}
\label{tab:clusters}
\begin{tabular}{l c c c c c c c c c}
\hline\hline
\noalign{\smallskip}
&
& \multicolumn{2}{c}{Geometric cluster centre}
& $R_\mathrm{cluster}$
& $Q$
& \multicolumn{4}{c}{Typical separations\tablefootmark{a} (in au)}
\\
\cline{3-4}
\cline{7-10}
Cluster
& \# 
& R.A.~($^\mathrm{h}:^\mathrm{m}:^\mathrm{s}$)
& Dec.~($^\circ:^\prime:^{\prime\prime}$)
& (au)
& param.
& mean $d_\mathrm{nn}$
& median $d_\mathrm{nn}$
& peak $d_\mathrm{nn}$
& TPCF
\\
\hline
\noalign{\smallskip}
\ngc-E      & \phn10  & 17:20:48.97  & $-$35:46:12.03  & \phn50\,000 & 0.58$\pm$0.2 & 20\,000    & 12\,200    & \phn2\,500 & \ldots \\ 
\ngc-I      & \phn21  & 17:20:53.70  & $-$35:47:05.86  & \phn58\,000 & 0.72$\pm$0.3 & \phn8\,250 & \phn6\,200 & \phn3\,000 & 1\,600 \\
\ngc-I(N)   & \phn79  & 17:20:55.57  & $-$35:45:22.88  & \phn82\,500 & 0.79$\pm$0.2 & \phn5\,600 & \phn4\,300 & \phn4\,500 & 4\,500 \\
\ngc-I(NW)  & \phn24  & 17:20:51.63  & $-$35:44:23.82  & \phn83\,000 & 0.64$\pm$0.3 & 11\,700    & \phn9\,300 & \phn8\,900 & \ldots \\ 
\hline
\noalign{\smallskip}
\ngc\        & 142     & 17:20:54.04  & $-$35:45:30.56  &  252\,000   & 0.61$\pm$0.1 & \phn9\,200 & \phn5\,300 & \phn5\,400 & -- \\
\hline
\end{tabular}
\tablefoot{
\tablefoottext{a}{The three first separations correspond to the mean, median and peak distance between nearest neighbours determined from the minimum spanning tree analysis. The last value corresponds to the peak determined from the two-point correlation function analysis.}
}
\end{table*}

%
\section{Physical properties of the clusters in \ngc}\label{sec:physical-properties}

In this section, we characterise the physical properties of the four clusters. We search for differences in the mass and density of the cores that constitute each one of the clusters, we study their spatial distribution and evaluate the fragmentation level.

%
\subsection{Core mass function}\label{sec:CMF}

In \S~\ref{sec:mass} (see also Table~\ref{tab:catalog} and bottom panels of Fig.~\ref{fig:distribution}) we determine the masses and densities of the different sources detected with ALMA. The mean, median and maximum mass of the cores in each cluster are listed in Table~\ref{tab:clusters_v2}. We find typical median masses of about 3~\mo, with a larger mean mass for clusters \ngc-I and \ngc-I(N). From the maximum core mass of each cluster we see that these two clusters still have the potential to form more massive stars. About 66\% (or 182~\mo) of the total mass in \ngc-I is concentrated in two single cores (sources 38 and 44), one of which appears already resolved in multiple fragments at higher angular resolution \citep[see][]{Brogan2016}. In contrast, \ngc-I(N) appears more fragmented with a larger population of low to intermediate-mass cores. The five most massive cores (sources 55, 84, 87, 91 and 107) encompass only 30\% (or 213~\mo) of the total mass. Finally, the masses of the fragments in clusters \ngc-E and \ngc-I(NW) suggest that they harbor mainly low/intermediate-mass fragments.

\begin{table*}[t!]
\caption{Masses, power-law index of CMFs, and mass segregation ratios for each cluster}
\label{tab:clusters_v2}
\begin{tabular}{l c c c c c c c c }
\hline\hline
\noalign{\smallskip}
& mean $M_\mathrm{core}$
& median $M_\mathrm{core}$
& Temperature
& n$_{\mathrm{H}_2}$
& $M^\mathrm{th}\!\!_\mathrm{Jeans}$
& $M^\mathrm{nth}\!\!_\mathrm{Jeans}$
& $\alpha$-CMF
& $\Lambda_\mathrm{MSR}$

\\
Cluster
& (\mo)
& (\mo)
& (K)
& (cm$^{-3}$)
& (\mo)
& (\mo)
& [$M$>2~\mo]
& [$M$>10~\mo]

\\
\hline
\noalign{\smallskip}
\ngc-E      & 5    & 3  & 22 & 3.5$\times$10$^{4}$ & 1.0  & 20  & $-$3.17  & \ldots \\
\ngc-I      & 14   & 3  & 23 & 4.0$\times$10$^{5}$ & 1.0  & 20  & $-$0.71  & 3.60 \\
\ngc-I(N)   & 9    & 5  & 19 & 3.2$\times$10$^{5}$ & 0.9  & 23  & $-$0.99  & 2.47 \\
\ngc-I(NW)  & 3    & 2  & 18 & 3.3$\times$10$^{4}$ & 0.9  & 24  & $-$1.32  & \ldots \\
\hline
\noalign{\smallskip}
\ngc        & 8    & 3  & 21 & 1.5$\times$10$^{5}$ & 1.5  & 34  & $-$1.10  & 2.68 \\
\hline
\end{tabular}
\end{table*}

\begin{figure}[ht]
\centering
\includegraphics[width=1.0\columnwidth, clip]{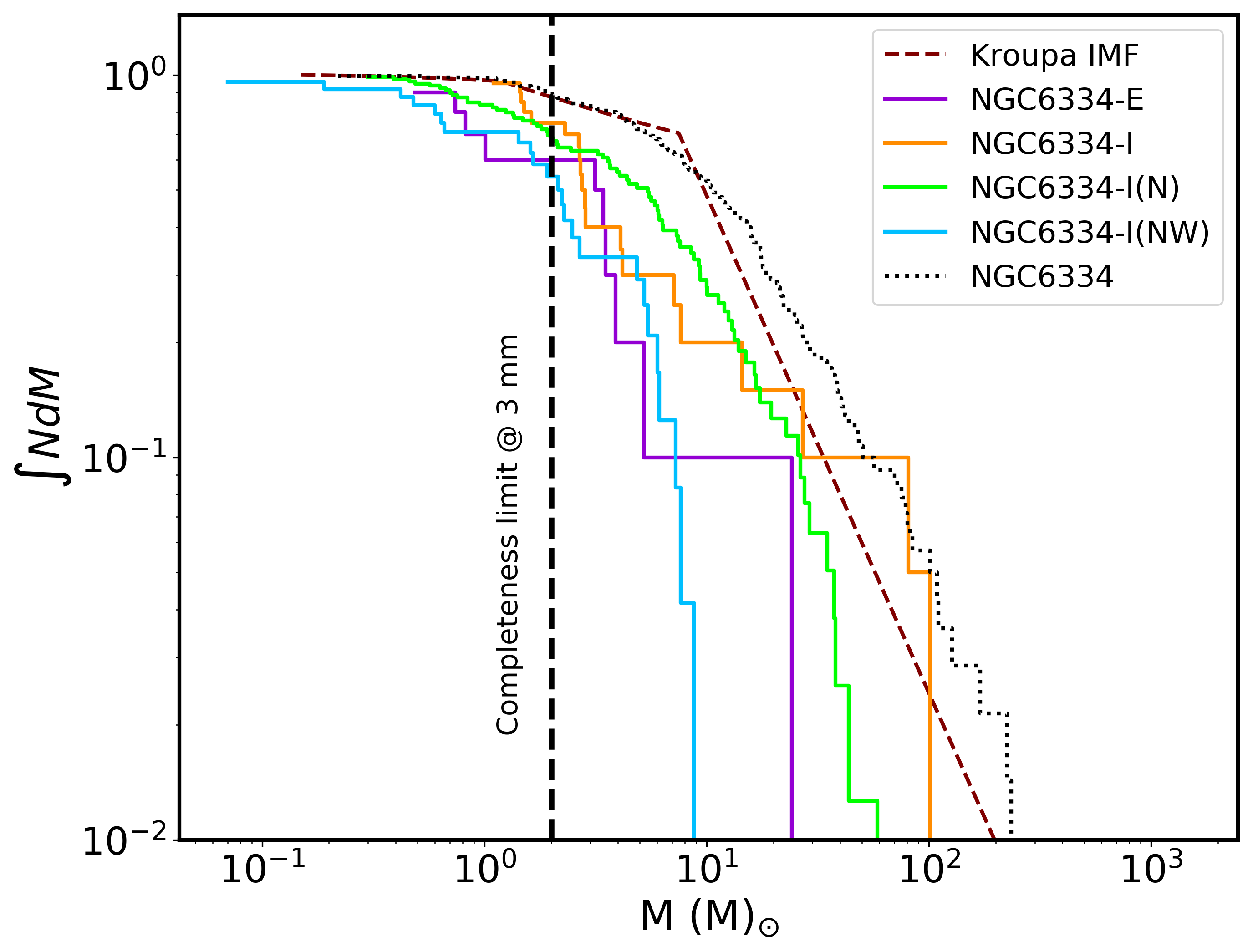}
\caption{Cumulative core mass function of continuum sources identified in \ngc-E (violet), \ngc-I (orange), \ngc-I(N) (lime) and \ngc-I(NW) (light blue). 
The CMF of the whole sample is indicated with a dotted black line. The red dashed line shows the Kroupa initial mass function shifted by factor of 15 to higher masses. }
\label{fig:CMF}
\end{figure}

\begin{figure}[t!]
\centering
\includegraphics[width=0.9\columnwidth, clip]{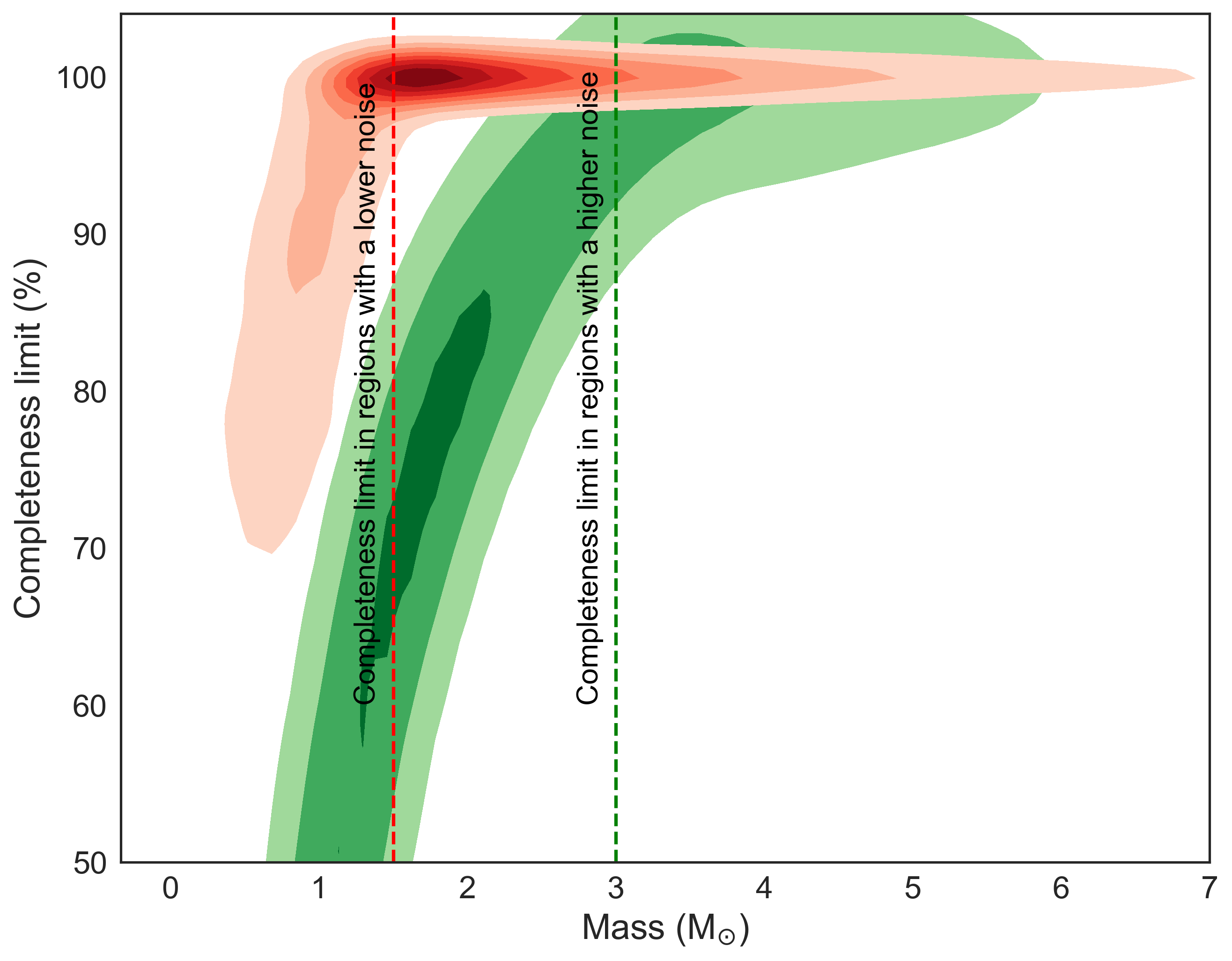}
\caption{Completeness level of the ALMA continuum map at 3~mm as a function of mass. The KDE shown in red indicates the detection probability of the cores in the noise-free regions. The KDE illustrated in green represents the detection probability for the cores located at the vicinity of bright sources. 
A completeness level of 90\% is obtained for cores with masses $>$ 1.5~\mo~and $>$ 3~\mo~for the regions with a lower noise (0.3~mJy~beam$^{-1}$) and a higher noise (0.9~mJy~beam$^{-1}$) respectively. These completeness limits are marked with vertical dashed and dotted lines for the noise-free and noisy regions, respectively.}
\label{fig:mass-completeness}
\end{figure}

The cumulative core mass functions (CMFs) for the whole region and each individual cluster are shown in Fig.~\ref{fig:CMF}. The CMFs of each cluster are compared to the initial mass function (IMF) as described in \citet{Kroupa2001}, are shifted towards higher masses for a better comparison. Clusters \ngc-I and I(N) seem to have an excess of high-mass cores compared to \ngc-E, \ngc-I(NW) and the IMF. The excess of massive cores is similar to that found in other high-mass star forming regions: G28.34+0.06 \citep{Zhang2015}, SgrB2 \citep{Sanchez2017}, seven IRDCs with 32 dense clumps \citep{Liu2018} and W43-MM1 \citep{Motte2018}. However, the fidelity of the analysis performed based on the core masses depends on the reliability of the extracted sources above the given threshold (see \S~\ref{sec:source-identification}). \\
\indent We determined the completeness level of our sample in terms of mass by generating random samples of clusters and evaluating which cores could be detected in our ALMA continuum image. We generated 10\,000 random clusters of 142 cores with core masses between 0.6~\mo\ and 15~\mo\ following an IMF $N(>M)\propto M^{-2.35}$ \citep[see][]{Salpeter1955, Kroupa2001}. For each run, we assigned a different temperature to each core in the range 20 to 100~K \citep[typical of dense cores, see e.g.][]{Sanchez-monge2013b}, and we repeated this process 10\,000 times. The lower limit of the assumed temperature range was chosen based on the average dust temperature as measured by \textit{Herschel}.
For each core with a mass and temperature, we derived the flux that would be detected at a distance of 1.3~kpc when observing at a frequency of 87.6~GHz. In the last step, we compared the computed fluxes with a given threshold and determine the probability for a source with mass $M$ to be detected. The completeness level is the ratio between the number of detected sources and the total number of sources at any given mass. In Fig.~\ref{fig:mass-completeness}, we show the completeness level for different masses when using thresholds of 0.3~m\jpb\ for the signal-free regions and 0.9~m\jpb\ for the cores which have been detected at the vicinity of the bright sources. We find a detection probability of 90\% for objects with masses $>1$~\mo~and $>2$~\mo, which increases up to $\sim$100\% for sources with masses $>1.5$~\mo~and $>4$~\mo~for the signal-free and noisy regions, respectively. In the following, we consider our mass completeness limit to be 2~\mo.\\
\indent We fitted a power-law function given by $dN/d\ln(M) \propto M^{\alpha}$ to the CMFs for the range of masses above the completenes level. The values of $\alpha$ for the whole sample and each cluster are listed in Table~\ref{tab:clusters_v2}. We obtain a value of $\alpha=-1.1\pm0.02$ for the whole sample, and from $-3.17$ to $-0.71$ for the individual clusters\footnote{We note that the shallower slope for cluster NGC6334-I indicates a larger fraction of massive cores, and this can be linked to the recent burst reported by \citep{Hunter2017}.}. The measured slope for the high-mass end of CMF in \ngc\ clusters is not consistent with the $-1.35$ value of the IMF \citep[see e.g.][]{Salpeter1955, Kroupa2001}. This result could challenge the idea that the shape of the IMF is directly inherited from the shape of the CMF and that there is a self-similar mapping from the CMF to the IMF \citep[see e.g.][]{Goodwin2007, Hopkins2013, Cheng2017}. However, the measured slope can be biased due to the uncertainty in temperature determination and the unresolved multiplicity in the massive cores. In the following we investigate how these uncertainties can affect the measured slope of the CMFs. 

\begin{figure}[ht!]
\centering
\includegraphics[width=0.9\columnwidth]{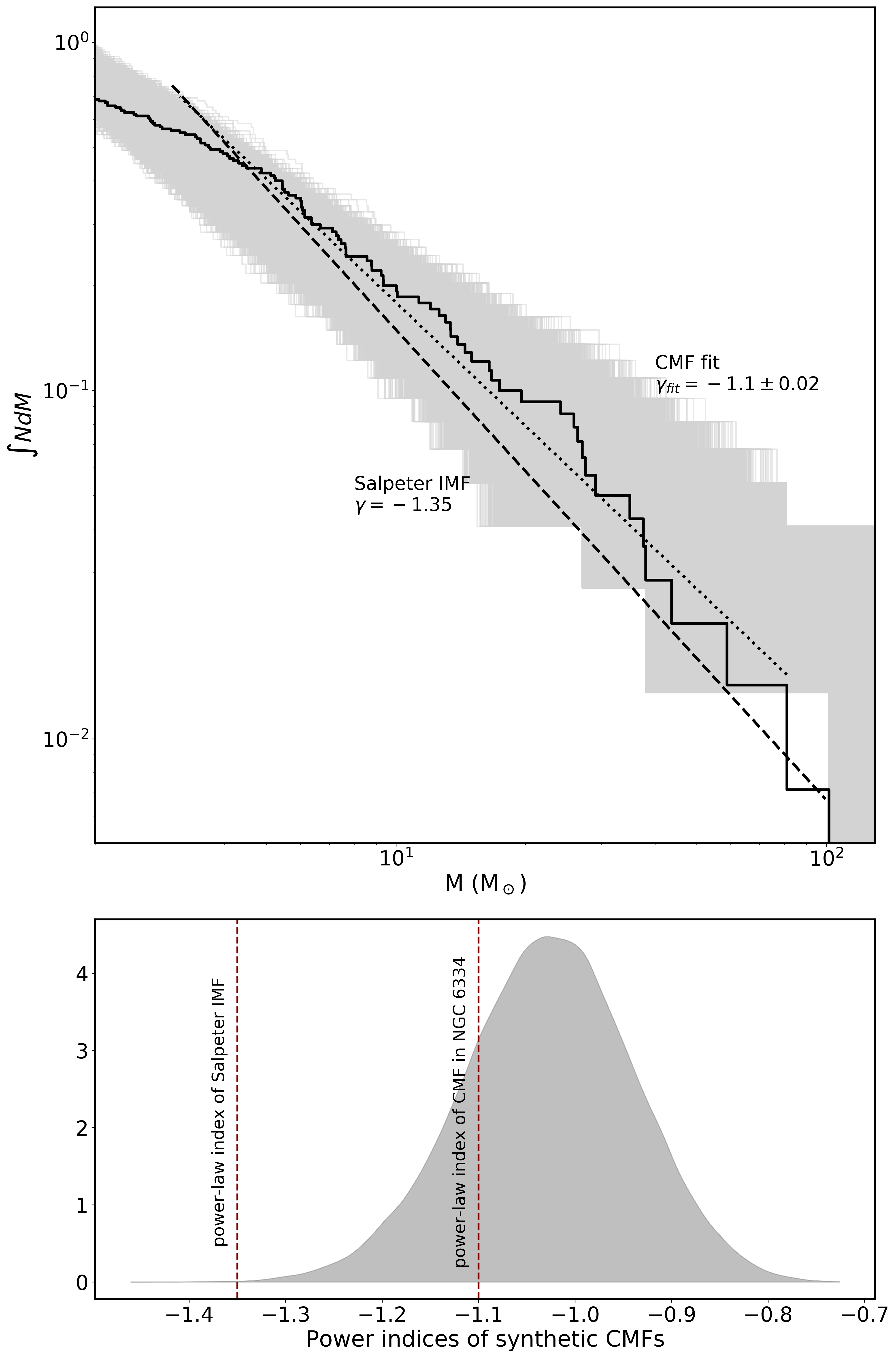}
\caption{\textit{Top:} \ngc\ cumulative core mass function shown in solid black line fitted by a single power-law $dN/d~\text{log}(M)\propto M^{-1.1}$ indicated with a black dotted line. The grey area represents the synthetic CMFs for 100\,000 groups of cores. The Salpeter IMF, shown with a black dashed line, shows a different slope compared to the CMF. \textit{Bottom:} KDE of the power-law indices of the synthetic core mass functions. The vertical dashed lines show the power-law index of Salpeter IMF and the power-law index of the CMF in the \ngc\ region.}
\label{fig:CMF-temperature}
\end{figure}

%
\subsubsection{Temperature bias in the determination of the CMF}\label{sec:T bias}

Far-infrared observations (e.g. \textit{Herschel}) that allow for the sampling of the whole spectral energy distribution, do not have enough spatial resolution to determine the temperature of individual cores. 
The study of how systematic errors in the assumed temperatures may affect the shape of the CMF is investigated by randomly modifying the assumed temperature for each core and determining its new mass. We considered temperatures of the cores to be uniformly distributed in the range 20--100~K, and randomly assigned them to the detected cores. Using Eq.~\eqref{eq:mass}, we computed the mass of each core and built the CMF. We repeated this process 100\,000 times, and we fitted the high-mass end of the CMF ($M>2$~\mo) with a power-law function. The results are shown in Fig.~\ref{fig:CMF-temperature}. We find the slope $\alpha$ to be in the range $-0.7$--$-1.46$, with a standard deviation of 0.10. The value of $\alpha$ derived in \S~\ref{sec:CMF} falls in the centre of this distribution. In summary, variations in the temperature may result in uncertainties of $\sim$45\% in the slope of the CMF. Out of the 100\,000 random tests used to produce Fig.~\ref{fig:CMF-temperature}, we find that in most cases the slope is slightly flatter than the slope of $-1.35$ that usually defines the IMF, although in some cases both numbers are consistent within the errors. The bottom panel of Fig.~\ref{fig:CMF-temperature} shows the distribution of power-law indices of the synthetic CMFs.\\
\indent We investigated a possible relation in temperature with measured fluxes that would reproduce the slope of the IMF. Instead of associating random temperatures to different cores, we assumed a proportionality between flux and temperature.
For this, we considered the empirical expression
\begin{equation}
T=T^\mathrm{obs}_0\left(\frac{S^\mathrm{obs}}{S_0^\mathrm{obs}}\right)^\gamma,
\end{equation}
where $T^\mathrm{obs}_0$ is the minimum assumed temperature for the cores (20~K in the present work), $S^\mathrm{obs}$ is the observed flux density of the cores (listed in Table~\ref{tab:catalog}), and $S^\mathrm{obs}_0$ is the minimum observed flux density (corresponding to 0.1~m\jpb\ in this work). A value of $\gamma=0.238$ is found to reproduce masses that are consistent with the IMF slope of $-1.35$ \citep[see e.g.][]{Salpeter1955}. 

In summary, we find a slightly flatter slope of the CMF suggesting the probable existence of an over excess of massive cores in the \ngc\ clusters. However, an accurate determination of temperatures for individual cores is required to firmly confirm this trend. We note that a similar bias may occur in other regions where an excess of high-mass cores has been found \citep[e.g.][]{Zhang2015, Sanchez2017, Motte2018}.

%
\subsubsection{Resolution bias in the determination of the CMF}\label{sec:resolution bias}

Another possible bias in the determination of the CMF is related to the limited angular resolution of the observations. With the actual resolution of the here-presented ALMA observations (i.e. 1300~au), the higher fragmentation level of a few sources reported by e.g. \citet{Hunter2014} and \citet{Brogan2016} remain unresolved. The blending of multiple (lower mass) sources may result in fragments for which we derive a higher mass, therefore biasing the CMF towards flatter slopes.\\
\indent We considered that each core detected with ALMA can fragment into smaller condensations, assuming Jeans fragmentation. Thus, the number of fragments for each core is given by
\begin{equation}\label{eq:N-fragments}
N_\mathrm{fragments}=\frac{M_\mathrm{core}\times \epsilon_\mathrm{frag} }{M_\mathrm{Jeans}}, 
\end{equation}
where $M_\mathrm{core}$ is the mass of the core (listed in Table~\ref{tab:catalog}), $M_\mathrm{Jeans}$ is the non-thermal or turbulent Jeans mass (see Eq.~\eqref{eq:Mjeans-nth} and \S~\ref{sec:fragmentation-clump}), and $\epsilon_\mathrm{frag}$ is a core formation efficiency which is expected to increase with the average density of the cores \citep[e.g.][]{Bontemps2010}. We considered the following empirical relation
\begin{equation}\label{eq:epsilon}
\epsilon_\mathrm{frag}=\epsilon_{0}\times \left(\frac{n_\mathrm{H_2}^\mathrm{obs}}{n_\mathrm{H_2}^\mathrm{max}}\right)^{p}, 
\end{equation}
where $n^\mathrm{obs}_\mathrm{H_2}$ is the measured Hydrogen volume density for each core (listed in Table~\ref{tab:catalog}), and $n^\mathrm{max}_\mathrm{H_2}$ is the maximum observed density. For any $\epsilon_0$ and $p$ we determined the number of fragments for each core. When the number of fragments is higher than 1, we assigned to each of them a mass equal to $M_\mathrm{Jeans}$. Finally, we built the CMF and determined the slope of the high-mass end. We find that $\epsilon_0=0.4$ and $p=1.3$ reproduce a slope of $-1.35$ in the CMF, equivalent to the IMF and suggesting that the resolution limitations can bias the CMF towards flatter slopes (i.e. excess of massive cores).\\
\indent In order to evaluate the feasibility of the fragmentation efficiency that can reproduce a CMF consistent with the IMF, we considered those cores of \ngc\ that have been observed at higher angular resolutions \citep[e.g.][]{Hunter2014, Brogan2016} and compared our predictions of fragmentation with the observed one. For source 43 we estimate 2 fragments, consistent with the 2 components detected by \citet{Brogan2016}. For source 44 we estimate 9 fragments, in agreement with the 7 components that have been resolved with a resolution of $0\farcs2$. Finally, those cores for which we did not predict further fragmentation also appear as single entities when observed at higher angular resolutions.\\
\indent As shown in the last two sections, there is a realistic range of temperature and fragmentation level that can reproduce the shape of IMF. Therefore, we conclude that the apparent discrepancy between our measured CMF and the IMF may be spurious and higher-angular resolution observations of the whole region, together with more accurate measurements of the temperature, e.g. via the observation of molecular line tracers, may help to better constrain the shape of the CMF.

%
\subsection{Mass segregation}\label{sec:mass-segregation}

\begin{figure*}[!ht]
\begin{center}
\includegraphics[scale=0.39]{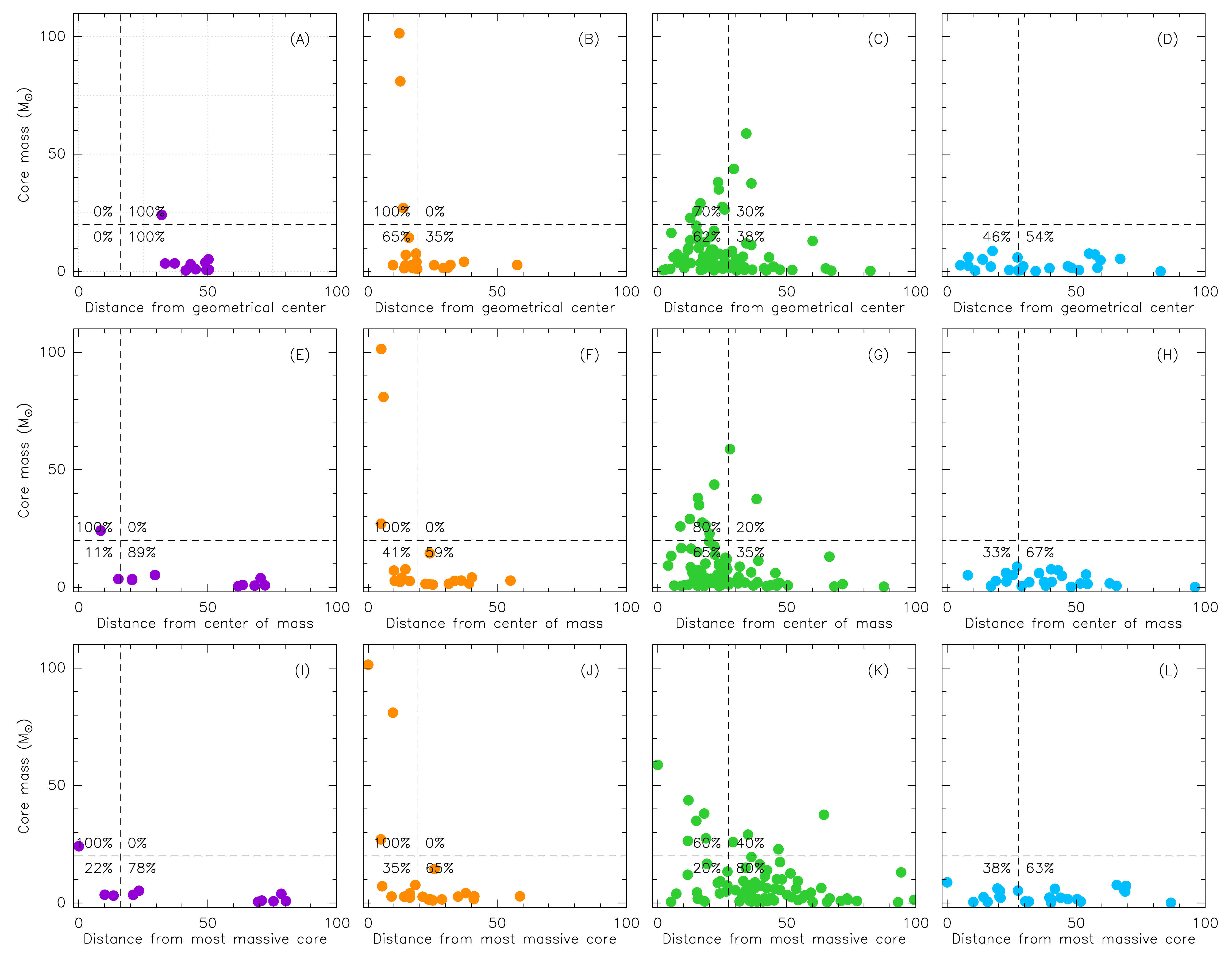}
\caption{Mass of each core against its distance from the geometrical centre of the cluster (top panels), the centre of mass (middle panels), and the most massive core of each cluster (bottom panels). Each column corresponds to a cluster: \ngc-E (first column, in violet), \ngc-I (second column, in orange), \ngc-I(N) (third column, in green), and \ngc-I(NW) (fourth column, in light blue). The vertical and horizontal dashed lines divide each panel in four quadrants with a mass threshold of 20~$M_\odot$ and a distance threshold corresponding to one third of the cluster radius. The numbers indicate the percentage of sources (above and below the mass threshold) that are located in the inner or outer part of the cluster.}
\end{center}
\label{fig:cluster-center}
\end{figure*}

Different observational and theoretical works \citep[e.g.][]{Chandrasekhar1942,Spitzer1969,Mcnamara1986, Zinnecker1993,Bonnell1998, Liu2015, Plunkett2018, Busquet2019} find that in young clusters containing high-mass stars, the most massive members tend to concentrate in the centre, while low-mass members are distributed throughout the whole cluster. This effect is referred to as mass segregation, and its study provides constraints for numerical simulations describing the formation of stellar clusters \citep[e.g.][]{Bonnell1998, Bonnell2006}. Moreover, observing young clusters leads us to investigate the origin of this effect, which can be either primordial or dynamical.\\
\indent We searched for evidence of mass segregation in the \ngc\ clusters. For this, we followed three different approaches. In the first method, we divided the cores of each cluster in low-mass ($M<20$~\mo) and high-mass ($M>20$~\mo) cores, and built individual MSTs for each one of the two populations. Assuming a star formation efficiency of 30\%, a compact source with a mass of $\sim$20~\mo\ ends up with forming a B-type star, thus allowing to separate between potential low-mass and high-mass stars. For cluster \ngc-I, the mean MST pathlength for high-mass cores (4\,700~au) is smaller than the low-mass ones (9\,700~au) suggesting that the massive objects are more closely located with respect to each other. In \ngc-I(N) we find similar pathlengths (6\,200~au) for both sets of cores. Due to the lack of cores with masses above 20~\mo~in \ngc-E and \ngc-I(NW), this analysis can not be performed in these two clusters.\\ 
\indent In the second method, we investigated the distribution of cores relative to the cluster centre, and how the mass distribution changes radially. We determined the centre of the cluster in three different ways: the geometrical centre (which is listed in Table~\ref{tab:clusters}), the centre of mass and the position of the most massive core in each cluster. The difference between the three centres is small in most of the cases, with a mean relative offset of $\sim$9.5\arcsec. In Fig.~\ref{fig:cluster-center}, we plot the mass of each core against its distance to the centre of the cluster. The panels of Fig.~\ref{fig:cluster-center} are divided in four quadrants determined by a threshold for mass and separation considered for mass segregation. Then the percentage of the sources that are located closer or further away than a given separation was computed for both the low-mass and high-mass cores. The actual thresholds adopted are 20~\mo\ to distinguish between high-mass and low-mass cores, and a 20\% of a radius to separate the inner and outer part of a cluster. The fraction of massive cores located close to the centre indicates the level of mass segregation in the cluster. As indicated in Fig.~\ref{fig:cluster-center}, 100\% of the high-mass cores (M $>$20~\mo) in \ngc-I  are located at a distance $<$20\,000~au to the centre showing a high degree of segregation, compared to the low-mass cores that appear more uniformly distributed. In the case of \ngc-I(N), 60\%--80\% of the massive cores are located in the inner regions. In contrary, \ngc-E and \ngc-I(NW) have no massive cores and therefore, no hint of segregation is found. 

\begin{figure*}[t]
\begin{center}
\includegraphics[scale=0.30]{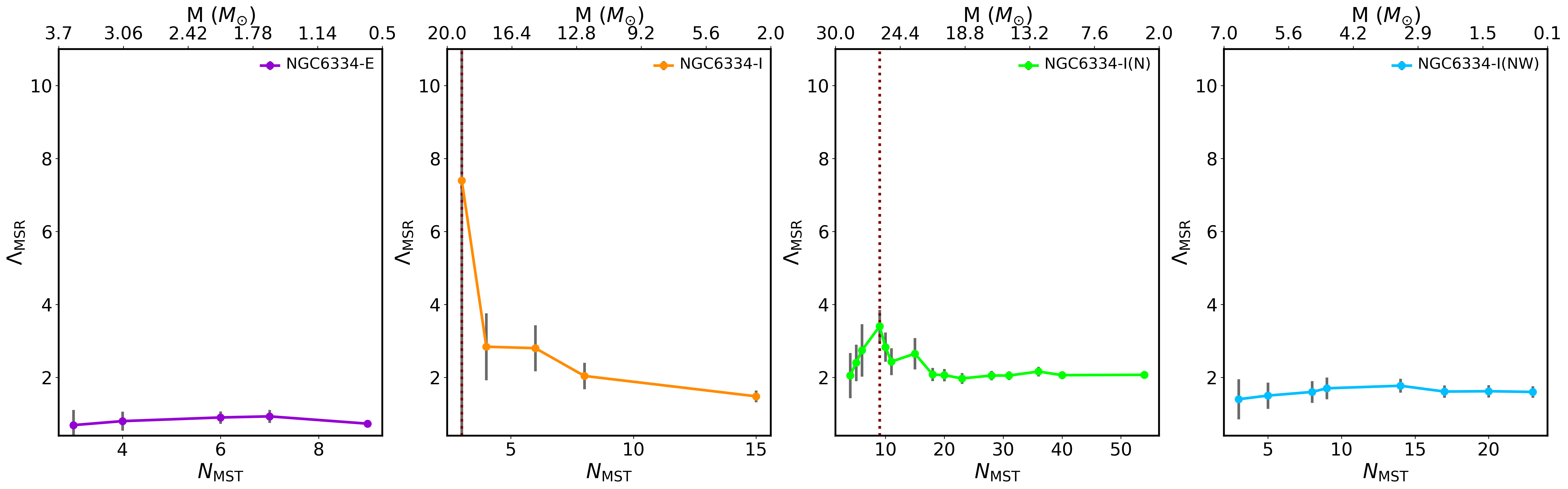}
\caption{The evolution of $\Lambda_\mathrm{MSR}$ with different number of sources, $N_\mathrm{MST}$. Each point of the figure corresponds to a specific mass range. The mass which corresponds to a sharp drop of $\Lambda_\mathrm{MSR}$ defines the segregation threshold. The dashed red lines show the $N_\mathrm{MST}$ at which the mass segregation picks out.
The grey vertical lines show the uncertainty of $\Lambda_\mathrm{MSR}$ as defined in Eq.~(\ref{eq:lambdaMSR}). }
\end{center}
\label{fig:mass-segregation-ratio}
\end{figure*}

\begin{figure*}[!h]
\begin{center}
\includegraphics[scale=0.40]{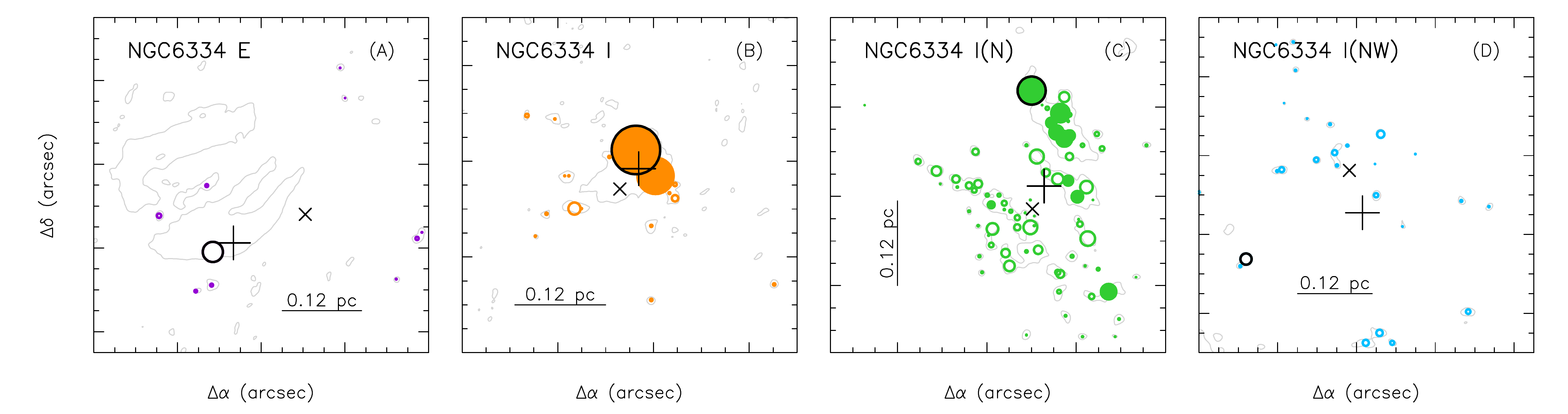}
\caption{Distribution of the cluster members relative to the centre mass marked with a plus sign, and relative to the geometrical centre marked with a cross. The segregated cores defined based on the mass-segregation ratio (see Fig.~\ref{fig:mass-segregation-ratio}) are shown with filled circles. The black circle shows the position of the most massive member in each cluster. The size of the circles correspond to the mass of the cores.}

\label{fig:mass-segregation}
\end{center}
\end{figure*}

In the last approach, we quantified the degree of mass segregation in each cluster using the mass segregation ratio ($\Lambda_\mathrm{MSR}$) defined by \citet{Allison2009} as
\begin{equation}
\label{eq:lambdaMSR}
\Lambda_\mathrm{MSR} = \frac{\langle l_\mathrm{random}\rangle}{l_\mathrm{massive}}\pm\frac{\sigma_\mathrm{random}}{\sigma_\mathrm{massive}},
\end{equation}
where $l_\mathrm{massive}$ is the mean MST length of the $N_\mathrm{MST}$ most massive objects in the cluster, with $\sigma_\mathrm{massive}$  the standard deviation. $\langle l_\mathrm{random}\rangle$ is the average of the mean MST lengths of randomly generated clusters with $N_\mathrm{MST}$ members uniformly distributed, for which its standard deviation is given by $\sigma_\mathrm{random}$. We used 500 randomly generated uniform clusters for a good determination of $\langle l_\mathrm{random}\rangle$. A value of $\Lambda_\mathrm{MSR}\approx1$ implies that the massive objects are distributed in the same way as a uniform cluster, while $\Lambda_\mathrm{MSR}>1$ corresponds to mass segregation in the cluster, i.e. the MST length of the $N_\mathrm{MST}$ most massive objects in the cluster is shorter, and they are more closely distributed compared to a uniform distribution. On the contrary, $\Lambda_\mathrm{MSR}<1$ suggests inverse-mass segregation, with the $N_\mathrm{MST}$ massive objects more widely distributed. In Fig.~\ref{fig:mass-segregation-ratio}, we present the mass segregation ratio $\Lambda_\mathrm{MSR}$ for different number of sources $N_\mathrm{MST}$ (or correspondingly, different masses). For \ngc-E, $\Lambda_\mathrm{MSR}$ fluctuates around unity suggesting no clear signs of mass segregation in this cluster. In contrast, cluster \ngc-I has a $\Lambda_\mathrm{MSR}$ that increases when $N_\mathrm{MST}<$ 3. This corresponds to a mass of 20~\mo, and indicates that cores with higher masses appear more segregated. In \ngc-I(N), despite the fluctuations in the level of segregation, there is an increase when $N_\mathrm{MSR}<9$, and therefore hints of segregation for cores with masses $>$ 25~\mo. Finally, for \ngc-I(NW), the level of mass segregation remains constant with no hints of segregation. It is worth noting that the constant levels of $\Lambda_\mathrm{MSR}$ are slightly above 1. This may suggest that the cores in the \ngc\ clusters are not uniformly distributed, but can be slightly concentrated (see Appendix~\ref{app:lambda-gaussian}).\\
\indent In summary, we find that the three methods considered reveal clear evidence of mass segregation in \ngc-I, and a possible segregation in \ngc-I(N). In \ngc-E and \ngc-I(NW), due to the lack of high-mass cores, no hint of segregation is found.
We note that possible biases (see \S~\ref{sec:T bias} and \S~\ref{sec:resolution bias}) may affect some of these results, and thus higher-angular resolution observations together with accurate temperature determination are necessary to more precisely characterise mass segregation. Finally, in Fig.~\ref{fig:mass-segregation}, we show the four clusters highlighting the position of the centre of mass (cross symbol) as well as the members of each cluster, with the most massive member indicated with a black circle, and with the segregated cores (based on the method shown on Fig.~\ref{fig:mass-segregation-ratio}) highlighted with filled circles.


%

\subsection{Fragmentation status}\label{sec:fragmentation}

As shown in \S~\ref{sec:cluster-identification}, the four clusters in \ngc\ show a number of fragments or cores with a range of masses and densities. The fragmentation process of dense clumps resulting in the formation of clusters is controlled by a combination of gravity, thermal and turbulent pressure as well as other effects such as the presence of magnetic fields. In the following, we investigate the effects of turbulent and thermal pressure in regulating the fragmentation of the clusters in \ngc, by determining the Jeans mass and comparing with the measured mass of the fragments \citep[see e.g.][]{Palau2015, Palau2014}.

%
\subsubsection{Clump fragmentation}\label{sec:fragmentation-clump}

The Jeans length of a gaseous clump is defined as

\begin{equation}
\label{eq:Jeans-length}
\lambda_J=c_{\mathrm{eff}}\left(\frac{\pi}{G\rho}\right)^{1/2}\left(\frac{n_{\text{H}_2}}{10^{5}~\text{cm}^{-3}}\right)^{-1/2},
\end{equation}
where $c_{\text{eff}}$ is the `effective' sound speed, $\rho$ is the density and $G$ is the gravitational constant. The Jeans mass is the mass of the gas contained in a sphere with a radius of $\lambda_J/2$:

\begin{equation}
\label{eq:Jeans-mass}
M_J=\frac{\pi^{5/2}}{6~G^{3/2}}~c_{\mathrm{eff}}^3~\rho^{-1/2}, 
\end{equation}

If we assume that thermal pressure is dominant in the fragmentation process, $c_{\text{eff}}$ corresponds to the sound speed of the gas and the Jeans mass is written as 

\begin{equation}
\label{eq:Mjeans-th}
\frac{M_{\text{Jeans}}^{\text{th}}}{M_{\odot}}=0.6285~\left(\frac{T}{10~{\text{K}}}\right)^{3/2}\left(\frac{n_{\text{H}_2}}{10^{5}~\text{cm}^{-3}}\right)^{-1/2},
\end{equation}
where $T$ is the kinetic temperature of the gas and n$_{\mathrm{H_2}}$ denotes the hydrogen volume density of the sphere. For this study, we obtained these values from the \textit{Herschel} temperature and column density maps \citep{Russeil2013}. \\
\indent If the internal pressure of the gas is dominated by turbulence, $c_s$ is replaced with the non-thermal component of the observed velocity dispersion $\sigma_{\text{1~D,nth}}$ measured from one the dense gas tracers observed with ALMA over the extent of the clusters ($\sigma_{\mathrm{nth}}\sim$0.7~km/s). The turbulent Jeans mass is then defined as \citep{Palau2015}
\begin{equation}
\label{eq:Mjeans-nth}
\frac{M_{\text{Jeans}}^{\text{nth}}}{M_{\odot}}= 0.8255~\left(\frac{\sigma_{\text{1D, nth}}}{0.188~\text{km}~\text{s}^{-1}}\right)^3 \left(\frac{n_{\text{H}_2}}{10^{5}~\text{cm}^{-3}}\right)^{-1/2}.
\end{equation}

The thermal and turbulent Jeans masses estimated for each cluster are listed in Table~\ref{tab:clusters_v2}. 
In the case of \ngc-E, \ngc-I and \ngc-I(N), the observed mean mass is considerably larger than the expected thermal Jeans mass. This suggests that turbulent motions dominate over the thermal pressure in these regions during the fragmentation process. Similar results are found in the massive star forming region G28.34-P1 \citep{Wang2011}, although other works suggest thermal Jeans fragmentation in other high-mass star forming clouds \citep[see e.g.][]{Palau2015}. By contrast, \ngc-I(NW) has a mean mass consistent with thermal Jeans mass. Considering that \ngc-I(NW) contains no massive core, this might suggest that in low-mass star forming regions the fragmentation process is regulated by the thermal pressure.

%
\subsubsection{Filament fragmentation}\label{sec:fragmentation-filament}

\begin{figure}[ht]
\centering
\includegraphics[width=1.0\columnwidth, clip]{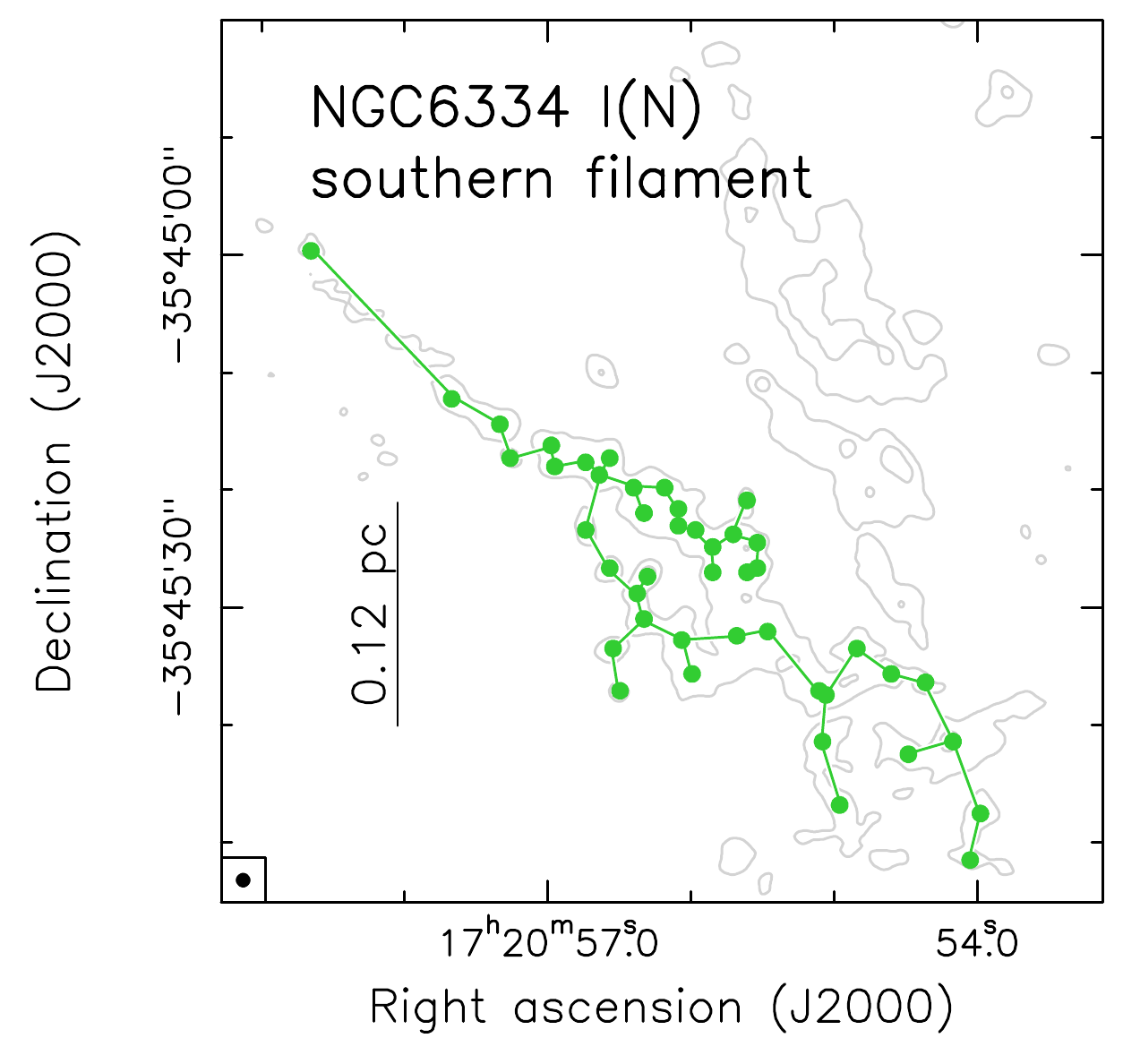}
\caption{Close-up view of the \ngc-I(N) cluster. The green circles mark the positions of the sources that belong to the southern filament. The green segments correspond to the minimum spanning tree of these sources. The extent of the filament is about 0.4~pc.}
\label{fig:filament-frag}
\end{figure}

As shown in Fig.~\ref{fig:alma3mm} (see also Fig.~\ref{fig:filament-frag}), we detected two elongated filament-like structures in \ngc-I(N), harboring most of the compact cores detected in the region. We aim at investigating the relevant fragmentation scale of the filament and test whether the fragmentation level is controlled by turbulent or thermal support. For this, we focus on the southern filament shown in Fig.~\ref{fig:filament-frag} which has an extent of $\sim$0.4~pc. While the super-resolution continuum map reveals more small-scale details, the combined 12~m with ACA image recovers more of the extended filamentary structure. Therefore, to measure the dust mass of the filament we used the combined continuum map. Integrating the flux of the ALMA combined map over the extent of the filament yields a total flux of 0.51~Jy. We subtracted the contribution from the cores embedded in the filament, and assumed an average temperature of 26~K (based on the \textit{Herschel} temperature map of Russeil et al. 2013) to derive a filament mass of 342~\mo. The mass-to-length ratio $M/l$ of the filament can be compared to the critical line mass $(M/l)_{\text{th-crit}}$ calculated using 
\begin{equation}
(M/l)_{\text{th-crit}}=2c_s^2/G=465\,\left(\frac{c_s}{1 \text{km}\,\text{s}^{-1}}\right)^2 M_{\odot}\,\text{pc}^{-1}
\end{equation}
to investigate the stability of the filament \citep{Ostriker1964b, Nagasawa1987, Inutsuka1992, Fiege2000a, Fischera2012}. If the line mass of the filament exceeds the critical value, the filament is gravitationally unstable against collapse. The mean line-mass of the selected filament in \ngc-I(N) is (M/l)=925~$M_{\odot}\,\text{pc}^{-1}$. As we neglect the inclination of the filament, the derived line-mass is an upper limit. For a thermally supported filament at a temperature of 26 K, the critical line-mass becomes (M/l)$_{\text{th-crit}}$=43 $M_{\odot}\,\text{pc}^{-1}$. Thus, the filament is highly super-critical, and therefore, self-gravity overcomes the internal pressure. This would lead the filament to undergo collapse. In this scenario, fragmentation into cores along the main axis of the filament may happen. The expected separation between fragments is estimated \citep[see e.g.][]{Trevino2019, Wang2014} as
\begin{equation}
\lambda_{\mathrm{cl}}= 22 \nu (4 \pi G \rho_{\mathrm{c}})^{-1/2}=
1.24\;\left(\frac{\sigma}{1\;\text{km}\;\text{s}^{-1}  }\right)\left(\frac{ n_c }{ \text{10}^{5}\; \text{cm}^{-3} }\right)^{-1/2}\;\text{pc},
\end{equation}
where $n_c$ is the particle number density of the gas in the centre of the filament. The fragment mass is therefore
\begin{equation}
M_{\mathrm{cl}}=(M/l)_{\mathrm{th-crit}}\times\lambda=575.3M_{\odot}~\left(\frac{\sigma}{1\;\text{km}\;\text{s}^{-1}  }\right)^{3}\left(\frac{ n_c }{ \text{10}^{5}\; \text{cm}^{-3} }\right)^{-1/2}
\end{equation}
The central density of the filament amounts to $n_c=1.2 \times 10^6$~cm$^{-3}$. Adopting a velocity dispersion of 0.7~km~s$^{-1}$ based on one of the observed dense gas tracers (Sadaghiani \et\ in prep.), the typical spacing and fragment mass are estimated to be 0.2~pc and 59~\mo\ respectively, while the thermal support (26~K and $\sigma_{\mathrm{th}}=$0.08~km/s) yields 0.002~pc and 0.08~\mo. \\
\indent Our ALMA observations of the dust continuum emission at 3~mm resolve the filament into a chain of 42 cores. The mean projected separation between the identified cores along the filament is 0.022~pc, which should be taken as a lower limit if projection effects are taken into account. The median mass of the cores embedded in the filament is 5~\mo. This suggests that turbulent motions within the filament contribute to the fragmentation process \citep[see e.g.][]{Seifried2015}. Turbulence in the filament can originate from, in addition to micro-scale turbulent cascades, the relative motions of the cores within the filament, which can increase the level of turbulence in the surroundings of the filament \citep[see e.g,][]{Clarke2017}. 


%
\section{Discussion}\label{sec:discussion}

%
\subsection{Hierarchical fragmentation in \ngc}\label{sec:hierarchical}

\begin{figure*}[!ht]
\begin{center}
\includegraphics[scale=0.4]{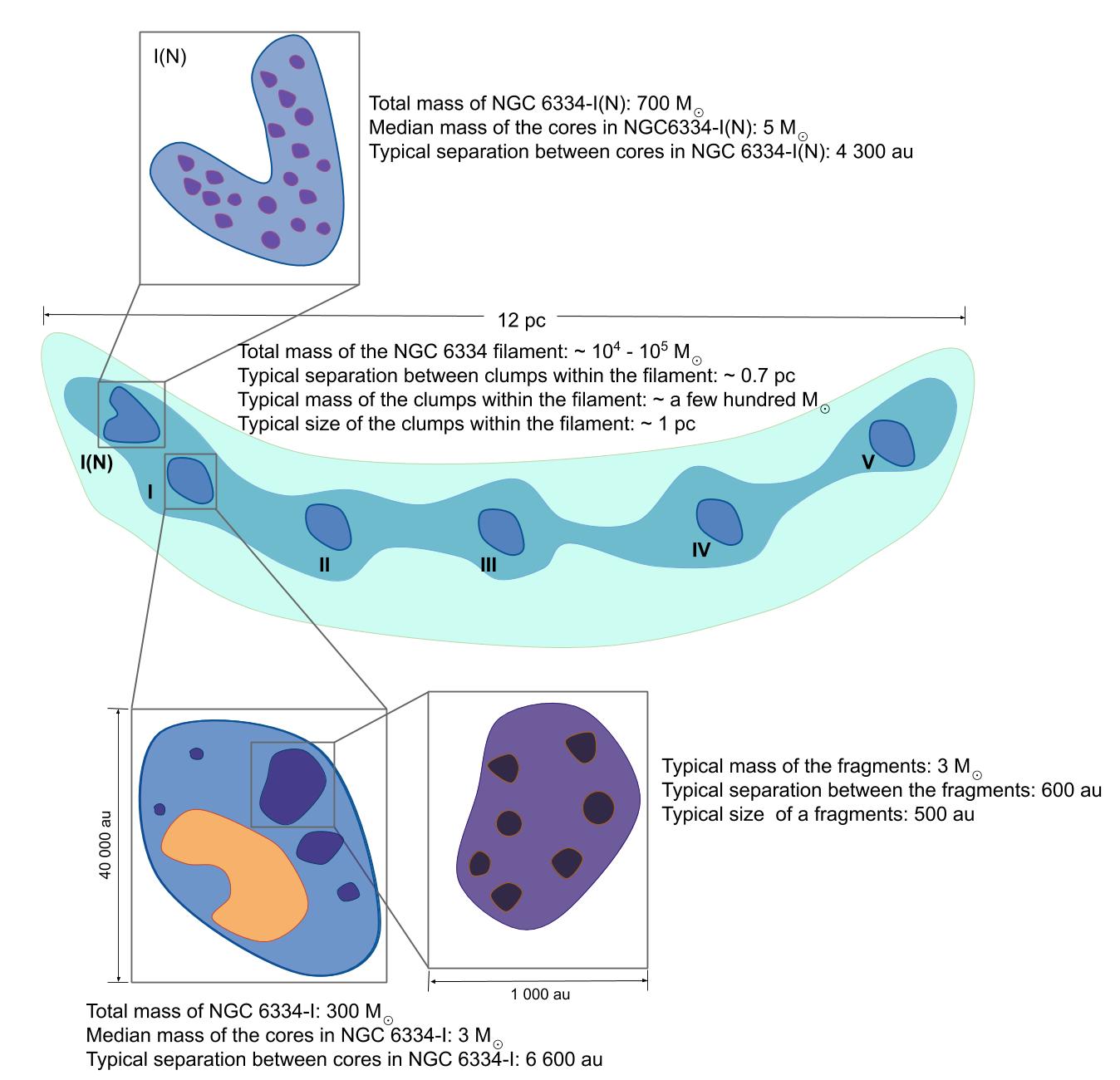}
\caption{Schematic view of the central part of the \ngc\ filamentary cloud. The distribution is based on \textit{Herschel} observations by \citet{Russeil2013}. The horizontal and vertical orientations correspond to the Galactic longitude and latitude coordinates. The sketch displays the hierarchical fragmentation seen in the cloud from the scale of the filament down to the scale of individual dense cores and protostars. The colors are scaled based on the density (from light blue to black). In this color scheme, the lighter colors (e.g. blue) correspond to the lower densities, while the darker colors (e.g. violet and black) represent the dense regions. The ionised gas is shown with orange. 
The values of masses and sizes are based on the observations presented in this work and from the literature \citep[][]{Brogan2016, Hunter2014}.}
\label{fig:fragmentation}
\end{center}
\end{figure*}

In this section we discuss the different levels of fragmentation found in the \ngc\ star-forming complex. We explore the region from the large-scale filament cloud (a few 10~pc) down to small scales where cores (a few 1000~au) reveal the sites of formation of individual stars or small systems. The different levels of structures are depicted in the sketch shown in Fig.~\ref{fig:fragmentation}. The whole molecular cloud extends for about 50~pc in an elongated filamentary structure aligned with the galactic disk \citep{Persi2008,Russeil2013,Sanchez2015}. Its central part covering an extent of about 12~pc has larger densities \citep[$\sim$10$^6$ $\mathrm{cm}^{-3}$, e.g.][]{Zernickel2013} and appears to be collapsing and contracting. Simultaneously, the filament seems to be fragmenting into a number of objects with sizes of 1~pc and (gas and dust) masses of a few hundred \mo\ \citep[e.g.][]{Russeil2013}. These objects, which are called clumps or embedded protoclusters, have been identified via \textit{Herschel} observations at sub-millimetre wavelengths and consist of a few tens of objects with a typical separation of a few ten thousands~au.\\ 
\indent The second level of fragmentation is found within the clumps, where a number of smaller dense condensations with sizes of a few hundred to a few thousand~au are found. These objects, called cores, are easily detected in our ALMA observations at millimetre wavelengths. Within \ngc-I and I(N) we detect strong grouping of ALMA dense cores at separations below 20\,000~au. We also find that the frequency of nearest neighbour separations of the cores increases down to our resolution limit. The mean typical separation of the cores is 3\,700~au, and is in agreement with turbulent pressure support regulating the fragmentation process. Particularly for \ngc-I(N), and in addition to the cores, we find two elongated filament-like structures with an extent of about 82\,000~au and widths of about 19\,000~au. This suggests that small-scale filaments can appear as substructures connecting the scales between the clumps and cores.\\
\indent Observations at sub-arcsecond angular resolution \citep{Hunter2014, Brogan2016} reveal that some of the cores (e.g. sources 43 and 44) are resolved into multiple fragments at a scale of a few hundred~au. The further fragmentation of the dense cores is in agreement with numerical simulations that predict that core fragmentation at scales between 500--5\,000~au is an efficient formation mechanism of bound clusters \citep{Offner2010}. Using Eq.~\ref{eq:Jeans-length},
the typical separation between the cores expected from the thermal support for the hot core in \ngc-I\ (source 42) is estimated to be 412~au. 
This value is in agreement with the observed typical projected distance between the fragments (620~au) of this hot core. This point suggests that thermal fragmentation becomes dominant at small scales, while at clump scales, non-thermal fragmentation seems to be at work. \\
\indent Thermal Jeans fragmentation at small scales is found in other regions \citep{Liu2019,Beuther2018,Pokhrel2018,Palau2018,Palau2015}. \citet{Palau2015} studied 19 regions in nearby molecular clouds at $\sim$20\,000~au scales and detected a sample of low-mass cores with masses consistent with the thermal Jeans mass. Similarly, \citet{Pokhrel2018} studied multi-scale structures in the Perseus molecular cloud from the scale of the cloud to the scale of dust and gas around protoclusters. They conclude that at all scales pure thermal support is dominant. Although this trend is not consistent with our observations at the clump scales, however, it confirms the role of thermal pressure at scales of a few hundred~au. A more recent study by \citet{Palau2018} towards the Orion Molecular Cloud 1 South (OMC-1S) also shows that fragmentation at scales of $\sim$1100~au is in agreement with thermal Jeans fragmentation. \\
\indent In summary, the multi-scale observations toward \ngc\ indicate that fragmentation is a scale-dependent process which is governed by turbulent pressure at large scales at which the massive clumps and cores are forming. However, at small scales thermal gravitational fragmentation is more important. 

%
\subsection{Evolutionary sequence}

The ALMA observations towards \ngc presented here reveal the presence of four well-defined groups or clusters: \ngc-E, \ngc-I, \ngc-I(N), and \ngc-I(NW). Each of them has different properties in terms of number of dense cores detected, number of infrared sources, presence of \hii~regions, separation between members of the clusters, and distribution of masses of the members of the clusters. In the following, we investigate if the different properties may be related to a different evolutionary stage of each cluster.\\
\indent The association of the protoclusters with hot \hii\ regions or just with cold dust objects is likely related to their evolutionary stage, since \hii\ regions appear at the end of the evolutionary sequence that forms massive stars \citep{Kurtz2005, Cesaroni2005}. Based on this criterion, \ngc-E already harboring a well-developed \hii~region, is classified as the most evolved of the four clusters. In a younger stage we can find \ngc-I which is associated with an ultracompact (UC) \hii~region \citep[see][for classification of \hii~regions]{Kurtz2000,Peters2010}. Following, we can place \ngc-I(N) in an earlier stage, since the \hii~region identified within it, is more compact and would classify as a still embedded hypercompact \hii~region. Finally, \ngc-I(NW) only harbors dense cores, with no clear presence of \hii~regions, indicating that this cluster is in an earlier evolutionary stage or that no massive stars are forming in it (in which case we can not clearly classify the cluster in an evolutionary stage). Complementing the presence and properties of \hii~regions, we can also make use of the infrared stellar content. The ratio of mm~sources per young stellar objects (infrared sources) provides a simple diagnostic for the evolutionary stage of each cluster. Based on this, \ngc-E with 2 mm sources per young stellar object, and \ngc-I with 3 mm sources per young stellar object are the most evolved clusters. \ngc-I(N) with 4 mm sources per young stellar object follows \ngc-I. It is worth noting that \ngc-I(NW) with 2 mm sources per young stellar objects does not follow the expected evolutionary trend, suggesting that this can be a relatively evolved cluster with no massive stars.
In summary, we can classify the identified clusters from less evolved to more evolved in the following order: \ngc-I(N) as the youngest, followed by \ngc-I and with \ngc-E as the most evolved.\\
\indent We make use of the evolutionary classification of the clusters to search for variations in relevant parameters likely associated with the evolutionary stage. In particular, we investigate the variation of typical separations between cluster members, its distribution throughout the cluster, the presence of mass segregation effects and properties on the mass distribution of the members. From the MST analysis and the derived mean and median distances between neighbours (see \S~\ref{sec:spatial-clusters} and Table~\ref{tab:clusters}), we find a small trend with increasing separation as the cluster is more evolved (excluding \ngc-I(NW) from the comparison). This is consistent with the scenario proposed by \citet{Pfalzner2013} where the separation between cluster members increases with time due to gas expulsion and stellar ejections. A comparison of the KDEs of the member separations in each cluster (cf.\ Fig.~\ref{fig:KDEs-MST}) shows that the cores are widely distributed in \ngc-E and \ngc-I, while the cluster members appear more tightly distributed in \ngc-I(N).
It is worth noting that for \ngc-E the observed compact sources may be not related to the original cluster, but represent a new population of cores that has recently formed with the expansion of the \hii\,region.\\
\indent The structure of the clusters is evaluated with the $Q$ parameter (see \S~\ref{sec:spatial-clusters}). Despite the uncertainties due to low number statistics in some clusters, we find large and close to 0.8 values for clusters \ngc-I, \ngc-I(N) and \ngc-I(NW) suggesting that these clusters are concentrated with a relatively smooth radial profile. In contrast, \ngc-E shows a lower $Q$ value suggesting some sub-structure within it. These results can be compared with other observational and numerical simulation studies. \citet{Sanchez2010} studied a sample of 16 open clusters finding that the less evolved clusters tend to exhibit fractal patterns, while older clusters have their stars distributed in centrally concentrated structures. A similar result is reported in the simulations of \citet{Maschberger2010} where the early evolutionary stage of a cluster in a turbulent environment shows sub-structure (low $Q$ values) and evolves into a radially concentrated structure (large $Q$ values) after a certain number of free-fall times). Thus, the $Q$ parameter seems to be a measure of evolution. However, we can not confirm this trend in our sample of four clusters due to the large uncertainties in the measurements of the $Q$ parameter. Observations of other clusters within the \ngc\ star-forming complex may help to better investigate the relation of the $Q$ parameter with the evolutionary stage of star-forming clusters still embedded in large amounts of gas and dust.

Regarding mass segregation, we find (see \S~\ref{sec:mass-segregation}) hints of segregation only for clusters \ngc-I and \ngc-I(N), with the massive members being more segregated for the more evolved cluster \ngc-I. This suggests a possible dynamical origin in the mass segregation \citep{Allison2009r}. In this scenario, massive stars lose their energy through two- and many-body interactions with low-mass stars and hence sink to the centre of the cluster. Based on N-body simulations performed by \citet{Allison2009r}, the dynamical segregation in young clusters can be possible only if they are born in a cool and clumpy distribution. According to this scenario, it is possible to speculate that \ngc-I(NW) will be segregated in a later stage of its evolution, while \ngc-E has lost the hints of segregation. On the other side, the relaxation time scale \footnote{The relaxation time scale is determined as $t_\mathrm{relax}=\frac{N}{6\log(N)}\frac{1}{\sqrt{G\rho}}$, where $N$ is the number of cluster members, $\rho$ is the average H$_2$ density of the cluster and $G$ is the gravitational constant \citep[][]{Spitzer1971}. The free-fall time scale is calculated as $t_\mathrm{ff}=\frac{3\pi}{\sqrt{32G\rho}}$.} for \ngc-I and \ngc-I(N) ($\sim$10$^6$~yr) is one order of magnitude larger than the free-fall time ($\sim$10$^5$~yr). This puts the scenario of dynamical segregation in doubt. Moreover, the presence of prominent filamentary structures in \ngc-I(N) together with mass segregation, challenge the scenario of dynamical segregation in this cluster. The dynamical evolution that may have created the observed mass segregation should have also dispersed or removed structures like the filaments, resulting in a more homogeneous distribution of cores. Therefore, the presence of filamentary structures may be an evidence for primordial mass segregation in \ngc-I(N).\\
\indent Finally, the CMFs of the different clusters show a steeper slope for the more evolved cluster \ngc-E compared to the younger clusters \ngc-I and \ngc-I(N). Despite the low statistics for cluster \ngc-E, we can consider that the members of the younger clusters still have large amounts of gas and mass, resulting in more massive objects. Contrary to that, the members of the most evolved cluster \ngc-E may have already exhausted or accreted the gas and dust mass, resulting in less massive cores. However, one has to be cautious since the members of cluster \ngc-E could represent, as mentioned above, a second generation of cores with lower masses.\\
\indent The analysis presented here may constitute the first step towards a comprehensive study of a larger sample of high-mass (proto)clusters. Increasing the number of studied regions in \ngc\ will produce a sample of clusters with different properties which are forming from an homogeneous high-mass star-forming filament. Therefore, such a study will provide us with statistically reliable results of the cluster properties and its relation to the evolutionary stage and environment.

%
\section{Summary}\label{sec:summary}

We observed a portion of the high-mass star forming complex \ngc\ with ALMA. We analysed the continuum emission at 86.7~GHz, at an angular resolution of 1\arcsec~(corresponding to 1\,300~au). Our main findings are summarised as follows:

\begin{itemize}

\item We identified 142 compact continuum sources distributed over the surveyed region of 9~arcmin$^2$ (or 1.4~pc$^2$) with a mass completeness level of 2~\mo. This results in an average density of 40 cores pc$^{-2}$. The cores have radii in the range 390 to 6800~au, masses in the range 0.1 to 100~\mo, and densities 10$^{7}$ to 10$^{9}$~cm$^{-3}$.

\item The ALMA compact sources seem to be grouped in different structures. We used seven different machine-learning algorithms and identified four different clusters in the region: \ngc-E, \ngc-I, \ngc-I(N) and \ngc-I(NW). Among them, \ngc-I(N) is the most populated cluster with 80 cores.

\item We used the MST technique to determine the typical distances between cluster members. The typical separations for the clusters are in the range from 4\,000~au to 12\,000~au.

\item We studied the CMF of the different clusters and found an excess of high-mass cores compared to the IMF, challenging the idea of a self-similar mapping from the CMF to the IMF. We evaluated if this difference may be due to biases in the determination of the CMF due to uncertain temperature values (necessary for the mass determination), and limited angular resolution observations (unresolved fragments of the clusters). We found that these effects may considerably affect the slope of the CMF explaining the difference with respect to the IMF. Temperature effects and resolution limitations may also affect the excess of massive cores found in other high-mass star forming regions.

\item We searched for evidence of mass segregation in the clusters adopting three different methods. Clusters \ngc-I and I(N), both harboring massive cores, show evidence of mass segregation with the most massive cores located in the central regions of the clusters.

\item The masses and separations of the cores in the clusters \ngc-I and \ngc-I(N) are consistent with fragmentation of the original dense clump being regulated by turbulent motions.
The cluster \ngc-I(NW), with no clear signs of the presence of massive cores or stars, has median masses of the cores consistent with thermal Jeans fragmentation. This may suggest that the fragmentation of the clusters in the \ngc~star forming complex may be dominated by turbulent motions (or magnetic fields) in clusters forming high-mass stars, while it is mainly thermal dominated for low-/intermediate-mass star forming clusters.

\item We combined information at different scales for the \ngc~region (from the large-scale filament of few 10~pc, down to cores of few thousand~au), to establish a hierarchical fragmentation picture of the region. Fragmentation in \ngc~seems to be governed by turbulent pressure at large scales, while at small scales, thermal fragmentation becomes more relevant.

\end{itemize}

In conclusion, we investigated a possible correlation between the evolutionary stage of the identified clusters and their physical properties. Based on the content of the infrared sources and \hii\ regions, we classified the clusters from younger to more evolved: \ngc-I(N), \ngc-I, and \ngc-E. Cluster \ngc-I(NW) could be younger than \ngc-I(N), or could be a more evolved cluster with no presence of high-mass stars. We found that the median separation between the cluster members shifts to larger distances as the cluster evolves. This trend is consistent with theoretical models where gas expulsion and stellar ejections determine the separations of the cluster members. 
We also found strong hints of mass segregation in \ngc-I and \ngc-I(N). The presence of filaments in \ngc-I(N) indicates that mass segregation in this cluster has a primordial origin, while for \ngc-I it is not possible to derive a firm conclusion, and dynamical segregation could have played a role.
Finally, we found an excess of massive cores in clusters \ngc-I and \ngc-I(N), while \ngc-E indicates a lack of high-mass dense cores. This finding likely suggests that in more evolved clusters the gas reservoir is already accreted. In contrast, younger clusters include larger amount of gas and dust reservoir in the form of massive dense cores. The study of a larger sample of clusters in \ngc\ is necessary to derive more reliable conclusions on the possible variation of cluster properties with evolutionary stage. 

%
\begin{acknowledgements}

We thank the referee for the comprehensive review of the manuscript and valuable comments that significantly helped to improve the text. M. Sadaghiani, {\'A}. S{\'a}nchez-Monge, P. Schilke, A. Aghababaei and D.~Seifried research is carried out within the Collaborative Research Centre 956 (subproject A6, C5 and C6), funded by the Deutsche Forschungsgemeinschaft (DFG) - project ID 184018867. S.~D.~Clarke acknowledges support from the ERC starting grant No. 679852 ‘RADFEEDBACK’. J.~M.~Girart is supported by the MINECO (Spain) AYA2017-84390-C2-2-R coordinated grant.
This paper makes use of the following ALMA data: ADS/JAO.ALMA\#2015.1.00230.S. ALMA is a partnership of ESO (representing its member states), NSF (USA) and NINS (Japan), together with NRC (Canada), MOST and ASIAA (Taiwan), and KASI (Republic of Korea), in cooperation with the Republic of Chile. The Joint ALMA Observatory is operated by ESO, AUI/NRAO and NAOJ.

\end{acknowledgements}

%

%
\begin{appendix}

%
\section{Finding charts}\label{app:charts}

Table~\ref{tab:catalog} lists the catalogue of continuum compact sources detected with ALMA at 87.6 GHz towards the \ngc\ surveyed region. The table lists an identifier for each detected source (column~1), the right ascension and declination coordinates (columns~2 and 3), the intensity peak and flux density at 87.6~GHz (columns~4 and 5) and the size of the source given as the major and minor axis of an ellipse and its position angle (column~6). Columns~7 and 8 indicate the presence of radio continuum and infrared counterparts. The next columns list physical parameters such as the temperature (column~9), the gas and dust mass of the core (column~10), the volume H$_2$ density (column~11) and the H$_2$ column density (column~12). Finally the association with a cluster and its membership level are listed in columns~13 and 14. The last column shows the association of the detected source with previous sources reported in the literature.

\begin{figure*}[ht!]
\centering
\begin{tabular}{c c}
\includegraphics[width=0.45\textwidth]{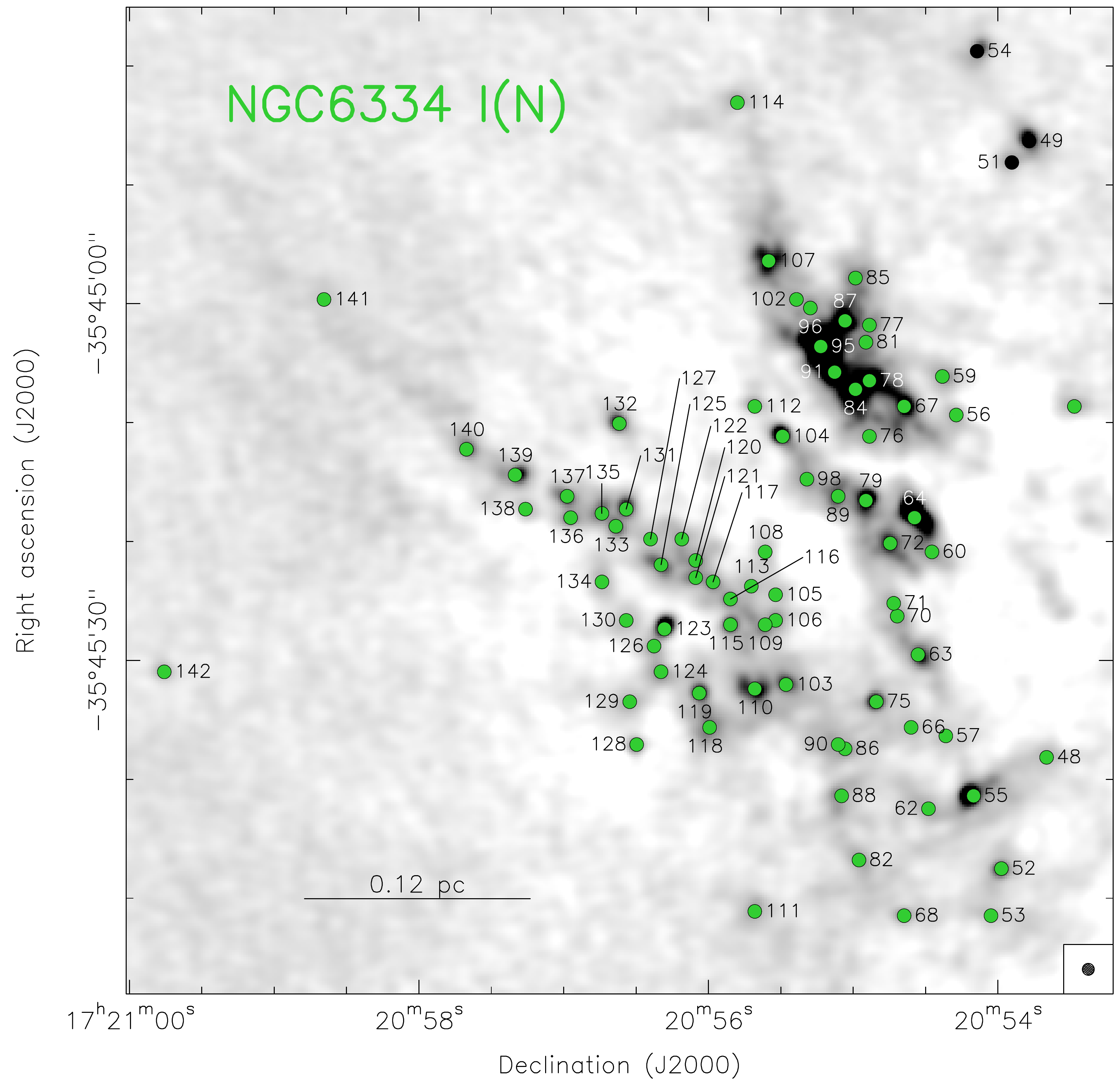} &
\includegraphics[width=0.45\textwidth]{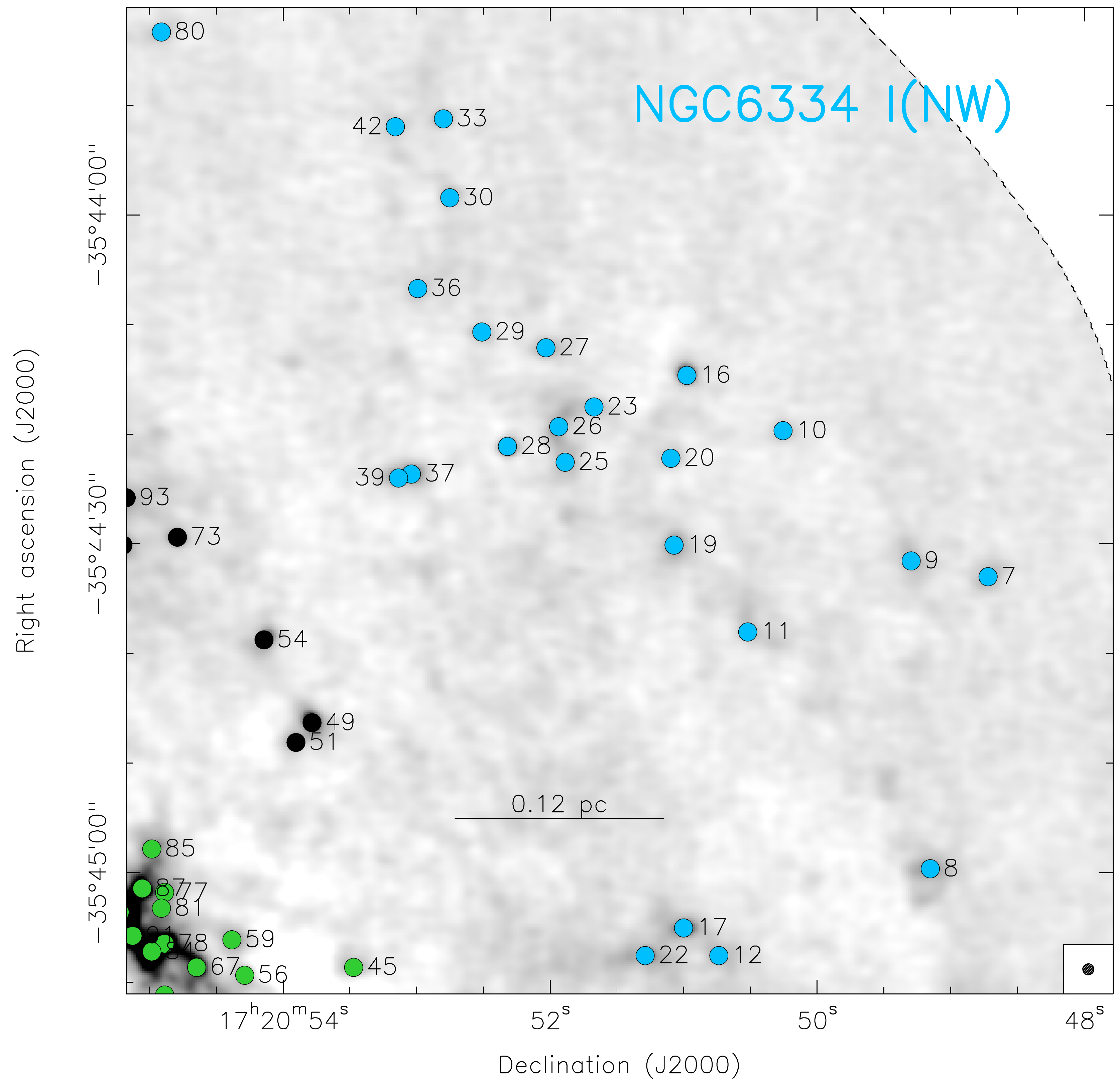} \\
\includegraphics[width=0.45\textwidth]{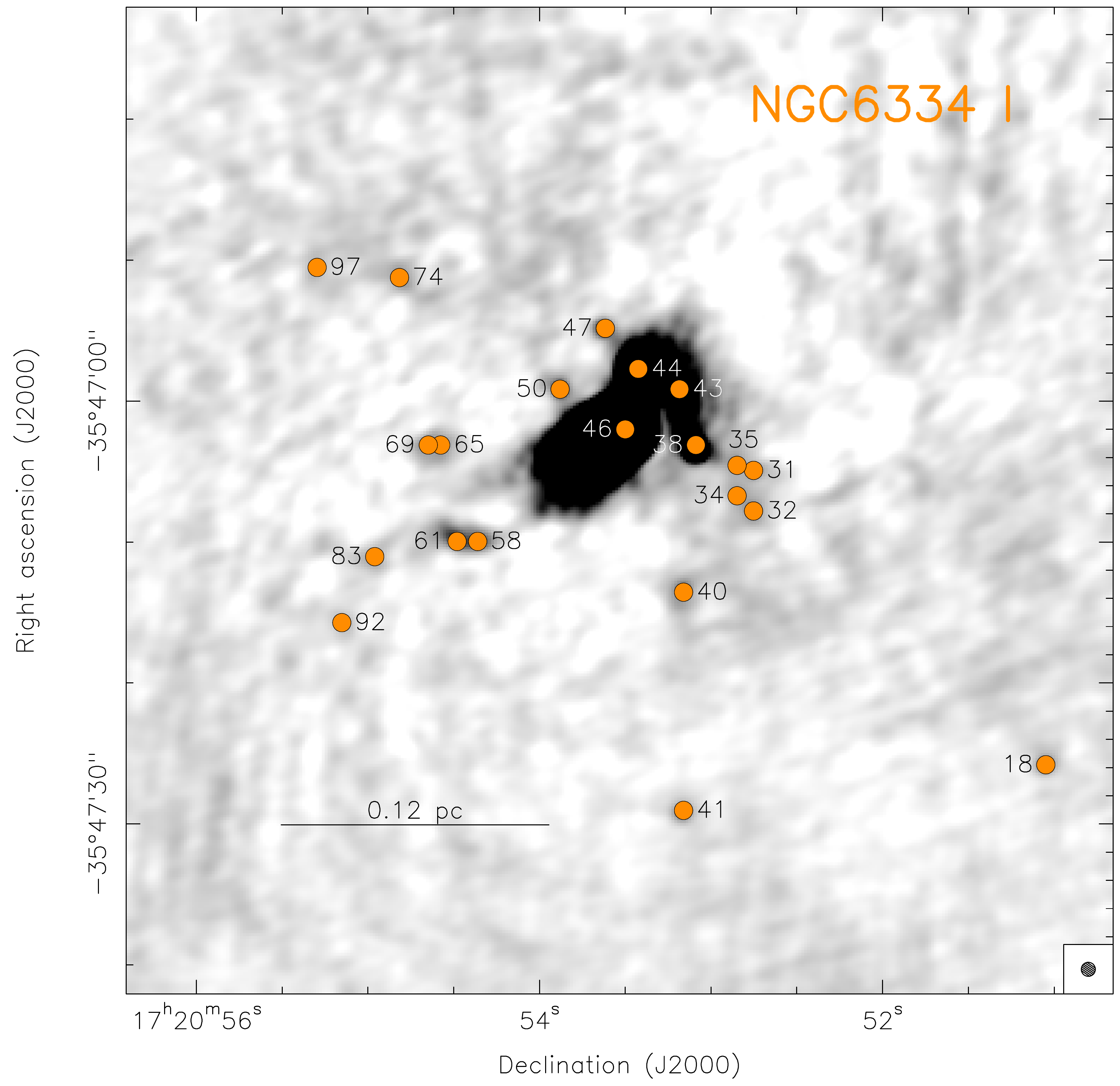} &
\includegraphics[width=0.45\textwidth]{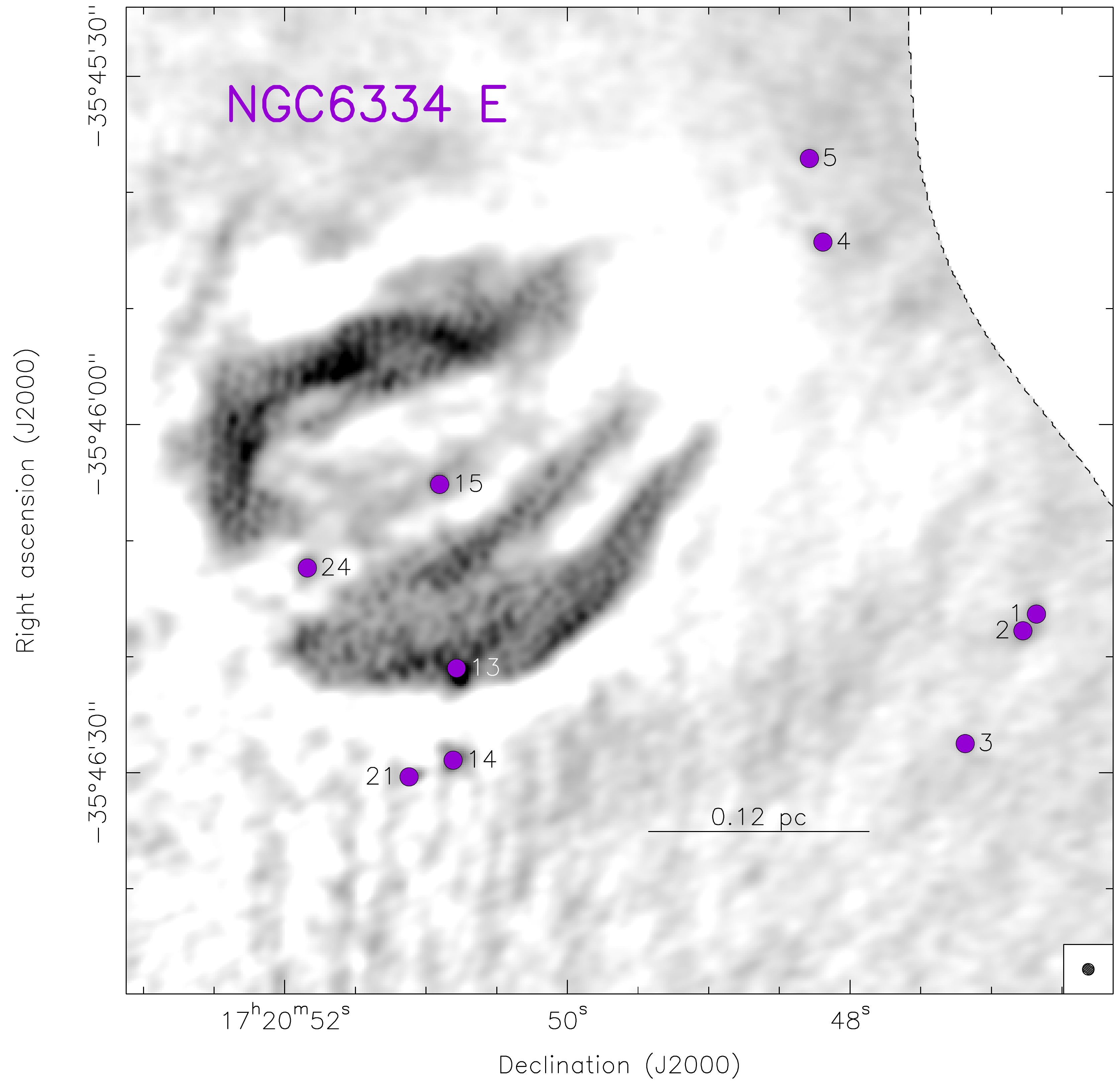} \\
\end{tabular}
\caption{Close-up view of the 4 different clusters identified towards the observed region with ALMA. The identifiers of each source are marked as listed in column~1 of Table~\ref{tab:catalog}.}
\label{fig:finding-charts}
\end{figure*}

In Fig.~\ref{fig:finding-charts} we show close-up views of the different regions of surveyed area (see Fig. 1 for an image of the whole region), and mark the identifiers of each source as listed in Table~\ref{tab:catalog}. 

\longtab[1]{
\begin{landscape}
\begin{longtable}{r c c c c c c c c c c c c c c c c l}
\caption{\label{tab:catalog} Properties of the ALMA continuum sources in \ngc} \\
\hline\hline

& \multicolumn{5}{c}{ALMA 87.6~GHz compact sources\tablefootmark{a}}
&
& \multicolumn{2}{c}{Counterparts\tablefootmark{b}}
&
& \multicolumn{4}{c}{Physical properties\tablefootmark{c}}
&
& \multicolumn{2}{c}{Membership\tablefootmark{d}}
& 
\\
\cline{2-6}\cline{8-9}\cline{11-14}\cline{16-17}
\multicolumn{1}{c}{\#}
& R.A.
& Dec.
& $I_\mathrm{\nu}$
& $S_\mathrm{\nu}$
& $\theta_\mathrm{S}$, P.A.
&
& free-free 
& IR
&
& $T$
& $M_\mathrm{g+d}$
& $n_\mathrm{H_2}$
& $N_\mathrm{H_2}$
&
& Cluster
& \%
& \multicolumn{1}{c}{Others\tablefootmark{e}}
\\
\multicolumn{1}{c}{(1)}
& (2)
& (3)
& (4)
& (5)
& (6)
& 
& (7)
& (8)
&
& (9) 
& (10)
& (11)
& (12)
&
& (13)
& (14)
& \multicolumn{1}{c}{(15)}
\\
\hline
\noalign{\smallskip}

\endfirsthead
\caption{Continued.}\\
\hline\hline

& \multicolumn{5}{c}{ALMA 87.6~GHz compact sources\tablefootmark{a}}
&
& \multicolumn{2}{c}{Counterparts\tablefootmark{b}}
&
& \multicolumn{4}{c}{Physical properties\tablefootmark{c}}
&
& \multicolumn{2}{c}{Membership\tablefootmark{d}}
& 
\\
\cline{2-6}\cline{8-9}\cline{11-14}\cline{16-17}
\multicolumn{1}{c}{\#}
& R.A.
& Dec.
& $I_\mathrm{\nu}$
& $S_\mathrm{\nu}$
& $\theta_\mathrm{S}$, P.A.
&
& free-free 
& IR
&
& $T$
& $M_\mathrm{g+d}$
& $n_\mathrm{H_2}$
& $N_\mathrm{H_2}$
&
& Cluster
& \%
& \multicolumn{1}{c}{Others\tablefootmark{e}}
\\
\multicolumn{1}{c}{(1)}
& (2)
& (3)
& (4)
& (5)
& (6)
& 
& (7)
& (8)
&
& (9) 
& (10)
& (11)
& (12)
&
& (13)
& (14)
& \multicolumn{1}{c}{(15)}
\\
\hline
\endhead
\hline
\endfoot
  1 & 17:20:46.680 & $-$35:46:16.32 & \phnn1.27 & \phnn$1.52\pm0.25$     & $1.70\times1.42$, \phn70
    && --    & f$^*$  && \phn50 & \phnn0.82 & \phn2.34 & \phn1.78 
    && E     & \phn86 & \\
  2 & 17:20:46.776 & $-$35:46:17.76 & \phnn1.64 & \phnn$2.71\pm0.43$     & $2.22\times1.68$, \phn25
    && --    & --    && \phn20 & \phnn3.90 & \phn5.74 & \phn5.43 
    && E     & \phn86 & \\
  3 & 17:20:47.184 & $-$35:46:27.48 & \phnn1.03 & \phnn$0.70\pm0.11$     & $1.70\times1.20$, \phn54
    && --    & --    && \phn20 & \phnn1.01 & \phn7.78 & \phn4.24 
    && E     & \phn86 & \\
     4 & 17:20:48.192 & $-$35:45:44.28 & \phnn0.70 & \phnn$0.92\pm0.16$     & $1.50\times1.38$, 309
    && --    & f$^*$  && \phn50 & \phnn0.50 & \phn1.81 & \phn1.27 
    && E     & \phn86 & \\
  5 & 17:20:48.288 & $-$35:45:37.08 & \phnn1.30 & \phnn$1.36\pm0.22$     & $1.50\times1.38$, 346
    && 4--100 & YSO~I  && \phn50 & \phnn0.73 & \phn2.63 & \phn1.85 
    && E     & \phn86 & \\
   6 & 17:20:48.620 & $-$35:48:26.69 & \phnn0.54 & \phnn$0.37\pm0.07$     & $0.66\times0.42$, \phn25
    && -- & f$^*$  && \phn50 & \phnn0.2 & \phn1.84 & \phn0.95 
    && --     & -- & \\       
  7 & 17:20:48.720 & $-$35:44:33.00 & \phnn0.74 & \phnn$1.12\pm0.18$     & $1.94\times1.26$, \phn53
    && --    & --    && \phn20 & \phnn1.61 & \phn4.35 & \phn3.36 
    && I(NW) & \phn86 & \\
  8 & 17:20:49.152 & $-$35:44:59.64 & \phnn1.48 & \phnn$3.78\pm0.60$     & $3.14\times1.86$, \phnn2
    && --    & --    && \phn20 & \phnn5.44 & \phn4.11 & \phn4.86 
    && I(NW) & \phn86 & \\
  9 & 17:20:49.296 & $-$35:44:31.56 & \phnn1.01 & \phnn$1.59\pm0.26$     & $2.00\times1.46$, 306
    && --    & --    && \phn20 & \phnn2.29 & \phn4.83 & \phn4.05 
    && I(NW) & \phn86 & \\
 10 & 17:20:50.256 & $-$35:44:19.68 & \phnn0.51 & \phnn$0.34\pm0.06$     & $1.02\times0.92$, 343
    && --    & --    && \phn20 & \phnn0.49 & \phn5.63 & \phn2.68 
    && I(NW) & \phn86 & \\
 11 & 17:20:50.520 & $-$35:44:38.04 & \phnn0.60 & \phnn$0.42\pm0.07$     & $1.26\times0.88$, 291
    && --    & --    && \phn20 & \phnn0.60 & \phn5.39 & \phn2.79 
    && I(NW) & \phn86 & \\
 12 & 17:20:50.736 & $-$35:45:07.56 & \phnn0.59 & \phnn$3.38\pm0.55$     & $4.52\times2.88$, 312
    && --    & --    && \phn20 & \phnn4.86 & \phn1.11 & \phn1.96 
    && I(NW) & \phn86 & \\
 13 & 17:20:50.784 & $-$35:46:21.00 & \phnn5.88 & \phn$44.77\pm6.79$     & $8.04\times5.90$, 324
    && --    & f$^*$  && \phn50 & \phn24.14 & \phn0.80 & \phn2.67 
    && E     & \phn86 & \\
 14 & 17:20:50.808 & $-$35:46:28.92 & \phnn2.13 & \phnn$6.51\pm1.01$     & $2.74\times2.56$, \phn31
    && --    & f$^*$  && \phn50 & \phnn3.51 & \phn2.05 & \phn2.64 
    && E     & \phn86 & \\
 15 & 17:20:50.904 & $-$35:46:05.16 & \phnn3.76 & \phnn$6.39\pm0.99$     & $2.64\times1.84$, \phn46
    && 9--100 & f$^*$  && \phn50 & \phnn3.45 & \phn3.46 & \phn3.72 
    && E     & \phn86 & \\
 16 & 17:20:50.976 & $-$35:44:14.64 & \phnn4.80 & \phnn$6.10\pm0.94$     & $1.88\times1.76$, 301
    && --    & --    && \phn20 & \phnn8.77 & \phn15.66    & 13.88    
    && I(NW) & \phn86 & \\
 17 & 17:20:51.000 & $-$35:45:05.04 & \phnn2.68 & \phnn$5.32\pm0.83$     & $3.12\times1.82$, \phnn4
    && --    & --    && \phn20 & \phnn7.65 & \phn5.98 & \phn6.98 
    && I(NW) & \phn86 & \\
 18 & 17:20:51.048 & $-$35:47:25.80 & \phnn1.55 & \phnn$1.97\pm0.32$     & $1.68\times1.58$, \phn69
    && --    & --    && \phn20 & \phnn2.83 & \phn7.17 & \phn5.66 
    && I     & \phn86 & \\
 19 & 17:20:51.072 & $-$35:44:30.12 & \phnn1.87 & \phnn$3.56\pm0.56$     & $2.30\times1.92$, \phn81
    && --    & --    && \phn20 & \phnn5.12 & \phn6.01 & \phn6.13 
    && I(NW) & \phn86 & \\
 20 & 17:20:51.096 & $-$35:44:22.20 & \phnn0.50 & \phnn$0.29\pm0.05$     & $1.02\times0.80$, 345
    && --    & --    && \phn20 & \phnn0.42 & \phn5.95 & \phn2.64 
    && I(NW) & \phn86 & \\
 21 & 17:20:51.120 & $-$35:46:30.36 & \phnn1.31 & \phnn$2.19\pm0.35$     & $3.14\times1.22$, \phnn6
    && --    & --    && \phn20 & \phnn3.15 & \phn4.67 & \phn4.41 
    && E     & \phn86 & \\
 22 & 17:20:51.288 & $-$35:45:07.56 & \phnn1.89 & \phnn$5.05\pm0.80$     & $4.44\times1.76$, \phnn2
    && --    & --    && \phn20 & \phnn7.26 & \phn3.57 & \phn4.87 
    && I(NW) & \phn86 & \\
 23 & 17:20:51.672 & $-$35:44:17.52 & \phnn0.39 & \phnn$1.73\pm0.29$     & $4.48\times1.54$, 354
    && --    & --    && \phn20 & \phnn2.49 & \phn1.51 & \phn1.92 
    && I(NW) & \phn86 & \\
 24 & 17:20:51.840 & $-$35:46:12.36 & \phnn1.57 & \phnn$3.63\pm0.57$     & $2.48\times1.86$, \phn73
    && --    & --    && \phn20 & \phnn5.22 & \phn5.57 & \phn5.86 
    && E     & \phn86 & \\
 25 & 17:20:51.888 & $-$35:44:22.56 & \phnn0.73 & \phnn$1.87\pm0.30$     & $2.38\times1.52$, \phn89
    && --    & --    && \phn20 & \phnn2.69 & \phn4.15 & \phn3.86 
    && I(NW) & \phn86 & \\
 26 & 17:20:51.936 & $-$35:44:19.32 & \phnn1.09 & \phnn$4.27\pm0.67$     & $5.00\times1.96$, \phn69
    && --    & --    && \phn20 & \phnn6.14 & \phn2.18 & \phn3.31 
    && I(NW) & \phn86 & \\
 27 & 17:20:52.032 & $-$35:44:12.12 & \phnn1.13 & \phnn$1.49\pm0.24$     & $1.82\times1.34$, \phnn2
    && --    & --    && \phn20 & \phnn2.14 & \phn5.81 & \phn4.48 
    && I(NW) & \phn86 & \\
 28 & 17:20:52.320 & $-$35:44:21.12 & \phnn3.40 & \phnn$3.65\pm0.57$     & $1.66\times1.54$, \phnn2
    && --    & --    && \phn20 & \phnn5.25 & 14.33    & 11.03    
    && I(NW) & \phn86 & \\
 29 & 17:20:52.512 & $-$35:44:10.68 & \phnn0.67 & \phnn$0.46\pm0.08$     & $1.22\times0.82$, \phnn6
    && --    & --    && \phn20 & \phnn0.66 & \phn7.18 & \phn3.48 
    && I(NW) & \phn86 & \\
 30 & 17:20:52.752 & $-$35:43:58.44 & \phnn1.02 & \phnn$0.99\pm0.17$     & $1.34\times1.28$, \phn52
    && --    & --    && \phn20 & \phnn1.42 & \phn6.64 & \phn4.27 
    && I(NW) & \phn86 & \\
 31 & 17:20:52.752 & $-$35:47:04.92 & \phnn0.76 & \phnn$2.85\pm0.46$     & $2.48\times1.94$, 350
    && --    & --    && \phn20 & \phnn4.10 & \phn3.46 & \phn3.94 
    && I     & 100    & \\
 32 & 17:20:52.752 & $-$35:47:07.80 & \phnn0.68 & \phnn$5.32\pm0.84$     & $3.66\times2.70$, \phn52
    && --    & --    && \phn20 & \phnn7.65 & \phn2.64 & \phn4.05 
    && I     & 100    & \\
 33 & 17:20:52.800 & $-$35:43:51.24 & \phnn1.15 & \phnn$1.15\pm0.19$     & $1.44\times1.28$, \phn14
    && --    & --    && \phn20 & \phnn1.65 & \phn7.08 & \phn4.69 
    && I(NW) & \phn86 & \\
 34 & 17:20:52.848 & $-$35:47:06.72 & \phnn0.86 & \phnn$1.60\pm0.26$     & $1.98\times1.58$, 317
    && --    & --    && \phn20 & \phnn2.30 & \phn4.32 & \phn3.76 
    && I     & 100    & \\
 35 & 17:20:52.848 & $-$35:47:04.56 & \phnn0.99 & \phnn$1.86\pm0.30$     & $1.98\times1.36$, \phn59
    && --    & --    && \phn20 & \phnn2.68 & \phn6.62 & \phn5.27 
    && I     & 100    & \\
 36 & 17:20:52.992 & $-$35:44:06.72 & \phnn0.53 & \phnn$0.13\pm0.03$     & $0.60\times0.56$, \phn90
    && --    & --    && \phn20 & \phnn0.19 & 10.48    & \phn2.96 
    && I(NW) & \phn86 & \\
 37 & 17:20:53.040 & $-$35:44:23.64 & \phnn2.31 & \phnn$4.19\pm0.66$     & $2.54\times1.98$, 358
    && --    & --    && \phn20 & \phnn6.03 & \phn5.76 & \phn6.29 
    && I(NW) & \phn86 & \\
 38 & 17:20:53.088 & $-$35:47:03.12 & \phn30.74 & \phn$56.31\pm8.50$     & $3.24\times3.06$, \phn34
    && --    & --    && \phn20 & \phn80.99 & 28.15    & 43.07    
    && I     & 100    & \scriptsize{MM4} \\
 39 & 17:20:53.136 & $-$35:44:24.00 & \phnn1.41 & \phnn$1.55\pm0.25$     & $1.98\times1.30$, \phn11
    && --    & --    && \phn20 & \phnn2.23 & \phn5.85 & \phn4.56 
    && I(NW) & \phn86 & \\
 40 & 17:20:53.160 & $-$35:47:13.56 & \phnn1.68 & \phnn$4.94\pm0.77$     & $3.14\times2.3$, \phn64
    && --    & f$^*$  && \phn50 & \phnn2.66 & \phn1.48 & \phn1.93 
    && I     & 100    & \\
 41 & 17:20:53.160 & $-$35:47:29.04 & \phnn1.61 & \phnn$1.99\pm0.32$     & $1.64\times1.52$, 388
    && --    & --    && \phn20 & \phnn2.86 & \phn7.81 & \phn6.01
    && I     & 100    & \\
 42 & 17:20:53.160 & $-$35:43:51.96 & \phnn0.46 & \phnn$0.44\pm0.08$     & $1.28\times1.14$, 314
    && --    & --    && \phn20 & \phnn0.63 & \phn3.74 & \phn2.22
    && I(NW) & \phn86 & \\
 43 & 17:20:53.184 & $-$35:46:59.16 & \phn21.42 & \phn$102.6\pm15.5$     & $6.00\times4.10$, 276
    && --    & --    && 100    & \phn27.07 & \phn2.37 & \phn5.74
    && I     & 100    & \scriptsize{MM2} \\
 44 & 17:20:53.424 & $-$35:46:57.72 & 144.17    & \phn$384.3\pm57.7$     & $3.72\times2.48$, \phn70
    && 0--35  & --    && 100    & 101.40    & 38.88    & 57.57    
    && I     & 100    & \scriptsize{MM1} \\
 45 & 17:20:53.472 & $-$35:45:08.64 & \phnn0.93 & \phnn$1.49\pm0.24$     & $2.48\times2.06$, 355
    && --    & --    && \phn20 & \phnn2.14 & \phn5.29 & \phn4.21 
    && I(N)  & 100    & \\
 46 & 17:20:53.500 & $-$35:47:02.01 & 340.20    & \phn$2073\pm311$       & $3.86\times2.70$, \phn31
    && 100    & f$^{*}$    && -- & -- & --  & -- 
    && I  & 100    & \scriptsize{MM3} \\
 47 & 17:20:53.616 & $-$35:46:54.84 & \phnn2.86 & \phnn$4.95\pm0.77$     & $2.84\times1.76$, 321
    && --    & --    && \phn20 & \phnn7.12 & \phn6.85 & \phn7.46 
    && I     & 100    & \scriptsize{MM7} \\
 48 & 17:20:53.664 & $-$35:45:38.16 & \phnn0.65 & \phnn$0.85\pm0.15$     & $1.76\times1.12$, \phnn2
    && --    & YSO~II && \phn50 & \phnn0.46 & \phn1.79 & \phn1.22 
    && I(N)  & 100    & \\
 49 & 17:20:53.784 & $-$35:44:46.32 & \phnn4.65 & \phnn$9.33\pm1.44$     & $2.90\times2.28$, 299
    && --    & --    && \phn20 & \phn13.42 & \phn8.39 & 10.56    
    && I(NW) & \phn71 & \scriptsize{SMA8} \\
 50 & 17:20:53.880 & $-$35:46:59.16 & \phnn2.89 & \phnn$5.10\pm0.80$     & $2.62\times1.86$, 307
    && --    & f$^*$  && \phn50 & \phnn2.75 & \phn2.73 & \phn2.95 
    && I     & 100    & \\
 51 & 17:20:53.904 & $-$35:44:48.12 & \phnn1.00 & \phnn$1.63\pm0.26$     & $1.90\times1.52$, \phn14
    && --    & --    && \phn20 & \phnn2.34 & \phn5.20 & \phn4.29 
    && I(NW) & \phn71 & \\
 52 & 17:20:53.976 & $-$35:45:47.52 & \phnn2.78 & \phnn$3.55\pm0.55$     & $1.90\times1.68$, \phn29
    && --    & YSO~I  && \phn50 & \phnn1.91 & \phn3.54 & \phn3.10 
    && I(N)  & 100    & \\
 53 & 17:20:54.048 & $-$35:45:51.48 & \phnn0.54 & \phnn$0.58\pm0.11$     & $1.98\times0.94$, 340
    && --    & --    && \phn20 & \phnn0.83 & \phn3.46 & \phn2.31 
    && I(N)  & 100    & \\
 54 & 17:20:54.144 & $-$35:44:38.76 & \phnn2.11 & \phnn$4.18\pm0.66$     & $2.70\times1.86$, \phn53
    && --    & --    && \phn20 & \phnn6.01 & \phn5.90 & \phn6.38 
    && I(N)  & \phn71 & \\
 55 & 17:20:54.168 & $-$35:45:41.40 & \phn12.11 & \phn$26.10\pm3.98$     & $4.86\times2.56$, \phn26
    && --    & --    && \phn20 & \phn37.54 & \phn9.13 & 15.74    
    && I(N)  & 100    & \scriptsize{SMA9} \\
 56 & 17:20:54.288 & $-$35:45:09.36 & \phnn0.79 & \phnn$3.78\pm0.60$     & $6.00\times2.04$, \phn52
    && --    & --    && \phn20 & \phnn5.44 & \phn1.36 & \phn2.33 
    && I(N)  & 100    & \\
 57 & 17:20:54.360 & $-$35:45:36.36 & \phnn0.46 & \phnn$2.25\pm0.37$     & $3.72\times2.00$, 333
    && --    & --    && \phn20 & \phnn3.24 & \phn1.72 & \phn2.28 
    && I(N)  & 100    & \\
 58 & 17:20:54.360 & $-$35:47:09.96 & \phnn2.20 & \phnn$2.69\pm0.42$     & $3.24\times1.18$, \phn12
    && --    & f$^*$  && \phn50 & \phnn1.45 & \phn2.14 & \phn2.02   
    && I     & 100    & \scriptsize{SMA10} \\
 59 & 17:20:54.384 & $-$35:45:06.12 & \phnn1.35 & \phnn$3.39\pm0.54$     & $2.50\times2.04$, 325
    && --    & --    && \phn20 & \phnn4.88 & \phn4.50 & \phn4.97   
    && I(N)  & 100    & \\
 60 & 17:20:54.456 & $-$35:45:20.88 & \phnn1.30 & \phnn$0.39\pm0.07$     & $0.94\times0.44$, 321
    && --    & --    && \phn20 & \phnn0.56 & 21.92    & \phn6.94
    && I(N)  & 100    & \\
 61 & 17:20:54.480 & $-$35:47:09.96 & \phnn2.69 & \phn$10.06\pm1.55$     & $4.88\times2.36$, 350
    && --    & --  && \phn20 & \phn14.47 & \phn3.98 & \phn6.58
    && I     & 100    & \\
 62 & 17:20:54.480 & $-$35:45:42.48 & \phnn0.82 & \phnn$4.42\pm0.70$     & $6.52\times2.28$, 319
    && --    & --    && \phn20 & \phnn6.36 & \phn1.19 & \phn2.24
    && I(N)  & \phn86 & \\
 63 & 17:20:54.552 & $-$35:45:29.52 & \phnn3.32 & \phn$12.09\pm1.86$     & $3.34\times2.46$, 319
    && --    & --    && \phn20 & \phn17.39 & \phn3.90 & \phn6.90
    && I(N)  & 100    & \\
 64 & 17:20:54.576 & $-$35:45:18.00 & \phn20.23 & \phn$53.91\pm8.15$     & $5.86\times2.62$, 300
    && --    & --    && \phn95 & \phn14.99 & \phn2.63 & \phn5.06
    && I(N)  & 100    & \scriptsize{SMA6} \\
 65 & 17:20:54.576 & $-$35:47:03.12 & \phnn1.13 & \phnn$2.71\pm0.43$     & $2.82\times1.72$, \phn51
    && --    & f$^*$  && \phn50 & \phnn1.46 & \phn1.45 & \phn1.57
    && I     & 100    & \\
 66 & 17:20:54.600 & $-$35:45:35.64 & \phnn0.57 & \phnn$0.87\pm0.15$     & $1.64\times1.26$, 346
    && --    & --    && \phn20 & \phnn1.25 & \phn4.42 & \phn3.12
    && I(N)  & 100    & \\
 67 & 17:20:54.648 & $-$35:45:08.64 & \phnn5.68 & \phn$32.37\pm4.92$     & $6.48\times4.10$, 283
    && 1--100 & --    && 208    & \phnn4.06 & \phn0.32 & \phn0.80
    && I(N)  & 100    & \scriptsize{SMA4} \\
 68 & 17:20:54.648 & $-$35:45:51.48 & \phnn1.28 & \phnn$0.95\pm0.15$     & $1.40\times1.02$, 355
    && --    & --    && \phn20 & \phnn1.37 & \phn8.38 & \phn4.93
    && I(N)  & 100    & \\
 69 & 17:20:54.648 & $-$35:47:03.12 & \phnn1.17 & \phnn$2.05\pm0.33$     & $2.36\times1.32$, \phn47
    && --    & f$^*$  && \phn50 & \phnn1.11 & \phn2.17 & \phn1.87
    && I     & 100    & \\
 70 & 17:20:54.696 & $-$35:45:26.28 & \phnn1.04 & \phnn$4.43\pm0.69$     & $4.08\times1.96$, 297
    && --    & --    && \phn20 & \phnn6.37 & \phn3.03 & \phn4.17
    && I(N)  & 100    & \\
 71 & 17:20:54.720 & $-$35:45:25.20 & \phnn0.82 & \phnn$0.47\pm0.08$     & $1.52\times0.62$, 320
    && --    & --    && \phn20 & \phnn0.68 & \phn8.39 & \phn3.91
    && I(N)  & 100    & \\
 72 & 17:20:54.744 & $-$35:45:20.16 & \phnn3.70 & \phn$20.23\pm3.09$     & $5.44\times3.86$, 328
    && --    & --    && \phn20 & \phn29.10 & \phn3.27 & \phn7.30
    && I(N)  & 100    & \scriptsize{SMA11} \\
 73 & 17:20:54.792 & $-$35:44:29.40 & \phnn2.38 & \phnn$3.00\pm0.47$     & $1.82\times1.60$, 341
    && --    & --    && \phn20 & \phnn4.31 & \phn9.17 & \phn7.67
    && I(NW) & \phn71 & \\
 74 & 17:20:54.816 & $-$35:46:51.24 & \phnn1.00 & \phnn$2.80\pm0.45$     & $2.68\times1.76$, 331
    && --    & f$^*$  && \phn50 & \phnn1.51 & \phn1.58 & \phn1.68
    && I     & 100    & \\
 75 & 17:20:54.840 & $-$35:45:33.48 & \phnn4.60 & \phnn$6.35\pm0.98$     & $2.08\times1.98$, \phn31
    && --    & f$^*$  && \phn50 & \phnn3.42 & \phn4.46 & \phn4.39
    && I(N)  & 100    & \scriptsize{SMA12} \\
 76 & 17:20:54.888 & $-$35:45:11.16 & \phnn1.18 & \phnn$5.95\pm0.93$     & $3.18\times2.66$, \phn44
    && --    & --    && \phn20 & \phnn8.56 & \phn3.82 & \phn5.37
    && I(N)  & 100    & \\
 77 & 17:20:54.888 & $-$35:45:01.80 & \phnn1.53 & \phnn$3.12\pm0.49$     & $2.48\times1.62$, \phn46
    && --    & --    && \phn20 & \phnn4.49 & \phn5.88 & \phn5.78
    && I(N)  & 100    & \\
 78 & 17:20:54.888 & $-$35:45:06.48 & \phnn9.80 & \phn$19.15\pm2.90$     & $2.60\times2.18$, \phn28
    && --    & --    && \phn20 & \phn27.54 & 22.23    & 25.69
    && I(N)  & 100    & \scriptsize{SMA2} \\
 79 & 17:20:54.912 & $-$35:45:16.56 & \phnn7.34 & \phn$18.01\pm2.74$     & $3.46\times2.86$, \phn87
    && --    & --    && \phn20 & \phn25.90 & \phn8.92 & 13.69
    && I(N)  & 100    & \scriptsize{SMA13} \\
 80 & 17:20:54.912 & $-$35:43:43.32 & \phnn0.58 & \phnn$0.14\pm0.03$     & $0.74\times0.42$, \phn45
    && --    & f$^*$  && \phn50 & \phnn0.08 & \phn4.82 & \phn1.32
    && I(NW) & \phn86 & \\
 81 & 17:20:54.912 & $-$35:45:03.24 & \phnn1.55 & \phnn$1.26\pm0.20$     & $1.38\times0.88$, 270
    && --    & --    && \phn20 & \phnn1.81 & 14.22    & \phn7.70
    && I(N)  & 100    & \\
 82 & 17:20:54.960 & $-$35:45:46.80 & \phnn0.97 & \phnn$1.34\pm0.22$     & $2.16\times1.22$, 306
    && --    & --    && \phn20 & \phnn1.93 & \phn4.73 & \phn3.77
    && I(N)  & 100    & \\
 83 & 17:20:54.960 & $-$35:47:11.04 & \phnn0.84 & \phnn$1.89\pm0.30$     & $2.06\times1.72$, \phn25
    && --    & --    && \phn20 & \phnn2.72 & \phn4.32 & \phn3.98
    && I     & 100    & \\
 84 & 17:20:54.984 & $-$35:45:07.20 & \phnn9.31 & \phn$26.46\pm4.01$     & $7.34\times3.84$, \phn89
    && --    & --    && \phn20 & \phn38.06 & \phn2.72 & \phn7.05
    && I(N)  & 100    & \scriptsize{SMA3} \\
 85 & 17:20:54.984 & $-$35:44:57.84 & \phnn1.61 & \phnn$8.35\pm1.29$     & $3.74\times3.14$, \phn24
    && --    & --    && \phn20 & \phn12.01 & \phn3.15 & \phn5.29
    && I(N)  & 100    & \\
 86 & 17:20:55.056 & $-$35:45:37.44 & \phnn1.15 & \phnn$6.49\pm1.02$     & $5.76\times2.92$, 347
    && --    & --    && \phn20 & \phnn9.33 & \phn1.45 & \phn2.91
    && I(N)  & 100    & \\
 87 & 17:20:55.056 & $-$35:45:01.44 & \phnn8.47 & \phn$30.40\pm4.61$     & $4.94\times3.88$, \phn29
    && --    & --    && \phn20 & \phn43.72 & \phn5.56 & 11.90
    && I(N)  & 100    & \scriptsize{SMA5} \\
 88 & 17:20:55.080 & $-$35:45:41.40 & \phnn1.37 & \phnn$3.05\pm0.49$     & $3.2\times1.82$, \phn66
    && --    & --    && \phn20 & \phnn4.39 & \phn3.48 & \phn4.04
    && I(N)  & 100    & \\
 89 & 17:20:55.104 & $-$35:45:16.20 & \phnn2.20 & \phnn$9.29\pm1.43$     & $3.68\times2.70$, \phn87
    && --    & --    && \phn20 & \phn13.36 & \phn4.53 & \phn6.99
    && I(N)  & 100    & \\
 90 & 17:20:55.104 & $-$35:45:37.08 & \phnn1.14 & \phnn$4.19\pm0.66$     & $3.74\times2.18$, 335
    && --    & --    && \phn20 & \phnn6.03 & \phn2.77 & \phn3.86
    && I(N)  & 100    & \\
 91 & 17:20:55.128 & $-$35:45:05.76 & \phn13.80 & \phn$24.31\pm3.68$     & $2.72\times2.04$, 293
    && --    & --    && \phn20 & \phn34.97 & 28.75    & 33.01
    && I(N)  & 100    & \scriptsize{SMA1a} \\
 92 & 17:20:55.152 & $-$35:47:15.72 & \phnn0.61 & \phnn$1.13\pm0.19$     & $1.94\times1.46$, 298
    && --    & --    && \phn20 & \phnn1.63 & \phn3.54 & \phn2.94
    && I     & 100    & \\
 93 & 17:20:55.176 & $-$35:44:25.80 & \phnn0.97 & \phnn$4.64\pm0.74$     & $3.70\times2.68$, 330
    && --    & --    && \phn20 & \phnn6.67 & \phn2.28 & \phn3.50
    && I(NW) & \phn71 & \\
 94 & 17:20:55.200 & $-$35:44:30.12 & \phnn1.73 & \phnn$3.79\pm0.60$     & $2.34\times2.18$, 311
    && --    & f$^*$  && \phn50 & \phnn2.04 & \phn1.9 & \phn2.1
    && I(NW) & \phn71 & \\
 95 & 17:20:55.224 & $-$35:45:03.60 & \phn33.16 & \phn$144.2\pm21.7$     & $10.44\times5.00$, 305
    && 0--52  & --    && 143    & \phn26.44 & \phn0.76 & \phn2.67
    && I(N)  & 100    & \scriptsize{SMA1b/d} \\
 96 & 17:20:55.296 & $-$35:45:00.36 & \phnn1.19 & \phnn$2.75\pm0.43$     & $2.20\times1.80$, \phnn6
    && --    & --    && \phn20 & \phnn3.96 & \phn5.40 & \phn5.24
    && I(N)  & 100    & \\
 97 & 17:20:55.296 & $-$35:46:50.52 & \phnn0.70 & \phnn$2.91\pm0.48$     & $3.88\times2.46$, \phn70
    && --    & --    && \phn20 & \phnn4.19 & \phn1.51 & \phn2.28
    && I     & 100    & \\
 98 & 17:20:55.320 & $-$35:45:14.76 & \phnn1.29 & \phnn$6.42\pm1.00$     & $3.46\times2.74$, 283
    && --    & --    && \phn20 & \phnn9.23 & \phn3.39 & \phn5.09
    && I(N)  & 100    & \\
 99 & 17:20:55.344 & $-$35:44:31.20 & \phnn1.15 & \phnn$2.45\pm0.39$     & $2.34\times1.84$, \phn30
    && --    & --    && \phn20 & \phnn3.52 & \phn4.18 & \phn4.25
    && I(NW) & \phn71 & \\
100 & 17:20:55.368 & $-$35:46:09.12 & \phnn1.06 & \phnn$9.06\pm1.41$     & $5.16\times3.10$, 351
    && --    & --    && \phn20 & \phn13.03 & \phn2.17 & \phn4.24
    && I(N)  & 100    & \\
101 & 17:20:55.392 & $-$35:46:13.08 & \phnn1.01 & \phnn$2.51\pm0.41$     & $1.98\times1.78$, 320
    && --    & f$^*$  && \phn50 & \phnn1.35 & \phn2.12 & \phn1.96
    && I(N)  & 100    & \\
102 & 17:20:55.392 & $-$35:44:59.64 & \phnn0.92 & \phnn$0.34\pm0.06$     & $0.96\times0.56$, 351
    && --    & --    && \phn20 & \phnn0.49 & 13.36    & \phn4.77
    && I(N)  & 100    & \\
103 & 17:20:55.464 & $-$35:45:32.04 & \phnn2.85 & \phnn$7.00\pm1.09$     & $3.38\times2.26$, 310
    && --    & --    && \phn20 & \phn10.07 & \phn5.24 & \phn7.01
    && I(N)  & \phn86 & \\
104 & 17:20:55.488 & $-$35:45:11.16 & \phnn6.59 & \phn$11.59\pm1.78$     & $4.62\times1.84$, 321
    && --    & --    && \phn20 & \phn16.67 & \phn7.27 & 10.32
    && I(N)  & 100    & \\
105 & 17:20:55.536 & $-$35:45:24.48 & \phnn0.53 & \phnn$1.69\pm0.28$     & $2.74\times1.92$, 332
    && --    & --    && \phn72 & \phnn0.62 & \phn0.56 & \phn0.62
    && I(N)  & \phn86 & \scriptsize{SMA15} \\
106 & 17:20:55.536 & $-$35:45:26.64 & \phnn1.29 & \phnn$0.79\pm0.13$     & $1.26\times0.86$, \phn35
    && --    & --    && \phn20 & \phnn1.14 & 10.75    & \phn5.47
    && I(N)  & \phn86 & \\
107 & 17:20:55.584 & $-$35:44:56.40 & \phnn3.63 & \phn$40.86\pm6.22$     & $13.04\times8.40$, 334
    && --    & --    && \phn20 & \phn58.77 & \phn0.55 & \phn2.82
    && I(N)  & 100    & \scriptsize{SMA17} \\
108 & 17:20:55.608 & $-$35:45:20.88 & \phnn0.51 & \phnn$0.50\pm0.08$     & $1.42\times0.82$, 270
    && --    & --    && \phn20 & \phnn0.72 & \phn6.07 & \phn3.21
    && I(N)  & \phn86 & \\
109 & 17:20:55.608 & $-$35:45:27.00 & \phnn1.47 & \phn$11.45\pm1.77$     & $4.82\times3.76$, \phn50
    && --    & --    && \phn20 & \phn16.47 & \phn2.31 & \phn4.78
    && I(N)  & \phn86 & \\
110 & 17:20:55.680 & $-$35:45:32.40 & \phnn3.77 & \phn$15.89\pm2.43$     & $4.86\times3.08$, 320
    && --    & --    && 139    & \phnn3.00 & \phn0.56 & \phn1.05
    && I(N)  & \phn86 & \scriptsize{SMA18} \\
111 & 17:20:55.680 & $-$35:45:51.12 & \phnn0.83 & \phnn$1.56\pm0.25$     & $2.00\times1.46$, \phn64
    && --    & f$^*$  && \phn50 & \phnn0.84 & \phn1.85 & \phn1.53
    && I(N)  & 100    & \\
112 & 17:20:55.680 & $-$35:45:08.64 & \phnn1.07 & \phnn$1.47\pm0.24$     & $1.72\times1.34$, 349
    && --    & --    && \phn20 & \phnn2.11 & \phn6.63 & \phn4.87
    && I(N)  & 100    & \\
113 & 17:20:55.704 & $-$35:45:23.76 & \phnn0.67 & \phnn$0.66\pm0.11$     & $1.38\times1.02$, \phn78
    && --    & --    && \phn20 & \phnn0.95 & \phn6.44 & \phn3.66
    && I(N)  & \phn86 & \\
114 & 17:20:55.800 & $-$35:44:43.08 & \phnn0.86 & \phnn$1.29\pm0.21$     & $1.84\times1.36$, 359
    && --    & f$^*$  && \phn50 & \phnn0.70 & \phn1.90 & \phn1.47
    && I(N)  & \phn86 & \\
115 & 17:20:55.848 & $-$35:45:27.00 & \phnn2.94 & \phnn$5.11\pm0.80$     & $3.06\times1.86$, 284
    && --    & --    && \phn20 & \phnn7.35 & \phn5.67 & \phn6.65
    && I(N)  & \phn86 & \scriptsize{SMA19} \\
116 & 17:20:55.848 & $-$35:45:24.84 & \phnn1.58 & \phnn$4.27\pm0.67$     & $3.08\times2.20$, 354
    && --    & --    && \phn20 & \phnn6.14 & \phn3.74 & \phn4.75
    && I(N)  & \phn86 & \\
117 & 17:20:55.968 & $-$35:45:23.40 & \phnn1.50 & \phnn$2.57\pm0.41$     & $1.84\times1.72$, \phn87
    && --    & --    && \phn20 & \phnn3.70 & \phn7.06 & \phn6.12
    && I(N)  & \phn86 & \\
118 & 17:20:55.992 & $-$35:45:35.64 & \phnn1.62 & \phnn$8.75\pm1.35$     & $3.32\times3.04$, \phn77
    && --    & --    && \phn20 & \phn12.59 & \phn4.22 & \phn6.53
    && I(N)  & \phn86 & \\
119 & 17:20:56.064 & $-$35:45:32.76 & \phnn3.57 & \phnn$6.97\pm1.08$     & $3.32\times2.08$, \phn88
    && --    & --    && \phn20 & \phn10.03 & \phn5.94 & \phn7.61
    && I(N)  & \phn86 & \scriptsize{SMA20} \\
120 & 17:20:56.088 & $-$35:45:21.60 & \phnn1.37 & \phnn$4.07\pm0.64$     & $3.60\times2.32$, \phn49
    && --    & --    && \phn20 & \phnn5.85 & \phn2.60 & \phn3.67
    && I(N)  & \phn86 & \\
121 & 17:20:56.088 & $-$35:45:23.04 & \phnn1.50 & \phnn$2.76\pm0.43$     & $2.32\times1.64$, \phn43
    && --    & f$^*$  && \phn50 & \phnn1.49 & \phn2.16 & \phn2.05
    && I(N)  & \phn86 & \\
122 & 17:20:56.184 & $-$35:45:19.80 & \phnn0.94 & \phnn$2.50\pm0.40$     & $2.36\times1.64$, 307
    && --    & --    && \phn20 & \phnn3.60 & \phn3.56 & \phn3.84
    && I(N)  & \phn86 & \\
123 & 17:20:56.304 & $-$35:45:27.36 & \phnn6.64 & \phnn$9.67\pm1.48$     & $2.98\times1.72$, \phn53
    && --    & --    && \phn20 & \phn13.91 & 12.75    & 14.12
    && I(N)  & \phn86 & \scriptsize{SMA21} \\
124 & 17:20:56.328 & $-$35:45:30.96 & \phnn1.29 & \phnn$3.83\pm0.60$     & $3.02\times2.18$, 307
    && --    & --    && \phn20 & \phnn5.51 & \phn3.52 & \phn4.40
    && I(N)  & \phn86 & \\
125 & 17:20:56.328 & $-$35:45:21.96 & \phnn1.51 & \phn$13.60\pm2.09$     & $5.92\times4.04$, 275
    && --    & --    && \phn20 & \phn19.56 & \phn1.80 & \phn4.28
    && I(N)  & \phn86 & \\
126 & 17:20:56.376 & $-$35:45:28.80 & \phnn0.94 & \phnn$0.76\pm0.12$     & $1.36\times0.84$, \phn87
    && --    & --    && \phn20 & \phnn1.09 & \phn9.59 & \phn5.00
    && I(N)  & \phn86 & \\
127 & 17:20:56.400 & $-$35:45:19.80 & \phnn0.57 & \phnn$2.55\pm0.41$     & $4.06\times1.92$, \phn25
    && --    & --    && \phn20 & \phnn3.67 & \phn1.80 & \phn2.45
    && I(N)  & \phn86 & \\
128 & 17:20:56.496 & $-$35:45:37.08 & \phnn1.51 & \phnn$1.38\pm0.23$     & $1.34\times1.26$, 344
    && --    & --    && \phn20 & \phnn1.98 & \phn9.69 & \phn6.14
    && I(N)  & \phn86 & \\
129 & 17:20:56.544 & $-$35:45:33.48 & \phnn0.73 & \phnn$1.71\pm0.28$     & $2.16\times1.60$, \phn21
    && --    & --    && \phn20 & \phnn2.46 & \phn4.11 & \phn3.73
    && I(N)  & \phn86 & \\    
130 & 17:20:56.568 & $-$35:45:26.64 & \phnn0.59 & \phnn$1.20\pm0.20$     & $1.86\times1.44$, \phn28
    && --    & --    && \phn20 & \phnn1.73 & \phn4.17 & \phn3.35
    && I(N)  & \phn86 & \\
131 & 17:20:56.568 & $-$35:45:17.28 & \phnn3.29 & \phnn$6.50\pm1.00$     & $2.68\times2.02$, 289
    && --    & --    && \phn20 & \phnn9.35 & \phn8.10 & \phn9.14
    && I(N)  & \phn86 & \scriptsize{SMA22} \\
132 & 17:20:56.616 & $-$35:45:10.08 & \phnn3.33 & \phnn$5.18\pm0.81$     & $2.30\times1.72$, 306
    && --    & --    && \phn20 & \phnn7.45 & 10.04    & \phn9.78
    && I(N)  & \phn86 & \scriptsize{SMA23} \\
133 & 17:20:56.640 & $-$35:45:18.72 & \phnn1.97 & \phnn$3.91\pm0.62$     & $2.40\times1.86$, 319
    && --    & --    && \phn20 & \phnn5.62 & \phn6.30 & \phn6.52
    && I(N)  & \phn86 & \\
134 & 17:20:56.736 & $-$35:45:23.40 & \phnn0.63 & \phnn$1.42\pm0.23$     & $2.32\times1.30$, 287
    && --    & --    && \phn20 & \phnn2.04 & \phn4.09 & \phn3.49
    && I(N)  & \phn86 & \\
135 & 17:20:56.736 & $-$35:45:17.64 & \phnn1.74 & \phnn$5.30\pm0.82$     & $3.02\times2.16$, \phn73
    && --    & --    && \phn20 & \phnn7.62 & \phn4.87 & \phn6.08
    && I(N)  & \phn86 & \\
136 & 17:20:56.952 & $-$35:45:18.00 & \phnn1.76 & \phnn$3.09\pm0.48$     & $2.30\times1.62$, 355
    && --    & f$^*$  && \phn50 & \phnn1.67 & \phn2.51 & \phn2.36
    && I(N)  & \phn86 & \\
137 & 17:20:56.976 & $-$35:45:16.20 & \phnn2.47 & \phnn$6.11\pm0.95$     & $3.38\times2.04$, 340
    && --    & --    && \phn20 & \phnn8.79 & \phn5.17 & \phn6.64
    && I(N)  & \phn86 & \\
138 & 17:20:57.264 & $-$35:45:17.28 & \phnn0.59 & \phnn$0.53\pm0.09$     & $1.36\times0.84$, 325
    && --    & --    && \phn20 & \phnn0.76 & \phn6.69 & \phn3.48
    && I(N)  & \phn86 & \\
139 & 17:20:57.336 & $-$35:45:14.40 & \phnn3.30 & \phnn$7.86\pm1.22$     & $3.54\times2.22$, 346
    && --    & --    && \phn20 & \phn11.31 & \phn5.40 & \phn7.43
    && I(N)  & \phn86 & \\
140 & 17:20:57.672 & $-$35:45:12.24 & \phnn1.96 & \phnn$4.22\pm0.66$     & $2.76\times1.86$, 308
    && --    & --    && \phn20 & \phnn6.07 & \phn5.70 & \phn6.26
    && I(N)  & \phn86 & \\
141 & 17:20:58.656 & $-$35:44:59.64 & \phnn0.43 & \phnn$0.27\pm0.05$     & $1.14\times0.86$, 336
    && --    & --    && \phn20 & \phnn0.39 & \phn4.59 & \phn2.17
    && I(N)  & \phn86 & \\
142 & 17:20:59.760 & $-$35:45:30.96 & \phnn0.85 & \phnn$0.56\pm0.09$     & $1.14\times0.96$, 347
    && --    & YSO~I  && \phn50 & \phnn0.30 & \phn2.89 & \phn1.46
    && I(N)  & \phn86 & \\
\hline
\end{longtable}
\tablefoot{
\tablefoottext{a}{Observed parameters of the ALMA compact sources identified as explained in \S~\ref{sec:source-identification}. In each column we have: R.A.\ (right ascension, column~2) in $^\mathrm{h}:^\mathrm{m}:^\mathrm{s}$; Dec.\ (declination, column~3) in $^\circ:^\prime:^{\prime\prime}$; $I_\nu$ (peak intensity at 87.6~GHz, column~4) in m\jpb; $S_\nu$ (flux density, column~5) in mJy; $\theta_\mathrm{s}$ (observed sizes of the major and minor axis obtained from the \texttt{SExtractor}) and P.A.\ (position angle, column~6) in arcsec$^2$ and degrees.} \\
\tablefoottext{b}{Information on the counterparts at different wavelengths. Radio continuum contribution ( free-free, column~7) to the ALMA 87.6~GHz continuum flux, determined from observations published in \citet{Hunter2014}, \citet{Brogan2016}, and \citet{Medina2018}. Association with infrared source (IR, column~8) from \citet{Willis2013}. f$^{*}$ denotes field stars.} \\
\tablefoottext{c}{Physical properties of the ALMA compact sources. In each column we have: $T$ (assumed temperature, column~9) in K; $M_\mathrm{g+d}$ (dust and gas mass, column~10) in \mo; $n_\mathrm{H_2}$ (volume density, column~11) in $10^8$~cm$^{-3}$; $N_\mathrm{H_2}$ (column density, column~12) in $10^{24}$~cm$^{-2}$.} \\
\tablefoottext{d}{Cluster membership for each ALMA compact sources: cluster to which the source is associated (column~13); and confidence level of membership in percentage (column~14, see \S~\ref{sec:cluster-identification}).} \\
\tablefoottext{e}{Other names for each source, from \citet{Brogan2016}, \citet{Hunter2014}.}
}
\end{landscape}}

%
\section{Clustering algorithms}\label{app:clustering}

%
\subsection{Technical details of cluster identification}\label{app:number-of-clusters}

In this section we use two different methods to determine the expected number of clusters existent in the catalogue of 142 ALMA compact sources of \ngc. In the first approach, known as the elbow-point analysis,  we run the K-Means algorithm for a range of $K$ values changing from 2 to 6 \citep{Macqueen1967,Mackay2003,Ivezic2014}. For each run, we calculate the within-cluster sum of squares (WSS). The WSS indicates the compactness of the clusters and tends to decrease as we increase $K$. The value at which the rate of decease shifts sharply (elbow-point) is taken as the number of clusters \citep{Charrad2014}. From this point on, adding another cluster does not improve the total WSS. As seen in Fig. B.1, the elbow-point corresponds to $K=4$. The value of $K$ corresponding to the elbow point is determined as the value with the larger separation to a straight line connecting the edges of the curve.\\
\indent In the second approach, known as Silhouette analysis, we run the K-Means algorithm for different numbers of clusters similar to the first approach. For each $K$, one Silhouette value is associated to each data point in a cluster, ranging from $-1$ to $1$. A high Silhouette value indicates dense and well configured clusters
\citep{Kaufman1990,Rousseeuw1987, Colombo2015}.
In order to find the optimal number of clusters, we compare the average Silhouette values for different number of clusters changing from 2 to 6. The configuration which has the highest average Silhouette value (depicted with a vertical dashed line in Fig.~\ref{fig:silhouette}) is the optimal one. As shown in Figure~\ref{fig:silhouette}, the highest average Silhouette value is obtained with $K=4$. The result is consistent with the outcome of  the first approach.


\begin{figure}[ht!]
\begin{center}
\includegraphics[width=1.0\columnwidth]{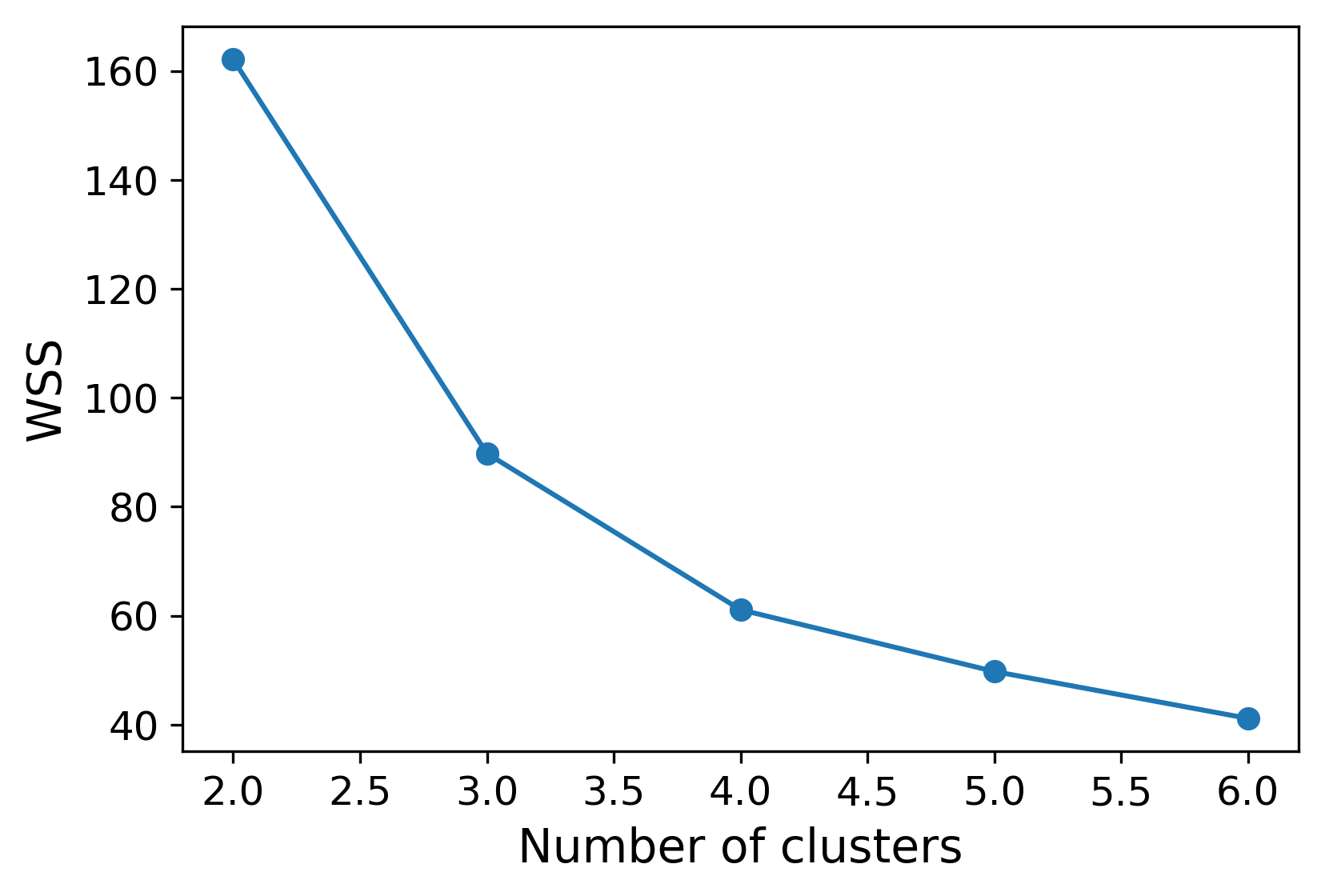}
\caption{Within-cluster sum of squares (WSS) or sum of squared distances of data points to the centre of a cluster for different number clusters. $K=4$ is coincident with the elbow point of the function and assumed to be associated with the optimal number of clusters. }
\label{fig:K-values}
\end{center}
\end{figure}

\begin{figure*}[t!]
\begin{center}
\includegraphics[scale=0.18]{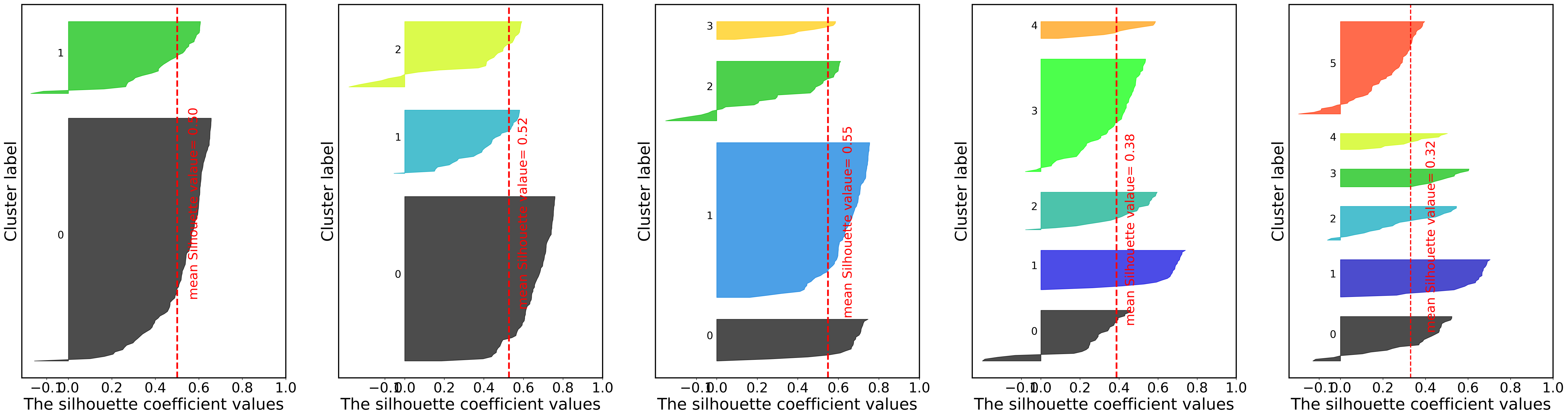}
\caption{Silhouette values for each data point for $K$ ranges from 2 to 6. The red dashed line denotes the average of the Silhouette values for each $K$. The values are 0.50 for $K=0$, 0.52 for $K=3$, 0.55 for $K=4$, 0.38 for $K=5$ and 0.32 for $K=6$. The highest mean value is obtained for $K=4$.}
\label{fig:silhouette}
\end{center}
\end{figure*}

%
\subsection{Description of different algorithms}\label{app:algorithms}

In the following we briefly describe the different algorithms that have been used to identify clusters in the observed region. The clustering algorithms are part of the machine-learning algorithms implemented in the \texttt{Python} package \texttt{scikit-learn}\footnote{See more information in the website \url{http://scikit-learn.org/stable/modules/clustering.html}}. 

\noindent
\textbf{K-means}: The procedure of the K-means algorithm is to randomly initialise the cluster centroids for a given number of clusters. In the next step, the algorithm goes through each of the data points and assigns them to the cluster that is the closest. Then, the average of all the points in the cluster will be calculated and the centroids will be moved to the average locations.

\noindent
\textbf{Mean-shift}: The steps which are followed by the Mean-Shift algorithm are similar to the K-Means method. The only difference is that instead of using the number of clusters as an input, the algorithm determines the probability density function from which the data is drawn. Then the centroids are placed at the maxima of the density function.

\noindent
\textbf{Spectral clustering}: The Spectral clustering algorithm makes use of the similarity matrix of the data to reduce the dimensionality of the space. Then, it is followed by the K-Means algorithm in the reduced dimensional space. 

\noindent
\textbf{Ward hierarchical and Agglomerative clustering}: The Ward hierarchical clustering and the Agglomerative clustering are the other two methods used to look for segmentation in the data. They try to link each data point by a distance measure to its nearest neighbour, creating a cluster. These two algorithms differ in the linkage method that they use to define the proximity between any two clusters. The Agglomerative hierarchical clustering uses the Average linkage which merges the clusters based on the minimum distance between two points in two different clusters. While, the Ward linkage computes the total WSS to determine the next two groups merged. 

\noindent
\textbf{Gaussian mixtures}: In practice each cluster can be mathematically represented by a parametric distribution like a Gaussian. Therefore, we can assign multivariant normal distribution components to the data points. Using the Gaussian mixture model algorithm each data point will be assigned to the component which yields the highest posterior probability. 

\noindent
\textbf{Birch algorithm}: The Birch algorithm operates in four phases. First, it builds a Characteristic Feature Tree for the data. In the second step, it reduces the input data to the subclusters and groups the crowded subclusters into a larger one. A branching factor limits the number of subclusters in each node. In step three, one of the other clustering algorithms (e.g. K-Means) is applied to the subclusters. A threshold gives us the flexibility to specify the diameter for the clusters. After this step, the major segmentations in the data are identified. In the last step, which is a refining step, the centroids of the clusters identified in step 3 are used as reference points to find the closest data points to them and define the new clusters. 

%
\section{Correlation between the mass segregation parameter $\Lambda_\mathrm{MSR}$ and the structural concentration of clusters}
\label{app:lambda-gaussian}

\begin{figure}[h!]
\begin{center}
\includegraphics[width=1.0\columnwidth]{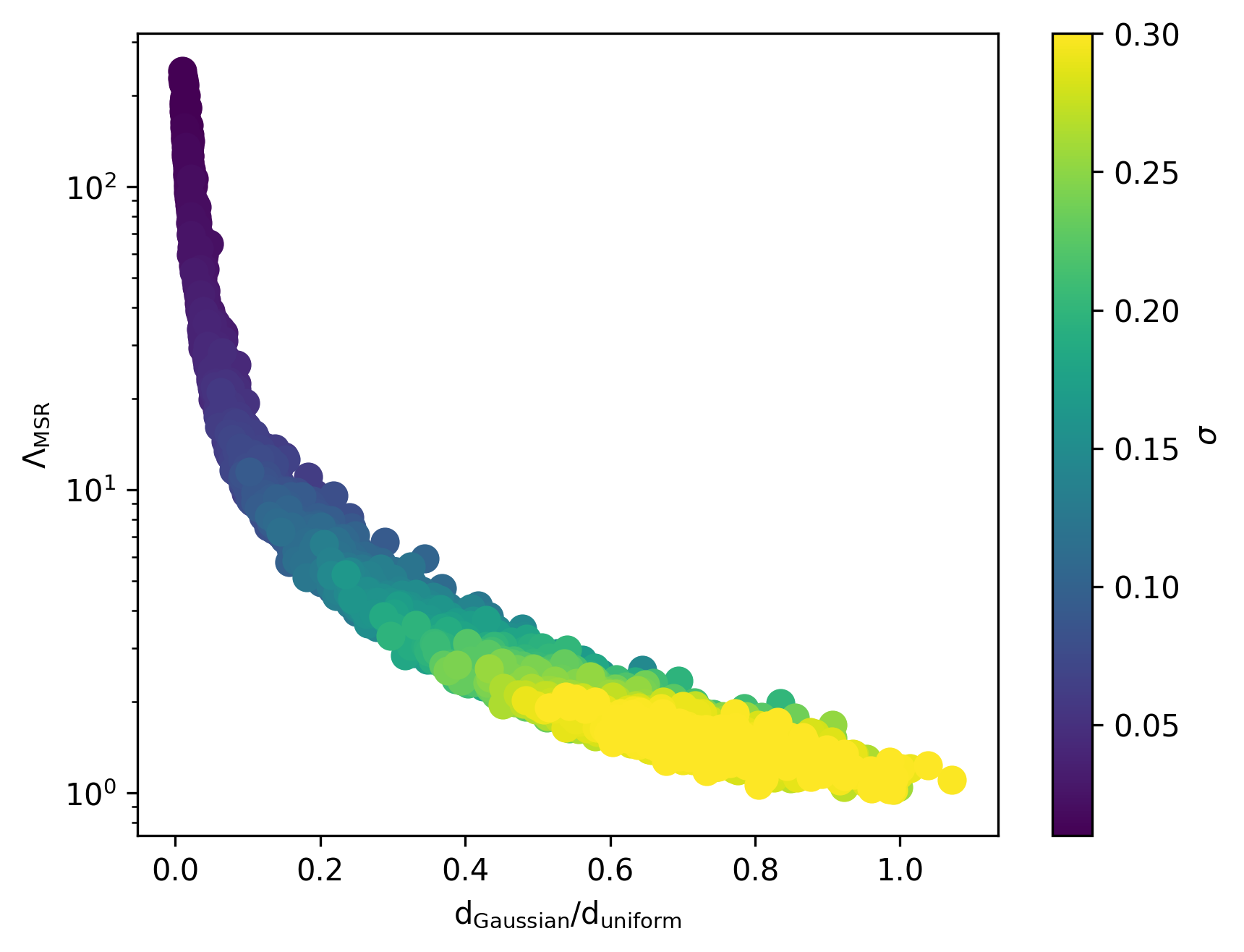}
\caption{Mass segregation parameter $\Lambda_\mathrm{MSR}$ against the size ratio of a Gaussian cluster compared to a uniform cluster. Violet to yellow circles
correspond to the randomly generated Gaussian clusters, each one defined by the Gaussian width $\sigma$ color coded in the color bar. Extended Gaussian clusters have a distribution of cores resembling that of a uniform cluster, and thus
$\Lambda_\mathrm{MSR}$ gets closer to 1.}
\end{center}
\label{fig:lambda-sigma}
\end{figure}

In this section, we investigate the correlation between the parameter $\Lambda_\mathrm{MSR}$ that has been used in \S~\ref{sec:mass-segregation} to study the mass segregation in the clusters, and the structural concentration of the members in the cluster. For this, we generate a series of clusters in which
the members follow a Gaussian distribution and compare them with a cluster with a uniform distribution. The concentration of the members in a Gaussian cluster is given by the width of the Gaussian $\sigma$. The larger the value of $\sigma$ is, the more dispersed the cores are (i.e. the Gaussian clusters resembles the uniform cluster).\\
\indent For each value of $\sigma$, we generate 50 Gaussian clusters containing 70 cores each. For each cluster, we run the MST analysis to measure the mean MST length scale. Then, we calculate the $\Lambda_\mathrm{MSR}$ parameter of the 70 cores, following Eq.~\ref{eq:lambdaMSR}. As the value of $\sigma$ increases, we expect that the distribution of cores in a Gaussian cluster becomes more similar to a cluster with uniformly distributed cores, i.e. the value of $\Lambda_\mathrm{MSR}$ approaches 1. In Fig.~\ref{fig:lambda-sigma}, we plot the value of $\Lambda_\mathrm{MSR}$ for each Gaussian cluster versus the ratio of its extent and the size of the uniform cluster. The size of the cluster is determined as done for the clusters in \ngc, i.e. as the distance between the farthest member to the geometrical centre of the cluster. As shown in Fig.~\ref{fig:lambda-sigma}, the parameter $\Lambda_\mathrm{MSR}$ converges to 1 as the Gaussian distribution becomes more disperse. This suggests that $\Lambda_\mathrm{MSR}$ is not only a measure of the mass segregation, but can also be indicative of the distribution and concentration of any number
of cores in the cluster. Constant values of $\Lambda_\mathrm{MSR}$ slightly above 1, as found in the clusters of \ngc, may indicate that the cores in the clusters are not uniformly distributed but slightly concentrated.
The correlation between the mass segregation ratio and the concentration of dense cores has also been recently reported by \citet{deVita2019}.
\end{appendix}
\end{document}